\documentclass[a4paper, 11pt, oneside]{Thesis}  
\graphicspath{Figures/}  
\usepackage[square, numbers, comma, sort&compress]{natbib}  
\usepackage{verbatim}  
\usepackage{vector}  
\hypersetup{urlcolor=blue, colorlinks=true}  

\usepackage{listings}
\usepackage{xcolor}

\definecolor{codegreen}{rgb}{0,0.6,0}
\definecolor{codegray}{rgb}{0.5,0.5,0.5}
\definecolor{codepurple}{rgb}{0.58,0,0.82}
\definecolor{backcolour}{rgb}{0.95,0.95,0.92}

\lstdefinestyle{mystyle}{
    backgroundcolor=\color{backcolour},   
    commentstyle=\color{codegreen},
    keywordstyle=\color{magenta},
    numberstyle=\tiny\color{codegray},
    stringstyle=\color{codepurple},
    basicstyle=\ttfamily\footnotesize,
    breakatwhitespace=false,         
    breaklines=true,                 
    captionpos=b,                    
    keepspaces=true,                 
    numbers=left,                    
    numbersep=5pt,                  
    showspaces=false,                
    showstringspaces=false,
    showtabs=false,                  
    tabsize=2
}

\usepackage{multirow}
\usepackage{multicol}
\usepackage{graphicx}
\usepackage[ruled]{algorithm}

\usepackage{svg}
\usepackage{hyperref}

\usepackage{amssymb}
\usepackage{pifont}
\newcommand{\cmark}{\ding{51}}%
\newcommand{\xmark}{\ding{55}}%
\begin{document}
\frontmatter      

\UNIVERSITY{{School of Computing and Information Systems \\ THE UNIVERSITY OF MELBOURNE }}    
%

%
\title  {Federated Learning Framework in Fogbus2-based Edge Computing Environments}

\date { 03 November 2022}
\subject    {Master of Computer Science}

\maketitle

\setstretch{1.3}  

\fancyhead{}  
\rhead{\thepage}  
\lhead{}  

\pagestyle{fancy}  


\addtotoc{Abstract}  
\section*{Abstract}
Federated learning refers to conducting training on multiple distributed devices and collecting model weights from them to derive a shared machine-learning model. This allows the model to get benefit from a rich source of data available from multiple sites. Also, since only model weights are collected from distributed devices, the privacy of those data is protected. It is useful in a situation where collaborative training of machine learning models is necessary while training data are highly sensitive. This study aims at investigating the implementation of lightweight federated learning to be deployed on a diverse range of distributed resources, including resource-constrained edge devices and resourceful cloud servers. As a resource management framework, the FogBus2 framework, which is a containerized distributed resource management framework, is selected as the base framework for the implementation. This research provides an architecture and lightweight implementation of federated learning in the FogBus2. Moreover, a worker selection technique is proposed and implemented. The worker selection algorithm selects an appropriate set of workers to participate in the training to achieve higher training time efficiency. Besides, this research integrates synchronous and asynchronous models of federated learning alongside with heuristic-based worker selection algorithm. It is proven that asynchronous federated learning is more time efficient compared to synchronous federated learning or sequential machine learning training. The performance evaluation shows the efficiency of the federated learning mechanism implemented and integrated with the FogBus2 framework. The worker selection strategy obtains 33.9\% less time to reach 80\% accuracy compared to sequential training, while asynchronous further improve synchronous federated learning training time by 63.3\%.
\clearpage  

%
%
\Declaration{

\addtocontents{toc}{\vspace{1em}}  

I, Wuji Zhu, declare that this thesis titled, `Federated Learning Framework in Fogbus2-based Edge Computing Environments' and the work presented in it are my own. I confirm that:

\begin{itemize} 
\item[\tiny{$\blacksquare$}]  This thesis does not incorporate without acknowledgement any material previously submitted for degree or diploma in any university; and that to the best of my knowledge and belief it does not contain any material previously published or written by another person where due reference is not made in the text.

\item[\tiny{$\blacksquare$}] Clearance for this research from the University’s Ethics Committee was not required.

\item[\tiny{$\blacksquare$}] The thesis is 25744 words in length (fewer than the maximum word limit in length, exclusive of tables, maps, bibliographies and appendices as approved by the Research Higher Degrees Committee).
\\
\end{itemize}

Signed:\includegraphics[scale=0.4]{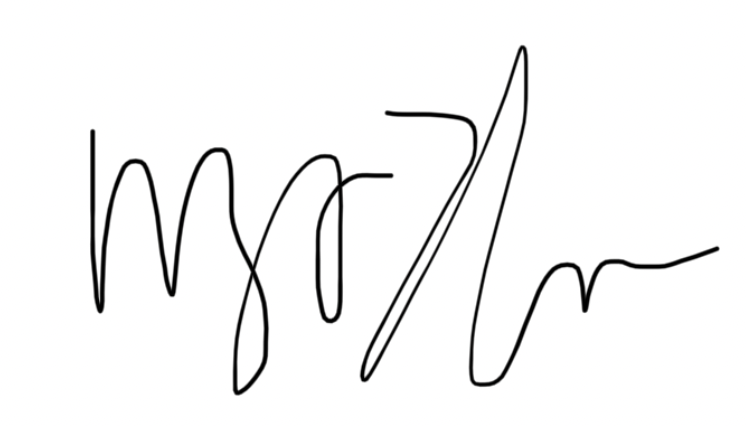}
\\ \rule[1em]{25em}{0.5pt}  
 
Date:      03 November 2022\\
\rule[1em]{25em}{0.5pt}  
}
\clearpage  


\setstretch{1.3}  
\acknowledgements{

I would like to thank my supervisor: Prof. Rajkumar Buyya, who introduced me to the interesting federated learning topic and provided valuable guidance for the research.

I also want to thank Mohammad Goudarzi, who is super experienced in research and provided me with many valuable suggestions to improve the study. He also provides guidance which I genuinely appreciate.

Lastly, I would like to thank my partner and my parents support me during the whole year's research period. They always encourage me when I face difficulties in research and I truly appreciate their support.
}
\clearpage  

\pagestyle{fancy}  

\tableofcontents  

\lhead{\emph{List of Figures}}  
\listoffigures  

\lhead{\emph{List of Tables}}  
\listoftables  

\setstretch{1.5}  
%

%


\mainmatter	  
\pagestyle{fancy}  


\lhead{\emph{Introduction}}  
\chapter{Introduction}
\section{Background of the research problem}

\subsection{Machine learning: the important role of data}
As part of artificial intelligence, machine learning builds prediction models based on historical data. Applications driven by machine learning include but are not limited to decision-making, speech recognition, natural language processing, computer vision, and recommendation systems \cite{ORI_1,ORI_2,ORI_3}. A large amount of data is the critical factor and foundation for the success of machine learning applications addressed above \cite{ORI_1, ORI_4}. For example, a very deep neural network for image classification, which is known as VGG \cite{ORI_5}, was trained on 1.2 million images with one thousand categories and led to extraordinary performance. Robust natural language processing model Generative Pre-trained Transformer 2 (GPT-2) utilized data from 8 million web pages \cite{ORI_6}. The access to a large amount of data provides GPT-2 with powerful language interpretation ability, enabling it to extend to language translators, text generators, chatting bots, and more. Recommendation systems rely heavily on the vast number of user profiles available, so they can compare and group similar users and make personalized recommendations. Without a sufficient response from clients, machine learning models can only group customers with minimal similarity, reducing the recommendation quality. The examples above suggest that strong machine learning models rely on a rich amount of data. In order to support the development of a more powerful machine learning model, accessing more data is crucial and inevitable.

\subsection{Hurdles when accessing data}
Although exploding amounts of data are continuously being generated from different sources, which provides opportunities for artificial intelligence applications, increasing awareness of data protection hinders the data collection process. General Data Protection Regulation (GDPR) \cite{ORI_7} was applied in 2018, which regulates and formalizes how organizations should utilize data collected. Similarly, the United States has the CALIFORNIA CONSUMER PRIVACY ACT (CCPA) \cite{ORI_8}, which also legislates practices of using customer data. A critical component of GDPR states that the utilization of data needs to be lawful, which is explained as having good reason to use the data \cite{ORI_8}. The lawfulness principle requires that whoever uses the data must inform the customer and receive consent for accessing such data. From this point, it can be seen that collecting vast amounts of data and using them for machine learning is becoming harder as more procedures have to be conducted. Moreover, even when collecting data is legally permitted, data silo issues will still cause inefficiency in using data. Data silo refers to the situation in which each site, such as hospitals, data stations, companies, organizations, and data storage of sensors, will have data collected independently. However, they cannot share the data with others sites easily due to different constraints. Constraints may include organizational regulations such that data cannot be shared with others or data is not allowed to leave the site in any form, or due to network constraints such that it is expensive to transfer data between sites. Data silo will lead to the situation that even when there is sufficient data collected at different sites, failing to access all data will still cause data insufficiency and inefficient utilization of data. A method is required to further support machine learning with a rich amount of data available through different sites that have constraints of accessing data. Such a method should not only grasp information and benefit from data from different sites but also keep data privacy that satisfies the requirements of data protection from legislation or organizations. Federated learning \cite{ORI_9} arises as a promising method to train models based on data from different sites while not breaking data protection rules from regulations and sites. This research project will look deeper into the better practice of federated learning, especially for resource management.

\subsection{Federated Learning: A promising solution to access sufficient data}
The core idea of federated learning is to communicate the model weights or parameters rather than the data used for training models \cite{ORI_9}. Federated learning first trains machine learning models over multiple devices on different sites locally. Since these data are trained locally, data privacy is protected, and there is no violation against the rules of GDPR \cite{ORI_7} or the rules of organizations themselves. After a few epochs of training performed at local spots of different sites, model weights are extracted and transmitted to a central place for aggregation. It is worth noting that at this step, data at each spot is still not being transmitted. Thus, the privacy of that data is still preserved by different sites. Afterward, the central place to which the extracted model information is transmitted will merge those weights from different sites by diverse algorithms, combined to a more robust model \cite{ORI_9} compared to models belonging to each site themselves. This process will be repeated for a few epochs until the combination results at the central place have little difference, which refers to global convergence. Since data is kept at its origin, risks of private data being abused are reduced \cite{ORI_10}, as well as the probability of data lost during transmission \cite{ORI_11,ORI_12}. Examples of applications of federated learning further prove its effectiveness. Google deployed federated learning on mobile phones to train a natural language processing model \cite{ORI_13}. The model is trained based on input the user types into the google keyboard while the actual data of words typed is never sent back to the clouds of Google, which keeps privacy. Moreover, with access to more data, Google developed a better prediction model to provide recommendations for possible words users might be interested in \cite{ORI_13}. Another example is federated learning used for targeting advertisements \cite{ORI_14}. While users are concerned that companies abuse their purchasing and browsing history, federated learning is applied to provide the right to choose whether to participate in assisting better advertisement service without sharing that information with the public. This results in more targeted and better-rewarded advertisements and preserves user privacy \cite{ORI_14}. Genome-wide association studies (GWAS) \cite{ORI_15} are a type of research based on gene features of patients to develop medical treatments. While the successful development of GWAS relies heavily on the number of gene sequences available for a particular type of disease, it met the obstacle of accessing enough data. Gene data is highly confidential since it can be maliciously used to detect patient defects and cannot be transmitted online based on regulations \cite{ORI_7}, \cite{ORI_15}. While federated learning is proven to be safe for the collaboration of data holders across sites and successfully accelerates the progress of GWAS \cite{ORI_15}. Considering these examples, it can be seen that federated learning can solve the data silo issue, which gives access to more data for machine learning researchers and developers. Such features will not only result in more robust models but also promise data security and data privacy for their owners. This research project will research better-federated learning practices with a focus on how federated learning can be better deployed on different distributed machines to result in more efficient and reliable machine learning model development.

\subsection{Federated learning: Heterogeneity challenge}
With the importance of federated learning discussed above, heterogeneity among different sites must be considered to deploy federated learning successfully on different sites. Heterogeneity among servers participating in federated learning ranges from systems' computing resources, networking characteristics, available data size, and availability \cite{ORI_16}. Different computing resources involve but are not limited to different CPU frequency, CPU utilization rates, RAM, and GPU. All these varied computational resources mentioned above result in different required training time for each site to train a fixed amount of epochs with a fixed amount of data. With different computing times among various sites participating in federated learning, often, fast computing sites will need to wait for slow computing sites, which constrains the efficiency of the training model in a distributed manner \cite{ORI_9, ORI_16}. Different Network characteristics can also influence the deployment of a federated learning \cite{ORI_17}. Because different servers will have various networking capabilities, including download and upload speeds, dropout rates, and available times, the time required to transmit the common model structure used by federated learning from a central place to different sites will vary significantly \cite{ORI_17}. In extreme cases, other sites can finish multiple training rounds when a model is transmitted to sites with minimal network capacity. Such differences can again cause some sites to wait for slower sites for the majority of the time, which is a waste of computation power. The different data sizes for training a model on each site will also challenge federated learning \cite{ORI_9}. A large training data model can produce a much more robust and accurate machine learning model compared to sites with limited training data \cite{ORI_9}. Thus, combining models of sites with various amounts of data will underperform compared to only combining models from sites with large amounts of data. However, simply dropping results from sites with limited data failed to explore the potential of all data available. From the challenges mentioned above, it can be seen that to successfully conduct federated learning, it is essential to address the heterogeneity among different participants of the training process. To address these challenges, intelligent decisions about participant selection for each round of training, network allocated for each participant as well as merging algorithms need to be carefully studied \cite{ORI_9, ORI_16, ORI_17}.  Research on better practices for federated learning with participants containing heterogeneous computing resources, helps machine learning models to access richer and more efficient sources of data since sites with low computation power can also be included, which is a motivation for this research.

\section{Resource management in heterogeneous computing environments}
As mentioned above, this research focuses the deployment of the federated learning in heterogeneous and distributed computing environments. The primary result includes but is not limited to selecting participants for federated learning at each epoch, smart allocation strategy for network resources, and selecting appropriate combination algorithms to merge models from different sites \cite{ORI_18}. All operations mentioned above require a resource management framework for the management and handling of the federated learning tasks. Decisions regarding which sites to choose for federated learning needs statistics describing computation resources available on each site, which is out of the scope of federated learning itself and needs assistance from the resource discovery framework. Furthermore, sending decisions as instructions to different sites regarding federated learning needs resource management framework's scheduling and job allocation capability \cite{goudarzi2022scheduling}. Thus, it is necessary to choose a proper resource management framework. Details of the comparison of different resource management frameworks will be provided in the chapter \ref{chapter:2}, and the framework chosen for this project is FogBus2 \cite{deng2021fogbus2,ORI_19}. FogBus2 is an integrated resource management framework that provides resource discovery support. This feature can be further exploited to provide statistics related to available system resources to select appropriate clients to participate in federated learning \cite{ORI_19}. Furthermore, FogBus2 is containerized, which supports easy deployment of federated learning codes on different machines \cite{ORI_19}. Lastly, FogBus2 have integrated scheduling and task allocation modules, which provides the potential for participant selection operation with minimal development \cite{ORI_19}. Thus, with the convenience FogBus2 provides, it stands as a good starting point for further investigation on good practices of federated learning and will be chosen for this project.

\section{Research Objective}
A sufficient amount of data is a crucial factor for the successful deployment of machine learning approaches in practice, while accessing the required data is an important challenge due to the following reasons: 1) Infeasibility to have direct access to large amounts of data in distributed systems, 2) Privacy regulations and awareness. Federated learning arises as an essential and necessary tool to allow machine learning approaches further benefit from the rich amount of available distributed data. However, for the practical deployment of federated learning approaches on large-scale distributed systems, the heterogeneous computational resources in distributed systems should be efficiently harnessed. Thus, the main goal of this research is to study federated learning approaches and the practical deployment of proper federated learning models on heterogeneous distributed systems. To achieve this, we design and implement a federated learning mechanism by extending the FogBus2 resource management framework in large-scale distributed systems. Particularly, I aim at addressing the following Research Questions (RQ):

RQ 1: \textbf{How to design and implement a federated learning mechanism to be efficiently integrated with the current properties of FogBus2 resource management framework so that new machine learning models can be designed and integrated on demand.}

RQ 2: \textbf{Based on an integrated federated learning mechanism on top of the FogBus2 framework, how to efficiently achieve the desired training accuracy in distributed and heterogeneous computing environments in a short time?}

\section{Research Scope}
In what follows, I briefly describe the main scope of this research:

- The main focus is on designing and implementing federated learning mechanisms on the FogBus2 resource management framework instead of other federated learning frameworks such as Pysyft \cite{ORI_20}, Tensorflow Federated \cite{ORI_21} or Federated AI Technology Enabler (FATE) \cite{ORI_22}.

- This study explores methods to minimize the time required to train a model shared by federated learning. Other optimization targets, such as minimum energy caused to train a model, minimum accuracy achievable in a federated scale, or minimum weighted accuracy among all participants or more, are not considered the target for this research project.

- The measure that I develop to increase time efficiency focus on the decisions based on real system parameters and logs provided by the FogBus2 framework. Methods designed and implemented in this research include: policies to select participants for federated learning for each round, and asynchronously aggregating machine learning models. Other methods which attempt to increase time efficiency from properties of machine learning models themselves, such as model compression or efficient model representation for transmission, are not considered.

\section{Contribution}
Although other methods attempt to improve time efficiency or computational energy efficiency based on system statistics, none of them have attempted to experiment with asynchronous federated learning. Furthermore, to the best of my knowledge, only a few works targeting optimizing the deployment of federated learning mechanisms provide the source code of their research. Also, among the works that open-sourced their work, it is very difficult to implement new worker selection policies or new models. The contributions of this research are summarised in the following:

- Designing and implementing the federated learning mechanism by extending the FogBus2 resource management framework.

- Implementing a lightweight mechanism so that participants' selection policies for federated learning and employed algorithms for merging models from different participants can be efficiently integrated.

- Implementation of an asynchronous participants selection policy based on real data obtained from FogBus2 resource management framework and underlying computational resources. This policy is proved in that it is more time efficient to reach a desirable accuracy than synchronous or other baseline methods.

- Conducting a large set of experiments on practically deployed federated learning mechanism on FogBus2 framework to evaluate the efficiency of design mechanism and algorithms.

\section{Thesis Organisation}
The rest of this thesis will be structured as follow. Chapter \ref{chapter:2} reviews relevant literature, which includes a general review of research on federated learning itself, followed by a specific review of research intended to optimize federated learning based on system parameters. After that, compares all different resource management frameworks that can provide system parameters and justify why FogBus2 is chosen for this project. Chapter \ref{chapter:3} provides a formal and formularized definition for the problem of the research. Chapter \ref{chapter:3} presents a discussion about algorithm design as well as the hypothesis made. After that, chapter \ref{chapter:4} discusses the experimental setup as well as the obtained results. Finally, conclusion chapter concludes this work and draws future directions for further progress. 

\lhead{\emph{Literature review}}  
\chapter{Literature Review}\label{chapter:2}

Terminology used in this thesis is shown in \ref{table:abbreviation} for reference. This section first gives a general review of federated learning itself in section \ref{section:2.1}. It will show the relevance between the success of federated learning with optimization based on system parameters. After that, section \ref{section:2.2} will mainly focus on optimization methods based on system statistics, which compare and suggest  several identified directions for further research. This section also shows the importance of resource management to assist research optimising federated learning. With the awareness of the importance of resource management, section \ref{section:2.3} compares resource management frameworks and justifies the choice of FogBus2 resource management framework used for this research.

\begin{table}[]
\caption{Abbreviation}
\label{table:abbreviation}
\resizebox{\textwidth}{!}{%
\begin{tabular}{|l|l|l|}
\hline
Terminology & Explanation & Abbreviation \\ \hline
Aggregation Server & Central site used to aggregate models from different parties & $AS$ \\ \hline
Worker & Site which contribute to federated learning via pushing the local model to aggregation servers & $w$ \\ \hline
Aggregation algorithms & Algorithms used to combine model results from different workers to a single model & $f_{aggr}$ \\ \hline
Worker selection algorithm & Algorithms used by aggregation server to select workers to participate in federated learning among all workers available. & $f_{sel}$ \\ \hline
Aggregation server model weights & The model weights of the model held by aggregation server (aggregate for i times) & $M_{asi}$ \\ \hline
Worker model weights & The weights of the model locally held by worker x (based on version i of the aggregation server and trained for j times) & $Mw_{x, i, j}$ \\ \hline
Worker averaging weights & Weights allocated for weighted federated averaging for worker x & $WEI_x$ \\ \hline
\end{tabular}%
}
\end{table}

\section{Review of federated learning}\label{section:2.1}
\subsection{Definition of federated learning}\label{subsection:definition}
Referring to the definition from the Federated Learning textbook \cite{ORI_9}, federated learning is a method to train a shared machine learning or deep learning model based on data from multiple parties. It is featured by pushing model parameters periodically between a central place for aggregation model weights and participants. For the purpose of this research, a complete federated learning process is partitioned into three stage, firstly, establish the connection, secondly, model training, and training evaluation lastly.

Since federated learning is a collaborative process, where the training will be conducted over multiple sites, the connection needs to be established before the start of federated learning training. A site can take three types of roles \cite{ORI_23, ORI_24}. Firstly, workers, this means sites that will contribute to the federated learning by training a commonly agreed model by data in their local spot. The model, not the raw data used for training, will be communicated to different sites, and those sites will further process the model. Secondly, the aggregation server. This means sites that collect model weights from workers and aggregate them through different algorithms into a single model. Since the aggregation server is the centre of the relationship that every worker must be able to communicate with it, it is also seen as a central point of control throughout the federated learning process. Lastly, peer, such role refers to sites that can communicate with each other, however not in none of them are aggregation servers of other parties. 

\subsubsection{Establish the connection}
\begin{figure}[h]
    \centering
    \includegraphics[width=\textwidth,height=\textheight,keepaspectratio]{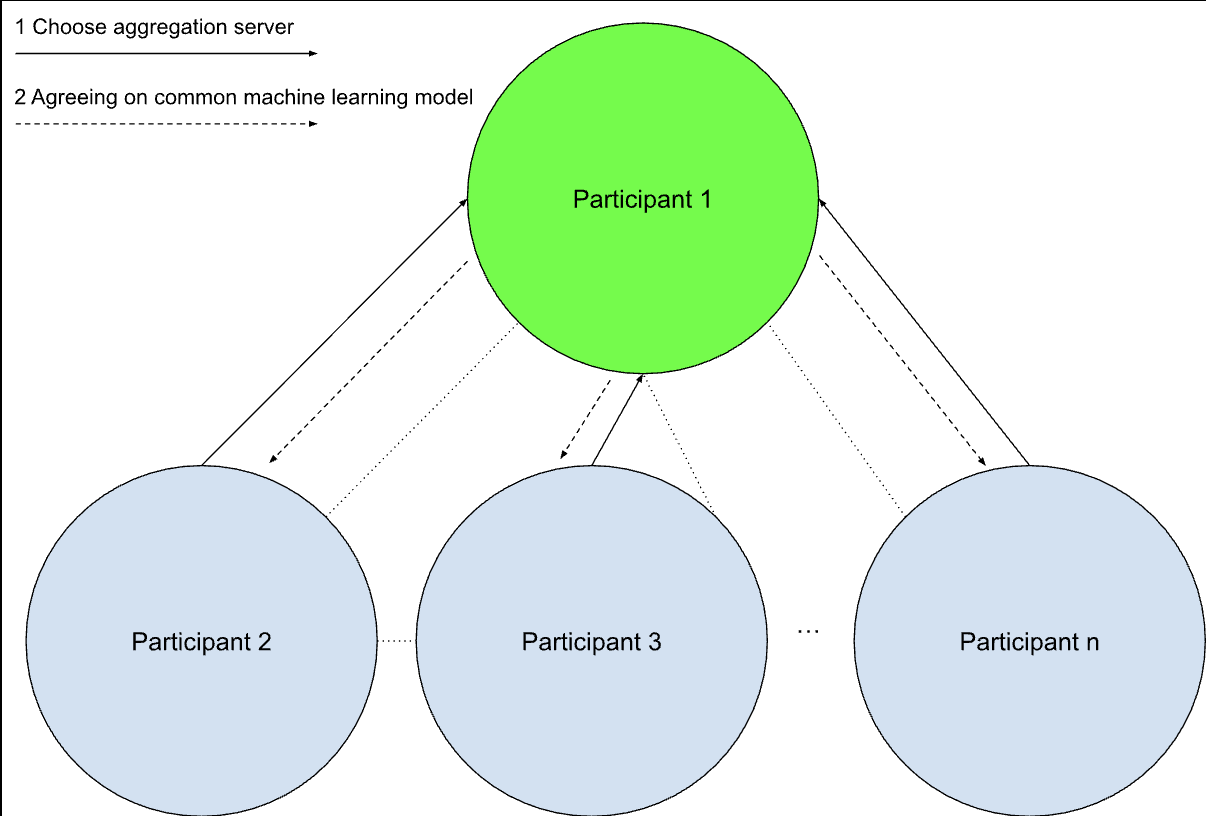}
    \caption{Federated learning relationship establishment}
    \label{fig:FL_Step1}
\end{figure}

The connection establishment process is the stage where different sites identify their roles as some workers, aggregation servers and peers. Also, the common machine learning model that every site will use for federated learning will also be agreed. The process of establishing the connection is shown in figure \ref{fig:FL_Step1}.

1. A set of sites that have a network connection with each other will select a central server for federated aggregation. Such connection is due to prior network conditions and is demonstrated in the dotted line in figure 1. Aggregation selection can be based on available computation resources \cite{ORI_23} or the number of participants a site can directly be connected to or manually assigned.

2. After the aggregation server is selected, which is green in the figure, every other participant becomes a worker, which is in blue. A lined arrow indicates a worker to aggregation server relationship.

3. After all sites identify their relationships with other sites, the aggregation server will send the description of the machine learning model that will be trained for federated learning to each worker. Information describing the model can be files containing model layers or purely a name as long as workers can uniquely identify the model aggregation server referring to.

\subsubsection{Federated learning model training}\label{subsubsection:train}
\begin{figure}[h]
    \centering
    \includegraphics[width=\textwidth,height=\textheight,keepaspectratio]{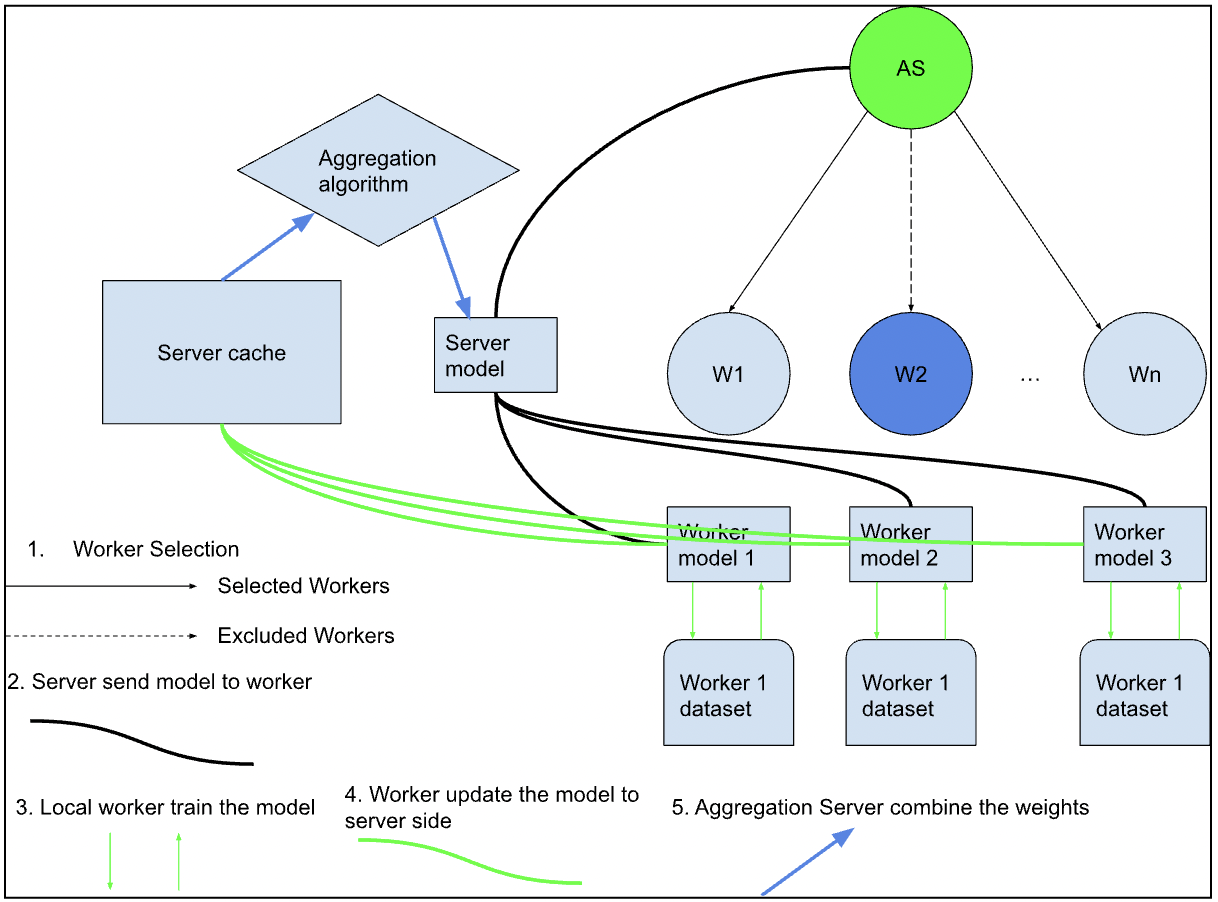}
    \caption{Federated learning training}
    \label{fig:FL_Step2}
\end{figure}

After the first step, all sites should have the common machine learning model ready for train based on the data set on their local site. The federated learning process is then illustrated in figure \ref{fig:FL_Step2}. The process is defined based on the Federated Learning textbook \cite{ORI_9} and surveys about federated learning \cite{ORI_23}, while here, the explanation will break the process into smaller and more detailed steps.

1. The aggregation server shown in the green circle chooses workers to participate in federated learning training. This process is referred to as worker selection. Worker selection is a trade-off between efficiency and accuracy. Intuitively, selecting more workers will allow the federated learning to access more data which causes better model performance. However, this means fast computing workers must wait for slower computing workers, which will require more time to train the model to the desired accuracy \cite{ORI_25}. In the diagram, circles shown in deep blue, including worker two, are excluded from the federated learning while other participants are included.

2. After a set of workers is selected for the training. The aggregation server will inform them to start the federated learning training. Correspondingly, workers will download the version of the model aggregation server provides. This is shown in thick black lines in figure 2. There are two methods to achieve this. One is an aggregation server sending information about the model directly to workers. This method is easy to implement and straightforward. However, this will cause network congestion at the aggregation server point \cite{ORI_26}. At the same time, the worker may not be available to receive the model due to network availability at the time the aggregation server wishes to send the model. An alternative method is the aggregation server to push the machine learning model to a database and gives download credentials to all selected workers \cite{ORI_9}, \cite{ORI_26}. Although it is harder to implement, this method is more friendly to workers with various network conditions. This is because workers can download the model at a time in their favour. At the same time, the aggregation server only needs to communicate with the database instead of multiple workers, which relieves network pressure.

3. As long as the model is downloaded to the worker side and overrides the model of the local worker, all workers will train the model and update model weights based on data originating on their sites. At this step, training the model is the same as other machine learning model training. Standard methods include but are not limited to stochastic gradient descent. Training is shown in green arrows in figure \ref{fig:FL_Step2}.

4. After workers finish training for a few epochs, they will send the weights of the model back to the aggregation server side. This step is shown in thick green lines. The aggregation server should allocate cache space beforehand to store workers' models.
Lastly, the aggregation server will merge response model weights from workers. Example algorithms include averaging model weights from all worker model weights \cite{ORI_9} and weighted averaging according to the amount of data each worker used for training \cite{ORI_26}. After the aggregation server finishes averaging, it is considered one epoch of federated learning training.

\subsubsection{Federated learning training evaluation}
After the federated learning training process mentioned above, the aggregation server will also evaluate the performance of the averaged model to decide whether to repeat the federated learning training step or not. The evaluation process is usually performed in two ways. The first method is the aggregation server use data available locally to test the performance of the aggregated model. Alternatively, the aggregation server can ask workers to download the latest model and evaluate the model performance locally, then average accuracy \cite{ORI_9}. After the evaluation process, if accuracy does not meet the expected accuracy, steps two and three will be executed repeatedly.

It is worth noting that throughout the whole process, data used to train models never leaves its origination, ensuring security concerns and passing data protection regulations such as GDPR \cite{ORI_7}.

\subsection{Classification of federated learning}\label{subsection:classification}
This research will explore asynchronous federated learning under one level of aggregation server and worker relationship. Moreover, this research will use horizontal federated learning rather than vertical federated learning or hybrid federated learning. This section will review the difference between different kinds of federated learning, which explains what the choices mentioned above are and why this kind of federated learning is chosen. 

\subsubsection{Classification by training data features available between sites}
As mentioned in \ref{subsubsection:train}, during the federated learning process, the aggregation server will select different workers to train a model and aggregate the results. Based on the types of features data each worker has, federated learning can be classified into horizontal federated learning, vertical federated learning and hybrid federated learning \cite{ORI_9}.

Horizontal federated learning means data held by different workers that they intend to use to train federated learning models have the same set of features, whereas those data are describing different instances. In this case, every worker is trained on the same model structure. Just different training data will result in different model weights. An example would be multiple banks training a model to predict the probability of customers failing to pay for the credit card. In this case, different banks which act as workers have similar features describing each customer, such as the number of deposits, name, address and more which refers to the same set of features. However, those banks hold different customer data, meaning data describing different instances. The benefit of horizontal federated learning is that workers commonly share the model with the same structure and different numerical weights. And merging different model weights can have general solutions such as averaging or weighted averaging instead of algorithm-specific measures. The drawback of horizontal federated learning is that all workers need the same interest in things they want to predict by the machine learning model since the model on the aggregation server is the same \cite{ORI_9}, which is less flexible and makes it harder to find sufficient workers \cite{ORI_26}.

Vertical federated learning, in contrast to horizontal federated learning, is when data held by different workers intend to use to train federated learning models to have a different set of attributes, whereas those data describe the same set of instances \cite{ORI_27, ORI_28, ORI_29, ORI_30}. In this case, models used by different workers are different since different model structures are required to correspond to different data features. An example is banks and supermarkets together training machine learning models regarding customers. Whereas supermarkets are interested in the purchase pattern of customers and banks are interested in future deposit amounts held by customers. Banks and supermarkets can identify the intersection of customers belonging to both of them and train more robust machine learning models since together, they can provide more features \cite{ORI_28}. The benefit of vertical federated learning is that a richer source of features available will result in a more potent model while pairing workers with different interests is more flexible and thus easier to find sufficient workers \cite{ORI_9}. The drawback of vertical federated learning is difficulties in merging model weights from different workers. Since features of data held by different workers are different, the corresponding structures of the model for each worker are different \cite{ORI_27, ORI_28, ORI_29, ORI_30}. Consequently, model-specific algorithms need to be designed to merge models with different structures, while algorithms used to merge are hard to reuse. 

Lastly, hybrid federated learning. This refers to workers who do not hold data referring to the same instance as vertical federated learning or data with the same set of features as horizontal federated learning. This method takes the idea from transfer learning \cite{ORI_9}, which intends to train models from workers with completely irrelevant data with transferable knowledge. Hybrid federated learning mainly focuses on algorithms and strategies to transfer knowledge, thus will not be considered in this research.

A comparison of different methods can be seen in table \ref{table:2.2}.
\begin{table}[]
\caption{Comparison of hierarchical, vertical and hybrid federated learning}\label{table:2.2}
\label{tab:my-table}
\resizebox{\textwidth}{!}{%
\begin{tabular}{|l|l|l|l|l|}
\hline
 & Data share same features & Data describing same entities & Merging Model Weights & Flexibility of Choosing workers \\ \hline
Hierarchical & YES & NO & General Reusable algorithms & LOW \\ \hline
Vertical & NO & YES & Machine learning model specific & HIGH \\ \hline
Hybrid & NO & NO & Machine learning model specific & VERY HIGH \\ \hline
\end{tabular}%
}
\end{table}

Since this research focuses on using system statistics to conduct federated learning in a time efficient manner, while focusing on vertical and hybrid federated learning is on algorithms used to merge model weights and time efficiency is more determined by machine learning model design \cite{ORI_9, ORI_27, ORI_28, ORI_29, ORI_30}, these two types of federated learning will be excluded from this research. Although hierarchical federated learning is not flexible in choosing workers with the same interests, through a standard pre-processing of data, similar features can still be extracted from data that diminish such concern. Thus, horizontal federated learning is the suitable type that will be used in this research.

\subsubsection{Asynchronous and synchronous federated learning}
Since this research will explore asynchronous federated learning. Thus, this section will review the difference between asynchronous federated learning \cite{ORI_31,ORI_32} and synchronous federated learning.

Referring back to point 5 of step 2, federated learning training in \ref{subsubsection:train}. The aggregation server will aggregate model weights once a sufficient number of workers finish transmitting their models to the cache of the aggregation server. However, this step does not require the aggregation server to wait until all selected workers respond, which will result in the asynchronous situation \cite{ORI_9, ORI_31, ORI_32}. There are three kinds of situation that can happen regarding worker $W$ sending the weights of the model trained by local data and the aggregation server $AS$ starting merging model weights from workers:

1. Model of $W$ arrives before $AS$ starts to aggregate model weights from workers, and when $AS$ starts to aggregate model weights from workers, it includes model weights $W$ in the aggregation process.
2. Model of $W$ arrives after $AS$ starts to aggregate model weights from workers, and $AS$ refuses to receive model weights of $W$ even when $W$ responds to $AS$. Model weights of $W$ will not be included in the aggregation process.
3. The $W$ model arrives after $S$ starts to aggregate model weights from workers. Although $S$ will not include model weights of $W$ for the current round of aggregation, $S$ will still receive that and use it for the next round of aggregation.

The combination of point one and point two is called synchronous federated learning, while the combination of asynchronous federated learning. The advantage of synchronous federated learning is the simplicity of merging algorithms since all worker model weights are based on the same version of server model weights. In contrast, model weights may be based on different versions of aggregation server weights for asynchronous federated learning. Thus extra design is necessary for the merging algorithms. The advantage of asynchronous federated learning is time efficiency compared to the synchronous version since the aggregation server does not have to wait for slow response workers before aggregation and proceeding to the next epoch. Instead, asynchronous federated learning can just proceed to the aggregation process while slow workers are still training the model locally and include a response from slow processing workers in a later round of aggregation. This feature is time efficient, which is inline with the research goal of this research. Thus asynchronous federated learning is chosen.

\subsubsection{Hierarchical federated learning and one level federated learning}
Hierarchical federated learning refers to the practice that multiple aggregation servers act as workers and report to another site which acts as an aggregation server \cite{ORI_33, ORI_34,ORI_35}. In contrast, one level of federated learning refers to only one aggregation server present in the whole federated learning network and only one level of aggregation server to workers relationship. The existence of hierarchical federated learning intends to deal with the network constraints between edge devices and central aggregation server since remote devices may have very limited download and upload speeds. Instead, hierarchical federated learning enables edge devices to communicate with a base station nearby, which is faster in communication and base station will act as an aggregation server. Base stations will then send the results to the cloud, where the cloud will aggregate results from different base stations \cite{ORI_34, ORI_35}. Hierarchical federated learning can provide a more communication schema and will improve time efficiency. However, including both asynchronous federated learning and hierarchical federated learning based on system statistics will confuse the effectiveness of these two methods separately. With the lack of research and benchmark for asynchronous federated learning based on system statistics only, this research will only focus on asynchronous federated learning algorithms and leave hierarchical federated learning as a separate topic. Furthermore, this research will ensure the extensibility of hierarchical federated learning from the implementation perspective (RQ 1) due to the potential of time efficiency boost of hierarchical federated learning.

\subsection{Aggregation algorithms}\label{subsection:aggregation_algo}
When the federated learning aggregation server decides to merge model weights from workers, no matter the asynchronous or synchronous version, algorithms have to be designed to conduct such an operation. The process of merging model weights from workers and using the aggregated results to update model weights of the aggregation server is called the aggregation process, where the algorithms used are called aggregation algorithms. Terms can be referred back to table \ref{table:abbreviation}. This section will review basic and commonly adopted aggregation algorithms, which are the foundation of all other aggregation algorithms. These algorithms are federated learning averaging (FedAvg) and weighted federated learning averaging (WFedAvg).

For mathematical definition purposes, both models of the aggregation server and models of workers are labelled by a version number.

For the aggregation server, let $M_{as}$ denote the model weights of the aggregation server. The version of Mas refers to the number of aggregation processes the aggregation server has conducted. For example, a newly created aggregation server will have the model with version zero since no aggregation has been conducted. $M_{asi}$ denotes the model weights of the aggregation server with version i.

For the workers, let $M_w$ denote the model weights. Since the worker needs to fetch the model weights of the aggregation server before training locally, the version of the server model fetched will also be recorded. With the same semantics of version calculation with the aggregation server, $M_{wx, i, j}$  denote the model weights of worker x, which fetched version i of the aggregation server model before the train and trained j epoch.

\subsubsection{Federated Averaging}
With terms defined, the most common algorithm used to merge worker model weights is federated averaging \cite{ORI_9, ORI_36}. As the name suggests, federated averaging means the new model weights are updated by the average of all worker model weights available in the cache \cite{ORI_36}. This is shown mathematically in equation \ref{equation:syn_aggregation}, with j denoting the agreed training epoch for each worker has to perform before responding to the aggregation server.

\begin{equation}\label{equation:syn_aggregation}
Mas_{i+1} = 1/n *\sum_{x=1}^{n} Mw_{x,i,j}
\end{equation}

The advantage of this method is simplicity and easy adoption in any scenario, as long as the aggregation server can store model weights from workers, such a method can proceed, and no extra parameter is required. Moreover, this formula also has an asynchronous version, as shown in equation \ref{equation:asyn_aggregation}.
\begin{equation}\label{equation:asyn_aggregation}
Mas_{i+1} = 1/n *\sum_{x=1}^{n} Mw_{x,i-k,j}  \forall k < i
\end{equation}

Note that the difference is that asynchronous federated learning does allow worker responses that are based on earlier versions of the aggregation model weights.

The drawbacks of federated learning averages are that purely averaging will diminish the effect of some high-performing workers when averaged with model weights from low-performing workers. The difference in model quality among workers is mainly due to two reasons: different training data sizes and time differences between each worker fetching the model weights from the aggregation server \cite{ORI_31, ORI_36}. For workers with different training data sizes, after the same number of epochs of training, workers with a more significant amount of training data will have a more accurate model compared to workers with less training data. Thus, purely averaging the result will cause the performance of more accurate models to drop since they will be combined with the performance of less accurate models. For different time differences between each worker fetching the model weights from the aggregation server, since the version of the aggregation server model is different at a different time when the worker fetches the model, the newer version will perform better. Since workers train the model with the base model of different performances, the performance of the resulting model will also vary a lot, and purely averaging again failed to consider such inequality between model performances. As a result, weighted federated learning arises as compensation.

\subsubsection{Weighted Federated Averaging}
Weighted federated learning intends to adjust the weights of different models such that models from workers with more data or higher performance due to computation power will be allocated larger weights. In comparison, models from workers with less performance will have lower weights \cite{ORI_9}. Let $WEI_x$ denote the weights allocated for worker x. Weighted federated learning averaging is shown in formula \ref{equation:syn_update} and \ref{equation:asyn_update}, with \ref{equation:syn_update} indicating the synchronous version and \ref{equation:asyn_update} indicating the asynchronous version.

\begin{equation}\label{equation:syn_update}
Mas_{i+1} = \sum_{x=1}^{n} Mw_{x,i,j}*WEI_{x}  \quad s.t. \sum_{x=1}^{n}WEIx = 1
\end{equation}

\begin{equation}\label{equation:asyn_update}
Mas_{i+1} = \sum_{x=1}^{n} Mw_{x,i-k,j}*WEI_{x}  \quad s.t. \sum_{x=1}^{n}WEIx = 1, k<i
\end{equation}

The advantage of weighted federated learning is the ability to incorporate the performance inequality between different models while extra calculation time is required to compute and design weighted functions. Three different weighted methods are suggested to deal with the time difference between when workers fetch model weights from the aggregation server \cite{ORI_36}. Let $i$ be the current version of the aggregation server model, and $xi$ be the version of the aggregation server model that worker $x$ fetched. The idea is that the larger difference between $xi$ and $i$ indicate that the model that worker $x$ is based on has lower performance. Thus, less weight is allocated to worker $x$ and vice versa. The three functions to achieve this are linear weight, polynomial weight and exponential weight \cite{ORI_36}. This is shown in equation \ref{equation:linear}, \ref{equation:polynomial} and \ref{equation:exponential}.

\begin{equation}\label{equation:linear}
WEI_x = 1/(i - xi + 1) \forall x
\end{equation}

\begin{equation}\label{equation:polynomial}
WEI_x = (i - xi + 1)^{-a} \quad \forall x
\end{equation}

\begin{equation}\label{equation:exponential}
WEI_x = exp(-a*(i-xi))) \quad \forall x
\end{equation}

The essay pointed out that the linear version of weighted federated averaging has the benefit of including weak and robust performance models. However, if a working model is based on a very old version of the aggregation server model, while others are based on a newer version that has much better performance, then including such a low-performance worker will lead to performance loss compared to the performance of excluding results from such a worker \cite{ORI_36}. For polynomial and exponential weighted federated learning, since weights will be biased more towards high-performing workers, which recently fetched the new version of the aggregation server model, it has the benefit of excluding low-performing workers, who will negatively influence the overall performance. On the other hand, such bias towards high-performance workers will neglect low-performing workers, which causes the aggregated model only benefit from part of the data available among all workers. The comparison between different aggregation algorithms is shown in table \ref{table:aggregate_algorithm}.

\begin{table}[]
\caption{Comparison of aggregation algorithm}
\label{table:aggregate_algorithm}
\resizebox{\textwidth}{!}{%
\begin{tabular}{|l|l|l|l|l|l|}
\hline
 & Formula & Extra parameter required & Influence from low performing workers & Includes all workers effort & Calculation Complexity \\ \hline
Federated averaging & \ref{equation:syn_aggregation}, \ref{equation:asyn_aggregation} & NO & HIGH & YES & O(n) \\ \hline
Linear Weighted Averaging & \ref{equation:linear} & YES & MEDIUM & MEDIUM & O(3*n) \\ \hline
Polynomial Weighted Averaging & \ref{equation:polynomial} & YES & LOW & LOW & O(3*n) \\ \hline
Exponential Weighted Averaging & \ref{equation:exponential} & YES & VERY LOW & VERY LOW & O(3*n) \\ \hline
\end{tabular}%
}
\end{table}

As shown in the comparison table, the more robust the bias aggregation algorithm has towards high-performing workers, the lower the negative performance influence they will suffer from low-performing workers. However, this will hinder the training to gain knowledge from data from low-performing workers, decreasing the total number of data federated learning training processes. The choice of merging algorithm is a trade-off between better-performance workers with less training data and lower-performance workers with more training data. This research hypothesises that linear and polynomial weighted averaging algorithms will outperform other aggregation methods. Since they stand as the middle point between significantly towards high-performing workers and extremely towards including all workers.

Moreover, the aggregation algorithms mentioned above only consider the version difference of the aggregation server model each worker holds. At the same time, more factors influence the performance of models from each worker \cite{ORI_9}. These factors include but are not limited to the size of each worker's available data, the floating point precision each worker supports, and randomness due to the stochastic gradient descent nature \cite{ORI_32, ORI_36}. Without considering these extra factors, the aggregation process will inaccurately be biased towards low-performing models instead of those which have better accuracy and cause longer training time required to achieve the desired performance. Thus, these statistics also need to be considered to provide more accurate heuristics for model performance among all workers and calculate model aggregation weights accordingly.

\subsubsection{Section summary}
To summarise the section, the federated learning procedure preserves data owners' privacy while allowing machine learning models to benefit from those data \ref{subsection:definition}. With the focus on improving time efficiency in federated learning in this research, this research will choose horizontal federated learning other than vertical or hybrid federated learning. This is because horizontal federated learning is less affected by the machine learning model applied and can focus on exploring effective worker selection strategies based on parameters regarding computing resources (Section \ref{subsection:classification}). Moreover, the research will explore both synchronous and asynchronous federated learning. The hypothesis is that asynchronous federated learning is more time efficient since fast computing workers do not have to wait for slow ones (Section \ref{subsection:classification}). Aggregation methods are essential for the successful update of the aggregation server model and affect the training time required to reach a desirable accuracy, which states the importance of comparison of difference aggregation algorithms (Section \ref{subsection:aggregation_algo}). Aggregation algorithms that are biased to the response from workers with more training data or newer versions of the aggregation server model will diminish the effects from other less performant workers. In contrast, algorithms with little or no bias risk being negatively influenced in terms of aggregated model performance by low-performing workers. Thus, a trade-off between these two kinds of methods is required, and difference aggregation algorithms need to be examined. The hypothesis assumes that linear or polynomial aggregation methods, which can balance the effects between workers with different performances, will outperform other aggregation algorithms (Section \ref{subsection:aggregation_algo}).

\section{Optimising federated learning based on system parameters}\label{section:2.2}
\subsection{Different kinds of optimisation strategy}
The performance difference between the theoretical federated learning model and actual deployment is mainly due to heterogeneity among computing resources available among workers \cite{ORI_37, ORI_38}. Methods to optimise federated learning based on system parameters includes but are not limited to worker selection, training epoch tunning and bandwidth resource allocation \cite{ORI_38}.

Worker selection refers to the practice of selecting a set of clients to participate in federated learning to achieve better time efficiency \cite{ORI_25}. Since, at an earlier stage, the model can improve efficiency with only a limited amount of data available. Thus, the worker selection algorithm can just select fast-responding and fast-computing workers in the earlier stage to allow the shared model to reach a certain level of performance. After the model converges or cannot improve accuracy with limited data from fast-responding workers, slow-computing workers can then be selected as a source to provide more training data, further improving the model \cite{ORI_25}. Such methods will save training time since slow-working workers are excluded from earlier training, which will save the overall time between each aggregation process and benefit from a rich data source in later epochs. This is the general idea of worker selection, and the exact selection strategy depends on the loss function of different research sets up as well as the set of system parameters available that can assist in calculating the time and energy required for each worker to train the model locally.

Training epoch tuning refers to the practice of deciding how many epochs each selected worker should train \cite{ORI_9}. For instance, a worker with a large set of data can train fewer epochs. In contrast, a worker with a smaller set of data needs to train for a more significant amount of epochs such that their responding model weights to the aggregation will have similar performance and will not cause heterogeneity and influence the efficiency of federated learning training. This type of tuning is dynamic according to worker performance at each epoch.

Bandwidth resource allocation deals with the architecture where one aggregation server, which is usually a local base station, communicates with multiple workers \cite{ORI_39}. In this case, the number of connections, as well as network input and output speed, is limited. Thus decision also needs to consider communication efficiency among workers. Such that some workers with less data or slower computing resources available will still have high priority in selection since they can fetch the model faster than others. The bandwidth allocation is widely adopted in autonomous vehicle federated learning networks, and the review will discuss the effectiveness of current research on such a topic.

It is worth noticing that there is an entirely different branch of federated learning optimisation which also uses system parameters. This type of research intends to model the optimisation of computation tasks as machine learning tasks and train the model via federated learning \cite{ORI_40, ORI_41}. Since the modelling of such a kind of problem requires system parameters, it has similarities with federated learning optimisation via system parameters interested in this research which is completely different indeed. An example is a research on the ultra-low latency vehicle communication \cite{ORI_40}. Under the requirement for high-performing networks, queued packages waiting to be processed and sent out must be controlled reasonably to ensure there is no substantial network congestion. Suppose the rise of the long queueing messages can be predicted. In that case, corresponding solutions can be applied, such as allocating more computing power or providing a higher priority in network transmission. However, the difficulty is that long queueing messages is a rare event. Thus, although it can be easily modelled as a poison process, it is hard to train an accurate model without sufficient event samples. Federated learning is then applied to provide sufficient training samples, which take system message queueing length as well as transmission latency for analysis purposes \cite{ORI_40}. As a result, long queueing message events can be better predicted, which improves communication efficiency. Another example is in the mobile edge network \cite{ORI_41}. This research intends to determine the appropriate proportion of computation to offload to the base station and mobile phones. It models the rise of new mobile phone computation tasks as a reinforcement learning problem so appropriate computation power will be left on the base station to handle future tasks arising on the mobile side \cite{ORI_41}. However, system parameters such as computation resources required for each task from different mobile phones are sensitive information. While different mobile phones are likely to require different computation power due to the behaviour pattern of their owner, traditional machine learning techniques are infeasible. Federated learning is then applied so sensitive information can be used to train the model to predict future computation resource requirements without leaking privacy. The research accurately predicts future mobile task computation requirements and allows the base station to receive offloading tasks accordingly. Furthermore, the successful offloading strategy improves the response time of mobile tasks under the coverage area \cite{ORI_41}. The two research mentioned above demonstrates the potential of optimising system time efficiency via federated learning. However, they did not optimise the federated learning itself but used federated learning to optimise system operation decisions. This is different from research interests to improve the efficiency of federated learning itself. Thus research in this direction will not be discussed further.

The rest of the section will review recent research to improve the efficiency of federated learning based on system parameters with a focus on worker selection algorithms, bandwidth allocation strategies and training epoch tunning. Moreover, a comparison will be made between strengths and weaknesses about how each research models the federated learning efficiency as an optimisation problem and the effectiveness of solutions suggested. That discussion will prove the necessity of a federated learning implementation that is highly reusable (RQ 1) and research on asynchronously conducted federated learning to improve efficiency (RQ 2).

\subsection{Research on system parameter based federated learning optimisation}
\subsubsection{Tunning federated learning aggregation frequency}
Tunning aggregation frequency is the method that tunes the number of epoch workers needs to train locally before responding to the aggregation server. 

First attempts to optimise federated learning based on system parameters tried to quantify all kinds of resources \cite{ORI_42}. In this research, the optimisation goal is to achieve maximum accuracy under a fixed resource budget. This is achieved by tuning the frequency of global aggregation. Since the resource budget is fixed, the total training epochs a worker can perform are fixed. Thus tunning the frequency of global aggregation can also determine the number of aggregations performed. The algorithm first marks all resources like computing power and network resource as different symbols and then calculates the number of aggregation rounds so that both resources from the aggregation server side and workers side will be exhausted \cite{ORI_42}. For example, if the aggregation server has a battery budget that can support five rounds of aggregation while the worker with a minimum battery budget can train for ten epochs, then the algorithm will ask all workers to respond to the aggregation server every two rounds of training. In this way, the battery budget of both the aggregation server and workers can be maximally utilised. With the number of training epochs workers have to perform before responding, the frequency of global aggregation can be deducted. This research is the first attempt to quantify participants' resources for federated learning while trying to achieve the best accuracy through fully utilising the computation power. According to the research experiment, this algorithm can achieve better accuracy by two to three per cent, showing the proposed algorithms' effectiveness \cite{ORI_42}. However, there are some drawbacks of such methods. Firstly, this algorithm failed to consider the asynchronous case, and the budget is calculated as the minimum budget among all workers, which will fail to explore all resources from other workers with more substantial computation power. Secondly, the algorithm intends to exhaust all available resources to achieve the best accuracy, which will cause workers with rich resources to calculate for a large number of epochs before responding to the aggregation server. This will then result in time inefficiency. Lastly, the algorithm failed to change according to the performance of workers. If the average accuracy of all workers slowly increases, then increasing the frequency of aggregation will allow the model to reach a steady accuracy faster, increasing the overall efficiency. A dynamic approach will be more suitable for the federated learning training process. Even with such disadvantages, the idea of quantifying computing resource still provide the potential of optimising federated learning through statistics regarding system resources.

\subsubsection{Worker selection policy}
On the other hand, the worker selection policy intends to improve the energy efficiency or training time of the federated learning process by selecting appropriate workers to balance resource efficiency and model accuracy.
A worker selection algorithm was first proposed that tried to select workers dynamically \cite{ORI_43}. The algorithm will convert all quantified resources from each worker to the time required to transmit and train the model. The algorithm assumes all workers can only communicate the model sequentially with the aggregation server and set up the total time allowed between each round of aggregation as a variable. After that, the worker selection algorithm will continuously add a worker to the selected worker set until the time required to allow all workers to finish training and transmit the model exceed the total time allowed for the current round of aggregation. The algorithm is then flexibly scaled according to model performance. If the model accuracy increases slowly, then the total time allowed for the current epoch will increase to allow more workers to participate in federated learning training hence obtaining a faster accuracy increase \cite{ORI_43}. The dynamic worker selection algorithm's flexibility allows it to successfully implement on different machine learning tasks such as MINST classification or CIFAR-10 classification. The design of dynamic changing policy will also be included and examined in this research. However, this algorithm has the potential for further improvements. Firstly, the algorithm does not support asynchronous again. Thus the phenomenon of fast computing workers waiting for slow workers will still exist. Furthermore, by the algorithm, the total time allowed for the aggregation round increase as the model performance increase, which will include workers that require longer time and will deteriorate the energy waste of fast computing workers. Secondly, although the modelling assumes all workers will communicate the weights of the model with the aggregation server sequentially, which will simplify the optimisation model, this ignores the potential of saving time by transmitting the model with workers parallelly. Suppose models are transmitted parallelly with the aggregation server. In that case, time will be saved, providing more time budget to allow slower workers to participate compared to the sequential transmission model. This will allow the federated learning training process to benefit from a more extensive range of data with a fixed total time budget, increasing the training efficiency.

Later research also deploys reinforcement learning to manage the federated learning worker selection process \cite{ORI_45}. While two previous research mentioned \cite{ORI_42, ORI_43} above tried to select as many workers as possible within a given resource budget, reinforcement learning considers the energy cost, which will trade-off between the maximum number of workers selected against the energy consumption \cite{ORI_45}. The reinforcement learning method considers the cost of each step as the energy that a set of workers will consume to perform the training based on CPU power and the time required to process. At the same time, the reward is calculated as the accuracy increase of the aggregated model. By solving the reinforcement learning model, a set of workers will be selected based on the trade-off between energy consumption and accuracy. This method has the advantage that the algorithm can be dynamically changed such that the cost related to a fixed amount of energy consumption can be increased if the algorithm wishes to focus more on energy consumption and vice versa. Moreover, this method is not a fixed algorithm. Since reinforcement learning will update the policy for selecting workers based on a response about worker performance and energy consumed in real-time, even if the worker selection failed to achieve optimal selection in earlier rounds, it will still adjust to a reasonable selection policy. The idea of dynamically adjusting worker selection policy according to model performance enables easy deployment of the algorithm with workers of various computing resources, which shows the potential to handle heterogeneity among workers (RQ 1) and will be considered in this research. A few points that can be improved is this method requires too many variables, such as distance between different workers and energy consumption each worker will consume, which is hard to obtain in reality. Furthermore, this research only focuses on the trade-off between energy consumption and model accuracy, while time is not considered.

Based on the research that considers energy consumption \cite{ORI_45} with the idea of simplifying system statistics, another worker selection research expresses the trade-off of accuracy and energy consumption as a linear formula instead of a reinforcement learning approach \cite{ORI_46}. Let $E$ denote the energy consumed based on a set of selected workers and $A$ be the accuracy achieved by the aggregation server model. Then the optimisation is based on the loss function shown in equation \ref{equation:loss_7}.

\begin{equation}\label{equation:loss_7}
Loss = alpha * E + (1-alpha) * -A \quad s.t.  0 \leq alpha \leq 1
\end{equation}

Same semantics as the reinforcement learning algorithm that a more significant alpha denotes more focus on energy consumed, and smaller alpha means accuracy is more important. At the same time, the simplified approach will only be based on energy consumed and accuracy, which are only two variables \cite{ORI_46}. This provides a further level of flexibility such that different methods for calculating energy consumption can be implemented based on specific scenarios. The extensibility is inline with the idea of this research (RQ 1) which will be considered, while this research will focus on time consumption rather than the energy consumed. After the simplified formula regarding worker selection \cite{ORI_46}, the research also tuned the frequency of aggregating models. However, the frequency tunning step is based on the assumption that each worker has available amounts of data that can train the model to a desirable accuracy alone. This research focuses on heterogeneous workers containing workers with minimal data available for training. Therefore, such an assumption cannot be satisfied, and the frequency tunning of federated learning will be based on other algorithm designs. Moreover, this research not only showcase the importance of worker selection, but also demonstrates that allowing a higher frequency of aggregation in an earlier round of federated learning and decreasing the frequency of aggregation as the model become more stable in later rounds is time efficient \cite{ORI_46}. Thus, this research will also consider tunning frequency to achieve better time efficiency.

Previous research regarding worker selection policies \cite{ORI_43, ORI_45, ORI_46} primarily depends on the resource that will be consumed to participate in the federated learning process by each worker. This generally assumes that more workers will allow the model to be exposed to more data hence will improve the accuracy of the shared model. In this way, worker selection policies are more biased towards fast-computing workers and hope more epochs of training conducted by fast-computing workers can result in better accuracy compared to fewer epochs of training that slow-computing workers can do. However, another research suggests the direction of model update gradients indicates the effectiveness and contribution of different workers \cite{ORI_49}. This research asks all workers to train for a few epochs and fetch their response rate. After that, the algorithm computes the average model weights in the first round of aggregation. Then the algorithm compares all model response rates with the averaged model by computing the norm difference \cite{ORI_49}. A more significant norm difference indicates that the model update direction is different from the majority trend, which suggests the corresponding worker has little or no contribution to the overall accuracy. On the other hand, workers who respond a weight with minimum norm difference with the averaged model indicate the model's effectiveness in improving the accuracy of the federated learning process \cite{ORI_49}. The suggested algorithm then conduct a worker selection process more biased towards worker with little norm compared to the averaged model. Experiments demonstrate that such an algorithm can achieve a desirable accuracy within fewer rounds of aggregation \cite{ORI_49}. While time is saved for federated learning training due to fewer epochs, more time is required since slower computing workers will be selected over those fast computing workers in this algorithm if they have less model norm difference. Thus there is a limited time efficiency boost.

One of the latest research also prevents slow workers not communicating with the aggregation server \cite{ORI_51}. Previous worker selection strategies have the idea of allowing fast computing workers to participate in federated learning training first and then add slow workers progressively in later rounds. However, this method has the drawback that slow computing workers cannot communicate with the aggregation server for a long time and can only participate in the federated learning in the last few rounds of training. This is wasting computing resources of slow computing workers, and the federated learning training process may not gain sufficient knowledge from data that can be provided by slow computing workers since there are insufficient rounds slow computing workers can contribute to the federated learning. Based on the such idea, the algorithm considered the time since the last time a worker fetched model weights from the aggregation server and contributed by responding to the updated model \cite{ORI_51}. The ranking formula for worker selection considers such time until the last contribution, together with other factors like computation time and energy consumed. It will select the top few workers for the worker selection procedure. By the algorithm, slow-computing workers will also be selected before fast-computing workers if they have not communicated with the aggregation server for a long time. Through the experiment, the average time elapsed through each iteration increased since slower models were also included in the early round, but training demonstrated a faster convergence as the mode was exposed to a more significant amount of data. Furthermore, accessing all computing resources available periodically allows fast computing workers to rest, reducing general energy consumption \cite{ORI_51}. The result suggests that access to a broader range of workers is beneficial for faster convergence with a fixed amount of epochs, which will be beneficial to the time efficiency if the issue of waiting for slow computing workers can be resolved.

One point of improvement for the research allowing slow-processing workers to participate in earlier rounds of federated learning training \cite{ORI_51} is that the number of slow-processing workers should be controlled. In this way, the worker selection algorithm can avoid the situation that too many slow-processing workers being selected in the same round, which will affect training efficiency. This issue is improved by the research modifying the worker selection process through clustering \cite{ORI_52}. The clustering methods will first cluster workers based on the time each worker requires to finish training and transmitting the model. After that, the algorithm will select workers from each cluster proportionally. Such a method will ensure that slow-processing clusters can be selected in the earlier round of federated learning training. At the same time, it ensures that the total training efficiency will be minimally affected since only partial slow computing of workers will be selected \cite{ORI_52}. The experiment tested different clustering methods and demonstrated that as long as the average time required by the slowest cluster of workers does not exceed twice the time required by the fastest cluster of workers, the proposed algorithm is more time efficient compared to worker selection that only selects fast workers at the starting rounds of federated learning. This again demonstrates the importance of letting a wider range of workers participate federated learning can allow fast convergence, which is inspiring for this research. The issue in this research is that if a worker is not selected from the cluster from any iteration, then it has to wait at least until the next iteration before it can contribute to the federated learning training process. This is because workers are selected proportionally from each cluster as defined by the algorithm. Waiting for workers who failed to be selected has the potential for further improvements if the algorithm allows all workers to participate simultaneously.

\subsubsection{Balancing strategy}
Most recent research attempts to accelerate the federated learning process by allocating tasks with different workloads to workers with different capacities to allow heterogeneous workers to finish tasks simultaneously  \cite{ORI_44, ORI_47, ORI_48, ORI_49, ORI_50, ORI_51, ORI_52}.

A joint optimisation algorithm is proposed, which considers the parallel model transmission model \cite{ORI_44}. In that research, the transmission power of workers and aggregation servers are adjusted to allow transmission efficiency. Firstly, the algorithm analysis the convergence of federated learning by assuming the gradient is uniformly Lipschitz continuous with respect to model weights and proposes the convergence time as a formula related to worker dataset size. By optimising the convergence time of federated learning training with respect to the worker chosen, a set of workers that will allow the fastest convergence will be selected \cite{ORI_44}. The proposed algorithm then fetches the transmission rate available on each worker and the aggregation server and proportionally divides the transmission power from the aggregation side so that all workers can receive the model simultaneously. As a result, workers with more robust network conditions will not suffer long waiting times since they will receive the model concurrently with workers with worse network conditions. The experiment has shown the effectiveness of proposed algorithms in that training time is reduced up to six per cent compared to random worker selection. However, here are a few points worth further improvement. Firstly, in that research, the algorithm only tries to balance the time workers receive the model. However, the different times required to train the model are also worth consideration. Although the simulated environment ensures the same computing power among all workers in the experiment, different computation resources available among workers can still cause waiting times for fast computing workers. Secondly, the worker selection process suggested by the algorithm needs to be solved by the Hungarian algorithm \cite{ORI_44}, which will change a lot in the worker selected if there are tiny changes in network conditions among workers if workers have similar network conditions. Thus, the worker selection process will be unstable under varying network conditions. This worker algorithm is only conducted once, and once the network condition of workers changes during federated learning training, there is no guarantee that the selected set of workers is optimal.

Instead of tuning the transmission rate to achieve the balance of worker time consumption through tuning the bandwidth, research also intends to tune the amount of data utilized by different workers \cite{ORI_47}. The idea is to tune the amount of data each worker will use to train to allow both slow-computing workers and fast-computing workers to finish processing simultaneously. The algorithm will also consider the time required for transmitting the model to achieve perfect balance \cite{ORI_47}. This method has the advantage of allowing all workers to participate in federated learning, exposing the federated learning training model to a larger dataset. Since more training data can result in faster convergence and hence be more time efficient, this research will also examine methods to incorporate all available workers. However, this method has the potential to fail in the scenario where the variance of worker computing speed varies too much. Suppose some compelling workers can finish the whole training process and respond to the aggregation server before others even start to train the model locally. Then by the balancing model strategy, slow workers will be asked not to train the model and just respond. This is because the balancing strategy will ask slow workers to train fewer data each epoch until it requires the same amount of time fast computing worker needs, which will set the training size of slow workers to zero if the difference between time required by different workers is too significant. Although the algorithm has limitations on handling largely varying workers, the idea of incorporating all workers is proven effective experimentally, reducing the total training time by five per cent \cite{ORI_47}. This suggests the necessity of finding a method that can both allow all workers to participate in federated learning while allowing all workers to explore their full potential rather than using just part of the data or computing resources available.

Other than tuning the amount of data used for training to achieve similar computation time among workers \cite{ORI_47}, model dropouts are also tried to allow workers to finish tasks simultaneously \cite{ORI_48}. The amount of calculation of computer vision convolution layer and fully connected layer can be precisely calculated while the number of computation dropouts can save can also be computed \cite{ORI_48}. The research proposed dropout method works as follows. First, the time required to conduct the calculation as well as transmit the model is calculated based on system statistics, while the worker requires minimum time to respond a model to the aggregation server is marked. After that, all other workers will define the dropout rate so that time saved by dropout will allow them to respond to the aggregation server similarly to the fastest worker \cite{ORI_48}. The advantage of this method is that all workers contribute full computation power and can utilize all data they have for training, avoiding the potential loss of not exploring all data that the method of the tunning amount of training data has \cite{ORI_47}. Moreover, with extensive networks, the dropout rate can vary a lot. This means the potential time that can be saved via dropout model weights can be largely adjusted according to each worker's computation power, allowing workers with larger heterogeneity to cooperate. This method, however, overly relies on the computation power of the aggregation server since, for each model weights response from the worker, the aggregation server has to pick up weights that the worker does not drop out and use those weights for aggregation, which is a timely process \cite{ORI_48}. 

Offloading refers to the act of distributing part of the computation task to the base station. This is then applied to balance the computation time required by federated learning workers \cite{ORI_50}. That research intends to offload part of the computation task to the aggregation server or nearby base station such that slow computing workers have fewer workloads compared to fast computing workers hence simultaneous completion can be achieved \cite{ORI_50}. This method is effective since offloading ensures all available computing resources can be fully utilised. While other methods mentioned above will allow the aggregation server to rest while workers are computing the model, such offloading policy allows the aggregation server to take part of the workers' jobs, which further explore available computing resource. The experiment of this research indicates that with the assistance of an aggregation server for training data, desirable accuracy can be achieved with a twenty-two per cent time improvement \cite{ORI_50}. However, to offload the computation task from the worker to the aggregation server, training data must be sent from the worker to the aggregation server. This requires a specific allowance to support sharing between the worker and aggregation server. Otherwise, without allowance of sharing data to the aggregation server from workers, this method will violate the critical principle of federated learning that data never leaves its source of origin. This research will only take the idea that keeping the aggregation server busy while workers are training to achieve better training efficiency but will not offload tasks to the aggregation server to preserve the privacy of data owners.

\subsubsection{Comparison of optimising federated learning by system parameter research}
Comparisons are shown in table \ref{table:big}. Here all methods reviewed through different perspective and terms are explained as follow:
\begin{itemize}
\item Implementation level: The methods proposed algorithm is examined. A prototype system refers to the algorithm that is deployed on different machines and verified. Simulation refers to only mathematically simulating the federated learning training process, such that the training time and accuracy of the underlying model are not tested on any data but simulated by mathematical models. In contrast, coded simulation mimics the network delay and computation difference on a single machine programmatically.
\item Update frequency: This feature refers to the frequency of proposed algorithms updating the optimisation strategy. While each epoch means the optimisation strategy will be updated each time the aggregation server finishes aggregating model weights, an example is the workers selection process is performed once every aggregation. Once means the optimisation policy is only calculated once and will keep steady throughout the federated learning training process.
\item Check for performance: This feature describes whether performance is analysed throughout the federated learning training process. Some federated learning algorithms only consider the energy required to conduct federated learning, while accuracy is assumed to grow positively as more training epochs are conducted.
\item Consider time constraint: This feature describes whether training time is considered a constraint throughout the training process. Some algorithms proposed only intends to achieve better accuracy of the shared model while training time is not considered, while other algorithms have a strong awareness of time efficiency.
\item Asynchronous: describe whether the algorithm can run asynchronously.
\item System parameter used: This is the raw statistics extracted from a system that are used to formulate federated learning optimisation algorithms. These statistics include:
\begin{itemize}
\item F: The frequency of the processor
\item P: Transmission power 
\item B: Bandwidth available for the worker
\item G: Channel gain
\item D: Distance between the aggregation server and worker
\item DS: Data size available from each worker
\item WL: Workload required to compute one epoch of training from the worker side
\end{itemize}
\item Federated learning parameter tunned:
\begin{itemize}
\item T: Total number of epochs trained throughout the federated learning training process
\item t: Number of epochs trained on workers between two aggregation processes from the aggregation server
\item F: The frequency of the processor
\item P: Transmission power
\item WS: Worker selection policy
\item DR: Model dropout ratio
\item DS: Data size used for training
\item OR: Offloading ratio
\item CLS: Size of cluster
\end{itemize}
\item Derived Parameter:
\begin{itemize}
\item TC: Time required for workers to communicate weights of the model
\item TU: Time required for update: Time required for training as well as loading the model \item and the aggregation process
\item TW: Time needs to wait until an aggregation server or a worker is available
\item EC: Energy required for local training
\item EU: Energy required to upload the model
\item L: Training loss
\item TL: Time since last aggregation with the aggregation server
\item TR: Transmission rate
\end{itemize}
\end{itemize}

\begin{table}[]
\caption{Recent research on system parameter-based federated learning optimisation}
\label{table:big}
\resizebox{\textwidth}{!}{%
\begin{tabular}{|l|lll|l|l|l|l|lllllll|lllllllll|llllllll|}
\hline
\multirow{2}{*}{Essay} & \multicolumn{3}{l|}{Implementation level} & \multirow{2}{*}{\begin{tabular}[c]{@{}l@{}}Update \\ frequency\end{tabular}} & \multirow{2}{*}{\begin{tabular}[c]{@{}l@{}}Check for \\ performance\end{tabular}} & \multirow{2}{*}{\begin{tabular}[c]{@{}l@{}}Consider \\ time constrain\end{tabular}} & \multirow{2}{*}{Asynchronous} & \multicolumn{7}{l|}{System parameters used} & \multicolumn{9}{l|}{Federated learning parameter tuned} & \multicolumn{8}{l|}{Derived parameter} \\ \cline{2-4} \cline{9-32} 
 & \multicolumn{1}{l|}{Prototype system} & \multicolumn{1}{l|}{Simulation} & Coded simulation &  &  &  &  & \multicolumn{1}{l|}{F} & \multicolumn{1}{l|}{P} & \multicolumn{1}{l|}{B} & \multicolumn{1}{l|}{G} & \multicolumn{1}{l|}{D} & \multicolumn{1}{l|}{DS} & WL & \multicolumn{1}{l|}{T} & \multicolumn{1}{l|}{t} & \multicolumn{1}{l|}{F} & \multicolumn{1}{l|}{P} & \multicolumn{1}{l|}{WS} & \multicolumn{1}{l|}{DR} & \multicolumn{1}{l|}{OR} & \multicolumn{1}{l|}{DS} & CLS & \multicolumn{1}{l|}{TC} & \multicolumn{1}{l|}{TU} & \multicolumn{1}{l|}{TW} & \multicolumn{1}{l|}{EC} & \multicolumn{1}{l|}{EU} & \multicolumn{1}{l|}{L} & \multicolumn{1}{l|}{TL} & TR \\ \hline
\cite{ORI_42} & \multicolumn{1}{l|}{YES} & \multicolumn{1}{l|}{YES} & YES & Epoch & YES & YES & NO & \multicolumn{7}{l|}{Generalise and treat all resources equally} & \multicolumn{1}{l|}{\cmark} & \multicolumn{1}{l|}{\cmark} & \multicolumn{1}{l|}{\xmark} & \multicolumn{1}{l|}{\xmark} & \multicolumn{1}{l|}{\xmark} & \multicolumn{1}{l|}{\xmark} & \multicolumn{1}{l|}{\xmark} & \multicolumn{1}{l|}{\xmark} & \xmark & \multicolumn{1}{l|}{\xmark} & \multicolumn{1}{l|}{\xmark} & \multicolumn{1}{l|}{\xmark} & \multicolumn{1}{l|}{\xmark} & \multicolumn{1}{l|}{\xmark} & \multicolumn{1}{l|}{\xmark} & \multicolumn{1}{l|}{\xmark} & \xmark \\ \hline
\cite{ORI_43} & \multicolumn{1}{l|}{NO} & \multicolumn{1}{l|}{YES} & YES & Epoch & NO & YES & NO & \multicolumn{1}{l|}{\cmark} & \multicolumn{1}{l|}{\xmark} & \multicolumn{1}{l|}{\cmark} & \multicolumn{1}{l|}{\xmark} & \multicolumn{1}{l|}{\xmark} & \multicolumn{1}{l|}{\cmark} & \xmark & \multicolumn{1}{l|}{\xmark} & \multicolumn{1}{l|}{\xmark} & \multicolumn{1}{l|}{\xmark} & \multicolumn{1}{l|}{\xmark} & \multicolumn{1}{l|}{\cmark} & \multicolumn{1}{l|}{\xmark} & \multicolumn{1}{l|}{\xmark} & \multicolumn{1}{l|}{\xmark} & \xmark & \multicolumn{1}{l|}{\cmark} & \multicolumn{1}{l|}{\cmark} & \multicolumn{1}{l|}{\cmark} & \multicolumn{1}{l|}{\xmark} & \multicolumn{1}{l|}{\xmark} & \multicolumn{1}{l|}{\xmark} & \multicolumn{1}{l|}{\xmark} & \cmark \\ \hline
\cite{ORI_44} & \multicolumn{1}{l|}{NO} & \multicolumn{1}{l|}{YES} & YES & Once & NO & YES & NO & \multicolumn{1}{l|}{\cmark} & \multicolumn{1}{l|}{\cmark} & \multicolumn{1}{l|}{\cmark} & \multicolumn{1}{l|}{\cmark} & \multicolumn{1}{l|}{\cmark} & \multicolumn{1}{l|}{\xmark} & \cmark & \multicolumn{1}{l|}{\xmark} & \multicolumn{1}{l|}{\xmark} & \multicolumn{1}{l|}{\cmark} & \multicolumn{1}{l|}{\cmark} & \multicolumn{1}{l|}{\cmark} & \multicolumn{1}{l|}{\xmark} & \multicolumn{1}{l|}{\xmark} & \multicolumn{1}{l|}{\xmark} & \xmark & \multicolumn{1}{l|}{\xmark} & \multicolumn{1}{l|}{\cmark} & \multicolumn{1}{l|}{\xmark} & \multicolumn{1}{l|}{\cmark} & \multicolumn{1}{l|}{\xmark} & \multicolumn{1}{l|}{\xmark} & \multicolumn{1}{l|}{\xmark} & \xmark \\ \hline
\cite{ORI_45} & \multicolumn{1}{l|}{NO} & \multicolumn{1}{l|}{YES} & YES & Epoch & YES & NO & NO & \multicolumn{1}{l|}{\cmark} & \multicolumn{1}{l|}{\xmark} & \multicolumn{1}{l|}{\cmark} & \multicolumn{1}{l|}{\xmark} & \multicolumn{1}{l|}{\cmark} & \multicolumn{1}{l|}{\cmark} & \xmark & \multicolumn{1}{l|}{\xmark} & \multicolumn{1}{l|}{\xmark} & \multicolumn{1}{l|}{\xmark} & \multicolumn{1}{l|}{\xmark} & \multicolumn{1}{l|}{\cmark} & \multicolumn{1}{l|}{\xmark} & \multicolumn{1}{l|}{\xmark} & \multicolumn{1}{l|}{\xmark} & \xmark & \multicolumn{1}{l|}{\cmark} & \multicolumn{1}{l|}{\cmark} & \multicolumn{1}{l|}{\cmark} & \multicolumn{1}{l|}{\xmark} & \multicolumn{1}{l|}{\xmark} & \multicolumn{1}{l|}{\xmark} & \multicolumn{1}{l|}{\xmark} & \xmark \\ \hline
\cite{ORI_46} & \multicolumn{1}{l|}{NO} & \multicolumn{1}{l|}{YES} & NO & Epoch & NO & YES & NO & \multicolumn{1}{l|}{\cmark} & \multicolumn{1}{l|}{\cmark} & \multicolumn{1}{l|}{\cmark} & \multicolumn{1}{l|}{\cmark} & \multicolumn{1}{l|}{\xmark} & \multicolumn{1}{l|}{\cmark} & \cmark & \multicolumn{1}{l|}{\xmark} & \multicolumn{1}{l|}{\xmark} & \multicolumn{1}{l|}{\cmark} & \multicolumn{1}{l|}{\cmark} & \multicolumn{1}{l|}{\cmark} & \multicolumn{1}{l|}{\xmark} & \multicolumn{1}{l|}{\xmark} & \multicolumn{1}{l|}{\xmark} & \xmark & \multicolumn{1}{l|}{\cmark} & \multicolumn{1}{l|}{\cmark} & \multicolumn{1}{l|}{\cmark} & \multicolumn{1}{l|}{\cmark} & \multicolumn{1}{l|}{\cmark} & \multicolumn{1}{l|}{\xmark} & \multicolumn{1}{l|}{\xmark} & \cmark \\ \hline
\cite{ORI_47} & \multicolumn{1}{l|}{YES} & \multicolumn{1}{l|}{NO} & NO & Epoch & NO & YES & NO & \multicolumn{1}{l|}{\cmark} & \multicolumn{1}{l|}{\xmark} & \multicolumn{1}{l|}{\cmark} & \multicolumn{1}{l|}{\xmark} & \multicolumn{1}{l|}{\xmark} & \multicolumn{1}{l|}{\cmark} & \xmark & \multicolumn{1}{l|}{\cmark} & \multicolumn{1}{l|}{\cmark} & \multicolumn{1}{l|}{\xmark} & \multicolumn{1}{l|}{\xmark} & \multicolumn{1}{l|}{\xmark} & \multicolumn{1}{l|}{\xmark} & \multicolumn{1}{l|}{\xmark} & \multicolumn{1}{l|}{\cmark} & \cmark & \multicolumn{1}{l|}{\cmark} & \multicolumn{1}{l|}{\cmark} & \multicolumn{1}{l|}{\cmark} & \multicolumn{1}{l|}{\cmark} & \multicolumn{1}{l|}{\xmark} & \multicolumn{1}{l|}{\xmark} & \multicolumn{1}{l|}{\xmark} & \xmark \\ \hline
\cite{ORI_48} & \multicolumn{1}{l|}{NO} & \multicolumn{1}{l|}{YES} & YES & Epoch & NO & NO & NO & \multicolumn{1}{l|}{\cmark} & \multicolumn{1}{l|}{\cmark} & \multicolumn{1}{l|}{\cmark} & \multicolumn{1}{l|}{\xmark} & \multicolumn{1}{l|}{\xmark} & \multicolumn{1}{l|}{\cmark} & \xmark & \multicolumn{1}{l|}{\xmark} & \multicolumn{1}{l|}{\xmark} & \multicolumn{1}{l|}{\xmark} & \multicolumn{1}{l|}{\xmark} & \multicolumn{1}{l|}{\xmark} & \multicolumn{1}{l|}{\cmark} & \multicolumn{1}{l|}{\xmark} & \multicolumn{1}{l|}{\xmark} & \xmark & \multicolumn{1}{l|}{\cmark} & \multicolumn{1}{l|}{\cmark} & \multicolumn{1}{l|}{\xmark} & \multicolumn{1}{l|}{\xmark} & \multicolumn{1}{l|}{\xmark} & \multicolumn{1}{l|}{\xmark} & \multicolumn{1}{l|}{\xmark} & \xmark \\ \hline
\cite{ORI_49} & \multicolumn{1}{l|}{NO} & \multicolumn{1}{l|}{YES} & YES & Epoch & NO & YES & NO & \multicolumn{1}{l|}{\cmark} & \multicolumn{1}{l|}{\cmark} & \multicolumn{1}{l|}{\cmark} & \multicolumn{1}{l|}{\cmark} & \multicolumn{1}{l|}{\cmark} & \multicolumn{1}{l|}{\xmark} & \cmark & \multicolumn{1}{l|}{\xmark} & \multicolumn{1}{l|}{\xmark} & \multicolumn{1}{l|}{\cmark} & \multicolumn{1}{l|}{\cmark} & \multicolumn{1}{l|}{\cmark} & \multicolumn{1}{l|}{\xmark} & \multicolumn{1}{l|}{\xmark} & \multicolumn{1}{l|}{\xmark} & \xmark & \multicolumn{1}{l|}{\cmark} & \multicolumn{1}{l|}{\cmark} & \multicolumn{1}{l|}{\cmark} & \multicolumn{1}{l|}{\cmark} & \multicolumn{1}{l|}{\cmark} & \multicolumn{1}{l|}{\xmark} & \multicolumn{1}{l|}{\xmark} & \xmark \\ \hline
\cite{ORI_50} & \multicolumn{1}{l|}{YES} & \multicolumn{1}{l|}{NO} & NO & Once & NO & YES & NO & \multicolumn{1}{l|}{\cmark} & \multicolumn{1}{l|}{\xmark} & \multicolumn{1}{l|}{\xmark} & \multicolumn{1}{l|}{\xmark} & \multicolumn{1}{l|}{\xmark} & \multicolumn{1}{l|}{\xmark} & \cmark & \multicolumn{1}{l|}{\xmark} & \multicolumn{1}{l|}{\xmark} & \multicolumn{1}{l|}{\xmark} & \multicolumn{1}{l|}{\xmark} & \multicolumn{1}{l|}{\xmark} & \multicolumn{1}{l|}{\xmark} & \multicolumn{1}{l|}{\cmark} & \multicolumn{1}{l|}{\xmark} & \xmark & \multicolumn{1}{l|}{\cmark} & \multicolumn{1}{l|}{\cmark} & \multicolumn{1}{l|}{\xmark} & \multicolumn{1}{l|}{\xmark} & \multicolumn{1}{l|}{\xmark} & \multicolumn{1}{l|}{\xmark} & \multicolumn{1}{l|}{\xmark} & \xmark \\ \hline
\cite{ORI_51} & \multicolumn{1}{l|}{YES} & \multicolumn{1}{l|}{NO} & YES & Epoch & YES & NO & NO & \multicolumn{1}{l|}{\cmark} & \multicolumn{1}{l|}{\xmark} & \multicolumn{1}{l|}{\cmark} & \multicolumn{1}{l|}{\xmark} & \multicolumn{1}{l|}{\xmark} & \multicolumn{1}{l|}{\cmark} & \xmark & \multicolumn{1}{l|}{\xmark} & \multicolumn{1}{l|}{\cmark} & \multicolumn{1}{l|}{\xmark} & \multicolumn{1}{l|}{\xmark} & \multicolumn{1}{l|}{\cmark} & \multicolumn{1}{l|}{\xmark} & \multicolumn{1}{l|}{\xmark} & \multicolumn{1}{l|}{\xmark} & \xmark & \multicolumn{1}{l|}{\xmark} & \multicolumn{1}{l|}{\xmark} & \multicolumn{1}{l|}{\xmark} & \multicolumn{1}{l|}{\xmark} & \multicolumn{1}{l|}{\xmark} & \multicolumn{1}{l|}{\cmark} & \multicolumn{1}{l|}{\cmark} & \xmark \\ \hline
\cite{ORI_52} & \multicolumn{1}{l|}{NO} & \multicolumn{1}{l|}{YES} & NO & Epoch & NO & YES & NO & \multicolumn{7}{l|}{Generalise and treat all resources equally} & \multicolumn{1}{l|}{\xmark} & \multicolumn{1}{l|}{\xmark} & \multicolumn{1}{l|}{\xmark} & \multicolumn{1}{l|}{\xmark} & \multicolumn{1}{l|}{\cmark} & \multicolumn{1}{l|}{\xmark} & \multicolumn{1}{l|}{\xmark} & \multicolumn{1}{l|}{\xmark} & \cmark & \multicolumn{1}{l|}{\cmark} & \multicolumn{1}{l|}{\cmark} & \multicolumn{1}{l|}{\xmark} & \multicolumn{1}{l|}{\xmark} & \multicolumn{1}{l|}{\xmark} & \multicolumn{1}{l|}{\xmark} & \multicolumn{1}{l|}{\xmark} & \xmark \\ \hline
\end{tabular}%
}
\end{table}

The following conclusion can be seen from the comparison table and review above. Firstly, optimising federated learning based on system parameters is effective and needs to be applied to increase time efficiency. Secondly, the workers selection process is essential to improve federated learning. However, resource loss is unavoidable no matter how the worker selection algorithm is designed as long as computing power among workers varies. Thirdly, asynchronous federated learning, which has not been examined before worth exploring since it has the potential to explore all computation resources available fully. Fourthly, in order to make the algorithms easily deployable under different sets of workers with various computing power, the algorithm has to have the ability to be dynamically adjusted according to worker performance. Fifthly, an implementation framework that allows different methods or algorithms to be tested under the same hardware setup is necessary for better comparison.

Firstly, optimisation based on statistics describing workers participating in federated learning training is effective \cite{ORI_42,ORI_43,ORI_44,ORI_45,ORI_46,ORI_47,ORI_48,ORI_49,ORI_50,ORI_51,ORI_52}. From comparison results regarding random worker selection or no optimisation, different research demonstrates a reduction in training time of federated learning between five per cent to twenty-two per cent. This is because the optimisation of federated learning in a heterogeneous environment can improve the utilisation rate of available resources compared to no optimisation, which is more time efficient. By saving training time of federated learning, the model with a desirable accuracy rate can be generated faster, increasing the response rate and giving back the workers' computing resources faster, which is energy efficient. Thus, optimising federated learning through system parameters is necessary.

Secondly, the worker selection process is effective. However, it is hard to explore all computation resources available to different workers if workers have heterogeneous computing resources \cite{ORI_43, ORI_45, ORI_46, ORI_51, ORI_52}. The effectiveness of worker selection can be seen from the experiment result and the reasoning behind worker selection. Experiment demonstrate less time required to reach desirable accuracy thorough worker selection \cite{ORI_43, ORI_45, ORI_46, ORI_51, ORI_52}. The time improvement is because worker selection allows more efficient training step conducted with a fixed time budget, and more efficient training step refers to letting fast computing workers train more epochs \cite{ORI_43} or select more energy efficient workers \cite{ORI_45} or partially take slow computing workers with fast computing workers \cite{ORI_46}. However, as long as the worker selection strategy excludes some workers from training based on the selection policy, computing resources of those unselected workers are wasted. Furthermore, although balancing strategies desire to let all workers finish tasks simultaneously \cite{ORI_44,ORI_47,ORI_48}, which minimises the effect of unbalanced waiting time. This is based on the constraining computation ability of fast computing workers, such as constraining the amount of data being utilised \cite{ORI_44} or available network bandwidth \cite{ORI_47}, which failed to explore the full potential of the computing capacity of those compelling workers. Thus, it can be concluded that as long as federated learning is conducted synchronously, which means workers not selected will not train or not utilise computation resources. It is hard to explore the full potential of computing power among workers. This leads to the next point, that asynchronous federated learning is important.

Thirdly, asynchronous federated learning has the potential to explore the full computation capacity of workers. While unselected workers have to wait until the next round of federated worker selection to contribute computing power to the federated learning training, asynchronous federated learning can allow all workers to participate in federated learning. This is because asynchronous federated learning is not concerned that slow computing workers will keep fast computing workers waiting. Instead, asynchronous federated learning can start aggregation when fast computing workers finish responding with their trained model weights. And when slow-responding workers finish training later, the aggregation server can then merge the results from slow workers. In this way, both slow-computing workers and fast-computing workers can contribute. The only concern is the extra complexity of handling asynchronous aggregation of the federated learning model, which requires a different algorithm design. This research will explore an algorithm to merge responses from different workers asynchronously. One direction that will be examined is weighted federated averaging methods \cite{ORI_9}, \cite{ORI_43}. Moreover, as shown in the table, none of the federated learning optimisation approaches above considers asynchronous federated learning, which emphasises the importance of exploring asynchronous federated learning based on system parameters (RQ 2).

Fourthly, the algorithm needs to change dynamically according to federated learning performance and available worker resources. This is essential since heterogeneous computing resources available among workers can vary a lot from place to place, while experiment setup is mainly different from reality. Moreover, even under a fixed set of workers, the computing resource available is subject to change during the federated learning training. The optimisation policy needs to be adaptively changed according to updates of system parameters from workers to keep the optimisation algorithm time-efficient and energy efficient.

Lastly, as shown in the table above, different methods will depend on different system parameters and derive different intermediate results for optimisation purposes. However, different research always compares with no optimisation or random worker selection \cite{ORI_42,ORI_43,ORI_44,ORI_45,ORI_46,ORI_47,ORI_48,ORI_49,ORI_50,ORI_51,ORI_52}, which partially proves the proposed algorithms' necessity. However, these methods are hard to compare with each other since it is hard to reproduce the experimental results due to the difference between the computation power of experimental devices. Moreover, the efficiency of those algorithms can also be based on code implementation rather than the algorithm itself. Consequently, implementing a federated learning framework that allows different algorithms to be added easily will support the comparison of the effectiveness of different algorithms  (RQ 1) is meaningful.

This suggests the direction of this research that implement a federated learning framework that can be easily extended to different worker selection algorithms (RQ 1) and explore optimisation based on system parameters through asynchronously merged results (RQ 2). The framework implementation needs to be extensible with an extension point for worker selection policy and tunning federated learning aggregation frequency and balancing strategy to allow different algorithms \cite{ORI_42,ORI_43,ORI_44,ORI_45,ORI_46,ORI_47,ORI_48,ORI_49,ORI_50,ORI_51,ORI_52} to be added easily to the framework.

\section{Implementation that can assist federated learning}\label{section:2.3}
\subsection{Existing federated learning framework}
First goal of the research is to implement a federated learning framework such that different machine learning algorithms can be easily added to. Based on this idea, this section will review widespread federated framework implementation, namely, Pysyft \cite{ORI_20}, TensorflowFederated \cite{ORI_21} and FATE \cite{ORI_22}.

Pysyft is a federated learning implementation provided by the team of PyTorch \cite{ORI_53}. The implementation of Pysyft is based on the same numerical library as PyTorch, while Pysyft provides further implementation on corporation and network between workers to achieve federated learning functions. Core ideas from Pysyft implementation that are inspiring for implementation in this research are the pointer idea to identifying models. The pointer idea uses an abstract class to store all necessary information to identify a remote model. Since in federated learning, the aggregation server has to request workers to perform training and request worker to response weights which need a method to refer to the model on the worker side. In the meantime, the worker needs to request the aggregation server to send the shared model weights to itself, which also needs information to identify the model interested among all different workers stored on the aggregation side. By holding a pointer relating to a remote model, different instructions can be sent regarding that model while no extra information is required. This is convenient for implementation, and such a pointer class can act as a placeholder representing a model that exists on different sites, improving the readability of code. This shows the convenience and necessity of a pointer class which will also be included in the implementation of this research. Alternative methods that allow remote sites such as workers or the aggregation server to send instructions about a model locally are remote procedure calls grouped with stub implementation by TensorflowFederated or FATE \cite{ORI_21,ORI_22}. Such methods will let a model expose all its functions as remote procedure call for others, and all other sites that wish to call those functions need to implement a stub to send the instructions. This method has the advantage that the code on the remote and local sides when calling a function is no different, which is user-friendly. However, whenever a new model needs to be implemented, all functions for that new model that wants to be remotely callable must go through the same remote procedure call and stub implementation, which fails to be highly extensible. Instead, for the pointer method, a new model can be easily added as long as it implements a pointer class so remote sites can refer to the model. This simple step makes the pointer method highly extensible and will be used in this research.
When the remote model can be referred to by pointer implementation, it is also essential to let the remote site know which action to perform. Pysyft implements this through an abstract syntax tr\cite{ORI_20}, while Tensorflow Federated and FATE already include this in remote procedure calls \cite{ORI_21}. Abstract syntax tree refers to the implementation practice of letting both workers and the aggregation server store the same module, which contains basic operations required for machine learning, PyTorch also uses this module as the foundation for numerical operation \cite{ORI_20}. However, this makes the development process of the new machine learning algorithm slow since expressing different actions, such as training or calculating gradient as a set of numerical operations remotely, is nontrivial. This research intends to make different models easily extensible to the framework, while the complex development process through AST will reduce the extensibility. Thus, this research will use the alternative solution, which is a remote procedure call. The remote procedure call implementation used by FATE and Tensorflow Federated express each model's action as a remote procedure call function \cite{ORI_21,ORI_22}. In this way, the remote procedure call related to linear regression training differs from the remote procedure call related to support vector machine training. Compared to the AST approach, each function is expressed as a single function call to remote sites, which is a simplification and will facilitate the implementation. This method can be further abstracted for this research. Since this research will identify a remote object through pointers, thus it is not necessary to implement different remote procedure call functions for the same operation related to different machine learning models. While FATE and Tensorflow Federated need different remote procedure call function names to distinguish between different machine learning models, this research can be differentiated by pointers. So that the remote procedure call operation has only to be defined once, and the new model can be easily extended by implementing those common operations.

Another factor that will be interested in this research is the transmission of models. The aggregation server and workers must communicate model weights regularly. There are two options available, one is directly transmitting the model through the same channel that instructions are sent, which Pysyft and Tensorflow Federated used \cite{ORI_21,ORI_22}. Alternatively, FATE suggests a method that the aggregation server will send its weights to a data warehouse. Whenever it wishes to send the weights to a worker or the worker is fetching the weights from it, instead of communicating with the aggregation server, the worker can download model weights from the warehouse that stores the copy of the weights of the aggregation server \cite{ORI_22}. The simplicity of implementation is the advantage of directly sending model weights via the channel used to send instructions since the same communication port can be used without additional development. The data warehouse method is more complicated in terms of development since it requires an extra component that will communicate with the aggregation server and workers, which requires extra setup and monitoring for network connections. However, the data warehouse method has the following advantage that will facilitate the federated learning training. Firstly, it relieves the network pressure at the aggregation side as well as the worker side. The aggregation server only needs to send out model weights once to the data warehouse, which will reduce network pressure compared to sending multiple copies to different workers. Since the data warehouse can be just responsible for model transmission, thus, workers can download the model weights of the aggregation server when their network is available, allowing workers to schedule model transmission tasks more flexibly. Secondly, since the data warehouse will be responsible for model weights transmission, the communication port the aggregation server used for sending instructions can deliver instructions more efficiently, increasing the time efficiency. For the reasons above, this research will implement a data warehouse structure where workers and the aggregation server will communicate the model.

As described by the decisions above, this research will use the pointer, which has all information required to identify a model remotely held by workers or the aggregation server. Moreover, the implementation will use remote procedure calls to make new models easily extensible. Lastly, this research will implement a data warehouse responsible for transmitting model weights, which is beneficial for communication efficiency and will provide more flexibility for workers.

\subsection{Existing resource management framework}
In order to implement the federated learning framework, it is necessary to choose a resource management framework as the foundation for implementation. Since this research intends to optimise federated learning based on system parameters, thus, the foundation framework should be capable of providing those system parameters. Moreover, this research will deploy the federated learning implementation on different machines, which states the importance of making the framework compatible with different operating systems. Lastly, as described in section 2.1.1, the connection establishment step requires the aggregation server to be aware of available computing entities within the network. This requires resource discovery functionalities. With those requirements, FogBus2 \cite{ORI_19} is chosen for this research.

Firstly, FogBus2 is chosen since it can provide system parameters \cite{ORI_19}. In FogBus2, the central controlling node will perform system profiling whenever a site with computing power is registered to the central controlling node \cite{ORI_19}. Throughout the process, CPU frequencies, processor availability, network conditions, ram availability and more system statistics can be easily accessed. Moreover, the framework will conduct a profiling process periodically such that updates on available computation power can be easily noticed. This feature is helpful for asynchronous federated learning in this research. This is because asynchronous federated learning will merge model weights from workers more often. After each merging, new worker selection can be conducted, resulting in a more frequent worker selection process. Frequent updates about system parameters among workers allow asynchronous federated learning to update worker selection accordingly. As discussed above, a more dynamic adaptive worker selection policy will result in more time-efficient federated learning. This suggests the frequent system parameter update is beneficial for more efficient federated learning practice. Lastly, system parameters are stored in a local database at the point of the central site. This further facilitates the process of obtaining system parameters. Since the aggregation server only needs to request to the local database in order to fetch those parameters and does not need to request from the central controlling node that might not be available at the time of requesting. The alternative framework can also provide system parameters. This includes a container enabled framework \cite{ORI_54} as well as one called FogPlan \cite{ORI_55}. However, FogPlan has no dynamic scheduling policy support, which means system parameters will not be automatically updated but requires a fetch command from the central controlling node. This requires the aggregation server to ask the central controlling node to ask remote workers for those system parameter updates, which is a longer process compared to reading from the local database provided by FogBus2 directly. The same issue also arises in the other container enabled framework \cite{ORI_54}. This demonstrates that FogBus2 is an excellent option for providing system parameters due to its simplicity of accessing system parameters and automatically updating them.

Secondly, FogBus2 is dockerised \cite{ORI_19}. If federated learning is implemented on top of FogBus2, it can easily be deployed on machines with heterogeneous environments and various operating systems. FogBus2 is runnable with minimal docker images required. For the central controlling node, the Master images need to be installed and run \cite{ORI_19}, while for all other workers, only the Actor image is required \cite{ORI_19}. This means that in order for workers to participate in federated learning, they only need to install part of the docker image, which is lightweight and will save memory. Moreover, docker has a relative faster start-up time compared to the virtual machine or other virtualisation tools. This allows workers to respond faster during the initialisation process, saving the overall training time. The time saved will contribute to the goal of this research to conduct federated learning more efficiently. Moreover, configuration over dockerised FogBus2 is relatively easy \cite{ORI_19}. Since federated learning intends to train machine learning models, which will require support from a machine learning library such as PyTorch \cite{ORI_53}. While manually installing machine learning dependencies on different operating systems require various configuration, through dockerised FogBus2, it can be easily achieved by adding dependencies into the configuration file. This facilitates the development process for new models since dependencies can be easily configured, supporting the research goal of making new models easily extensible. An alternative framework that can enable federated implementation easily deployable on different operating systems includes FogPlan \cite{ORI_55} and another centralised container-enabled framework \cite{ORI_56}. Although FogPlan has integrated docker support, which also provides portability, it needs to install all components to operate. This means both the aggregation server and workers must fully install FogPlan to be deployable on a heterogeneous environment. In contrast, FogBus2 only needs minimal necessary components installed on the aggregation server and worker instead of fully installing to allow different federated learning participants to communicate with each other. The fewer installation requirements FogBus2 provides over FogPlan will accelerate the federated learning process and make it more suitable for this research. The other centralised dockerised framework can support protability \cite{ORI_56}. However, it does not have any resource discovery components, which require additional development to obtain system parameters. This is less convenient compared to FogBus2.

Lastly, FogBus2 provides a resource discovery module that will facilitate the connection establishment process of federated learning. The connection establishment process of federated learning is when different participants connect with each other and know their role in the federated learning framework. This requires network configuration to allow different sites to communicate with each other while message handlers on each site need to forward the message to the federated learning training agent. Moreover, this requires the participant who starts the federated learning process to be aware of potential computing power available within the network so it can send initiative commands to those sites to invite them to participate in federated learning. FogBus2 encapsulates the process of finding available computing resources available among all sites that run the Actor component \cite{ORI_19}. Moreover, it has an integrated round-robin task allocation policy, allowing all available sites with sufficient computing power to conduct federated learning to participate in federated learning training, which is convenient for testing and verifying federated learning implementation. The alternative solution is letting one site listen for incoming requests for sites that wish to participate in federated learning, and sites need to send the registration information to the central point to express their interest in participating. However, this method has the disadvantage that each site needs to send the message separately, which requires additional setup for each new site added. This extra process is not scalable, which hinders extensibility. Other resource management frameworks, such as FogPlan \cite{ORI_56} or the dockerised resource framework \cite{ORI_54} can also provide the resource discovery module and operate similarly to FogBus2. However, the comparison above stated that those two framework fails other features such as portability or system parameter profiling.

In conclusion, FogBus2 is an integrated resource profiling module that can provide the system parameters required, dockerised to support easy extensibility on heterogeneous operating systems and a resource discovery module to facilitate relationship establishment of federated learning in one single framework. Other resource management frameworks such as FogPlan \cite{ORI_54} or other dockerised frameworks \cite{ORI_55,ORI_56} can provide only part of the requirements for federated learning implementation in this research, which makes FogBus2 more suitable and thus will be chosen.

\lhead{\emph{Methods}}  
\chapter{Methods}\label{chapter:3}

\section{Federated learning in FogBus2}\label{section:fl_implementation}

\begin{figure}[h]
    \centering
    \includegraphics[width=\textwidth,height=\textheight,keepaspectratio]{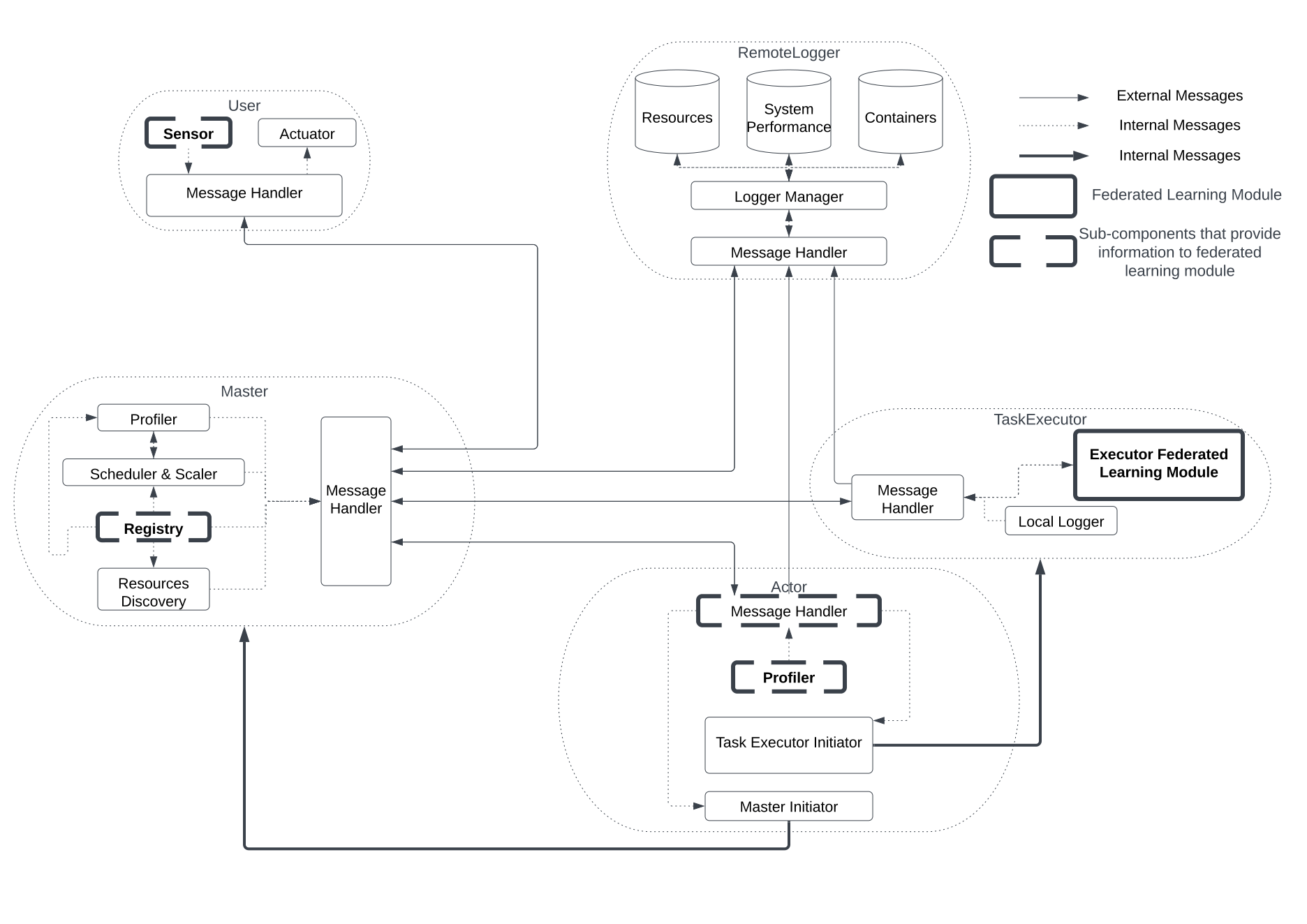}
    \caption{Federated Learning in FogBus2}
    \label{fig:3-1}
\end{figure}

The relationship between the federated learning module and FogBus2 framework is shown in figure \ref{fig:3-1}. In the figure, dashed bold boxes mean sub-components of FogBus2 that will provide information to the federated learning module. Moreover, the federated learning module overrides the executor module originally in FogBus2. This section will first introduce how modules in FogBus2 can provide information to support the federated learning module and how the federated learning module exists as tasks in the Executor module of FogBus2.

\subsection{Sensor}
The sensor sub-component in the User module of FogBus2 is used to grasp input from users and forward it to the Master module, which will be further forwarded to the executor within the TaskExecutor module. Executors are classes that run federated learning tasks as functions where function inputs are inputs from users stored in a dictionary. The sensor sub-component is then programmed to ask hyper parameters from users for the federated learning training process. While there can be various hyper parameters specific to different federated learning algorithms deployed, in this research, hyper parameters collected from users include:
1) The model shared by participants corporate in the federated learning training,
2) Total number of aggregations to perform on the aggregation server side,
3) Number of training epochs each worker has to complete before contributing the local model weights to the federated learning aggregation process,
4) Whether the federated learning training is conducted synchronously or asynchronously,
5) The learning rate initially used by different workers to update the model.
After the federated learning application starts in FogBus2 and all task executors are ready, the sensor sub-component collects those hyper parameters from users is the first process of federated learning training. After that, different executors will conduct federated learning training based on those hyper parameters.

\subsection{Registry}
In the federated learning implementation, different participants must regularly communicate with each other to transmit messages and model weights. This requires that different Executors running federated learning tasks know the IP address and port number of other federated learning Executors before federated learning training starts. In FogBus2, different tasks within an application are linked to each other according to a dependency graph. In the dependency graph, a task may be the parent of one or more tasks and be the child of one or more tasks. Furthermore, results from parent classes will be forwarded to their child classes as input to task function calls on child classes. This feature is used to transmit the IP address and port number between different Executors of federated learning. In this research,  the aggregation server is responsible for starting the federated learning training process by creating the federated learning model and calling selected workers to start training. This makes the Executor runs the aggregation server needs to know the network address of all other Executors running as workers so that it can send instructions to others in the first place. Other Executors running as a worker only need to wait until the aggregation server contacts them, so the network address of the aggregation server is available, and they can send messages back. Thus, the implementation let the task that operates as the aggregation server be a child task of all other tasks which run as workers. Moreover, the implementation lets the returning results of tasks that run as workers be the network address they are listening to. These results will be inputs of the task running the aggregation server. In this way, the aggregation server can have the network address of all other workers before federated learning training starts. Since this research only examines federated learning mechanisms of one level server to worker architecture rather than hierarchical federated learning, there are multiple workers and only one aggregation server. Consequently, the dependency graph is defined as multiple worker tasks being parent tasks of the task running aggregation server.

\subsection{Message Handler}
Since Executors running as workers needs to return the network address they are listening to, they first need to start a socket server listening on that address. However, in the original design of FogBus2, Executor, which is responsible for executing tasks, is not aware of their network address since they are only responsible for calculation. In order to make the Executor aware of the IP address for tasks running federated learning training, the implementation takes the IP address from the Actor module. The Actor module is responsible for starting the docker container of the TaskExecutor module in place, while the TaskExecutor module is the Executor sub-component. Thus, the Actor is physically on the same machine as the TaskExecutor, which makes the IP address of the Message Handler sub-component of the Actor module the same as the IP address of the Executor. The implementation adds a tag noting whether input data from users are related to federated learning or not. When the Actor module calls the TaskExecutor module to start tasks, it will add the IP address of its Message Handler to the input dictionary if the tag indicating tasks are related to federated learning. In this way, federated learning tasks can use that IP address to start a socket server and provide the correct IP address as results, which is forwarded to child tasks. Since the Executor running federated learning tasks need to communicate with each other regularly, the implementation lets them communicate with a separate port instead of the port used by the Message Handler sub-component to avoid conjestion. The port is subject to the port availability of the machine running TaskExecutor. In this way, the TaskExecutor can start the socket server listening on the same IP address and different port compared to the Message Handler of the Actor, initialising it for federated learning purposes.

\subsection{Profiler}
The Profiler sub-component in FogBus2 is responsible for collecting statistics related to available system resources. Federated learning optimisation depends on parameters describing available system resources. Since the Actor initialises the TaskExecutor, the Profiler sub-component within Actor is physically on the same machine as the Executor sub-component within TaskExecutor. Thus, data from Profiler of Actor also describe available resources for the Executor running federated learning tasks. The aggregation server is the site responsible for making optimisation decisions, so it requires system parameters from all workers. The implementation uses the property that parent tasks will pass results to child tasks to pass profiling information from worker Executors to the aggregation Executor. This is achieved by adding profiling data to Executor running as worker results if tasks are tagged as relating to federated learning. The TaskExecutor that runs the aggregation server will then receive the profiling information, which can be exploited to implement different federated learning optimisation mechanisms accordingly.

\subsection{Federated Learning Modules (Executor)}
After properly initialising the aggregation server Executor and multiple worker Executors, the necessary information is ready for federated learning tasks. In particular, worker Executors will start the socket server listening for instructions from the aggregation server. At the same time, the aggregation server Executor will own the network addresses of socket servers on workers, as well as profiling information describing available computing resources for each. The aggregation server task will then call functions from the federated learning module to define the federated learning process and execute the training accordingly. It is worth noting that Executors run as workers still kept alive by holding the socket server alive on a separate thread after returning the socket server address. This is different from other tasks in FogBus2 that will terminate after returning results for child classes. This is because worker tasks still need to regularly listen to instructions from the aggregation server to perform federated learning training.

\section{Federated Learning Modules}
\begin{figure}[h]
    \centering
    \includegraphics[width=\textwidth,height=\textheight,keepaspectratio]{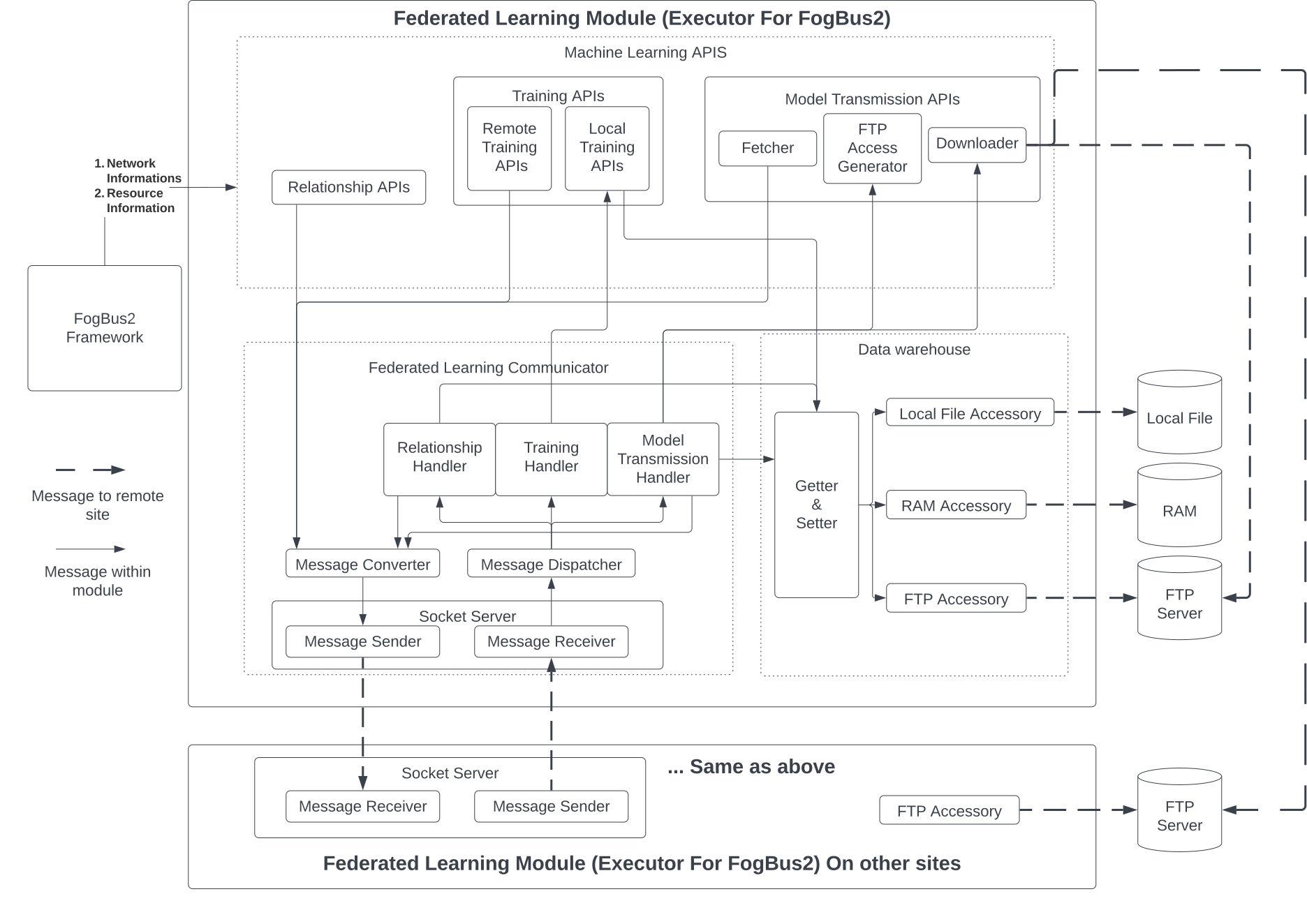}
    \caption{Federated learning module structure}
    \label{fig:3-2}
\end{figure}
Figure \ref{fig:3-2} shows the structure of federated learning implementation in this research. The federated learning implementation module takes the network address of other Executors and statistics of available computing resources among those Executors from the FogBus2 framework. Furthermore, tasks within the FogBus2 framework call functions from the implementation to define the federated learning training process.

The federated learning module is divided into three sub-modules: 1) Machine Learning APIs, 2) Federated Learning Communicator, and 3) Data warehouse. The Data warehouse sub-module allows easy storage and data access required by federated learning. Moreover, the federated learning communicator is used for communicating messages between different participants of federated learning training and handling them correspondingly. Lastly, machine learning APIs are functions that encapsulate federated learning logic such that calling those APIs is sufficient to define federated learning training. Overriding those APIs enable new machine learning model to be trained by federated learning.

\subsection{Data Warehouse Sub-Module}
\label{subsection:dw}

\begin{figure}[h]
    \centering
    \includegraphics[width=\textwidth,height=\textheight,keepaspectratio]{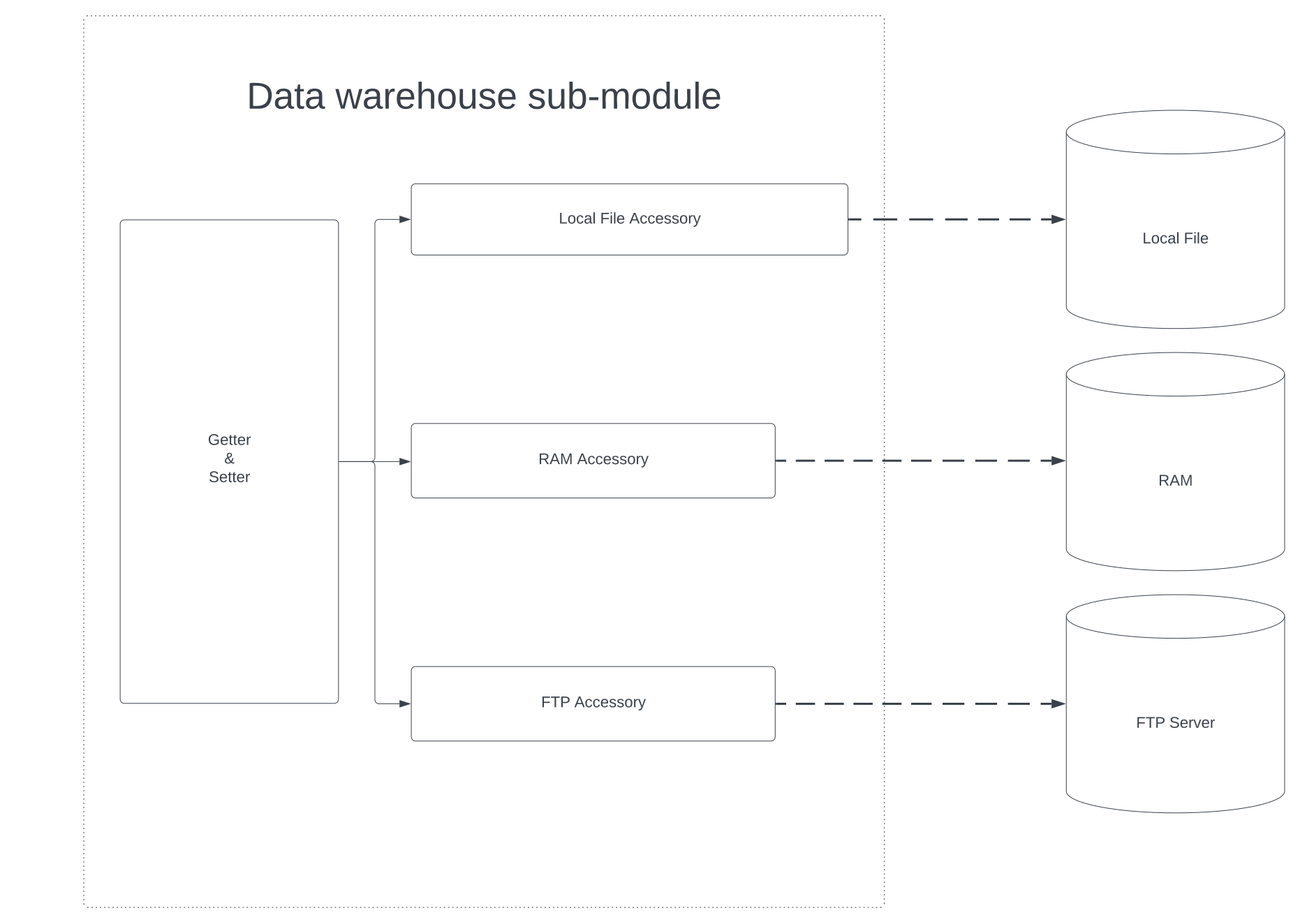}
    \caption{Data warehouse structure}
    \label{fig:3-3}
\end{figure}

Figure \ref{fig:3-3} shows the structure of the data warehouse sub-module.This module is responsible for providing an interface to access and store all kinds of data related to federated learning training. In federated learning training, data that needs to be stored includes 1) Machine learning classes, 2) Parameter weights of machine learning models, 3) Parameter weights of machine learning models of other participants, and 4) Data used for training. These four kinds of data can be placed in different storages, including RAM, a remote repository or files on local storage. However, writing to and retrieving data from those storages will have different implementations, and the data warehouse sub-module encapsulates various implementations. The data warehouse provides the getter and the setter functions such that all kinds of data can be accessed by providing a unique ID. Moreover, if data is saved to the warehouse for the first time, the sub-module will return an ID that uniquely identifies that data.

When data is saved to the data warehouse for the first time via the setter function, the warehouse will store the unique ID as a key. After that, it will store the type of storage media and all necessary information to access data for that storage type as value. It will return the unique ID as function output so that later the ID can be used to retrieve the data. When a unique ID is provided to the getter function, the data warehouse will first use the ID to retrieve the saved credentials and storage type used to store the data corresponding to the provided ID. Then it will use those credentials to retrieve the actual data. While setters allow data to be stored on a specified type of storage, the default storage for machine learning model weights and training data is the local disk. Furthermore, machine learning classes are stored on RAM as default.

This system design allows extension for different storage types. Defining a new storage type can be done by defining the methods to write and retrieve data from it and then adding that to the data warehouse. Later, when using getter, specify that the storage media will enable saving and retrieving data from it.

Due to the design that only a unique ID is sufficient to retrieve data, the machine learning model, which is one type of data, can be referred to only by the ID locally. Referring to a remote machine learning model needs to specify the network address as well. This gives rise to the idea of the Pointer class, used by federated learning training participants to identify a model on a remote site uniquely. The Pointer class consists of the data warehouse's network address and unique ID for it. For example, when the aggregation server asks a remote worker to conduct training, multiple worker network addresses can be saved, and each worker can own several machine learning models. The aggregation server can provide the address to uniquely identify the worker and the unique ID to identify the model on that worker.

\subsection{Communication Sub-Module}
\begin{figure}[h]
    \centering
    \includegraphics[width=\textwidth,height=\textheight,keepaspectratio]{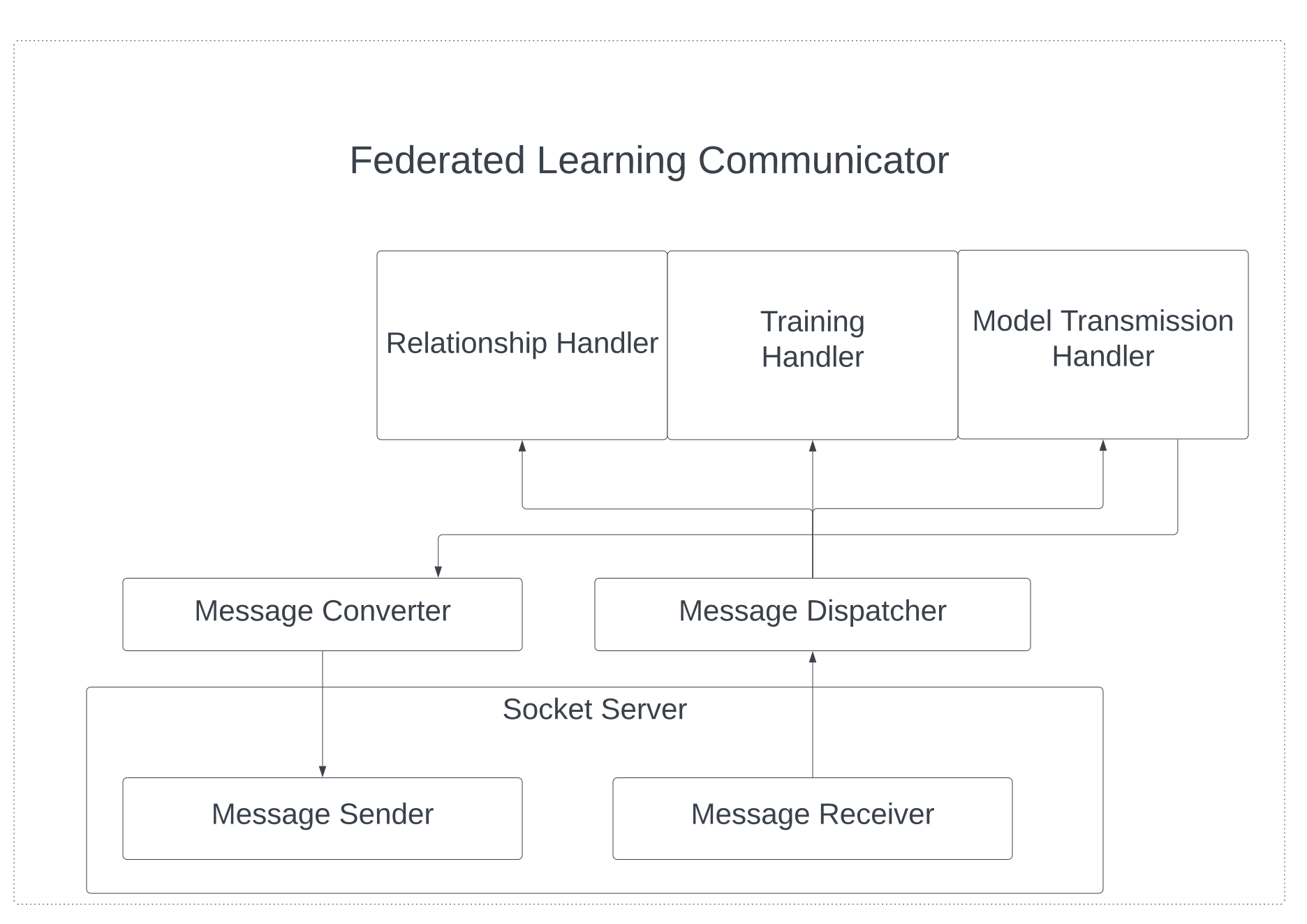}
    \caption{Federated Learning Communicator}
    \label{fig:3-4}
\end{figure}

Figure \ref{fig:3-4} shows the architecture of the federated learning communicator. Since federated learning training involves frequent message communication between different parties, the federated learning implementation provides its communicator. The federated learning communicator sub-module consists of a socket server, a message converter, a message dispatcher and handlers. The socket server is a server that listens to incoming messages via a receiver and sends out messages by the sender. All messages that go through the socket server are binary data. The message converter is responsible for converting messages into binary formats so that data in binary format can be transmitted between the local message sender and remote message receiver. When the message receiver receives a message, it reads the heading five characters, which indicates the topic of the message. The message dispatcher then uses these five characters to forward the message to corresponding handlers. There are three handlers in the implementation. Firstly, the relationship handler is responsible for handling incoming requests for establishing a federated learning relationship. For example, if the aggregation server asks a remote site to be the worker of itself, then the relationship handler on that remote site is responsible for handling related messages. Secondly, training handlers are responsible for handling messages related to federated learning training. This includes the aggregation server asking a worker to start training, and the worker acknowledges to the server that local training is complete. Lastly, model transmission handlers are responsible for sending a request to fetch the weights of a remote model and send back the credentials required to download the weights.

Noting that the weights are not transmitted directly through the socket connection between the message sender and receiver. This is because model weights are large compared to other messages. If sending those weights over the communication channel of federated learning, other messages have to wait for the weights to finish transmission. This long waiting time will influence the time efficiency if sent over the communication channel of federated learning. Alternatively, in the implementation of this research, when a site receives the request to fetch a local model weight, the site will save the weights to an FTP server and send back a one-time login credential. The remote site which fetches the model weights can use that credential to log in to the FTP server and download through file transmission protocol.

\subsection{Federated Learning API Module}
\begin{figure}[h]
    \centering
    \includegraphics[width=\textwidth,height=\textheight,keepaspectratio]{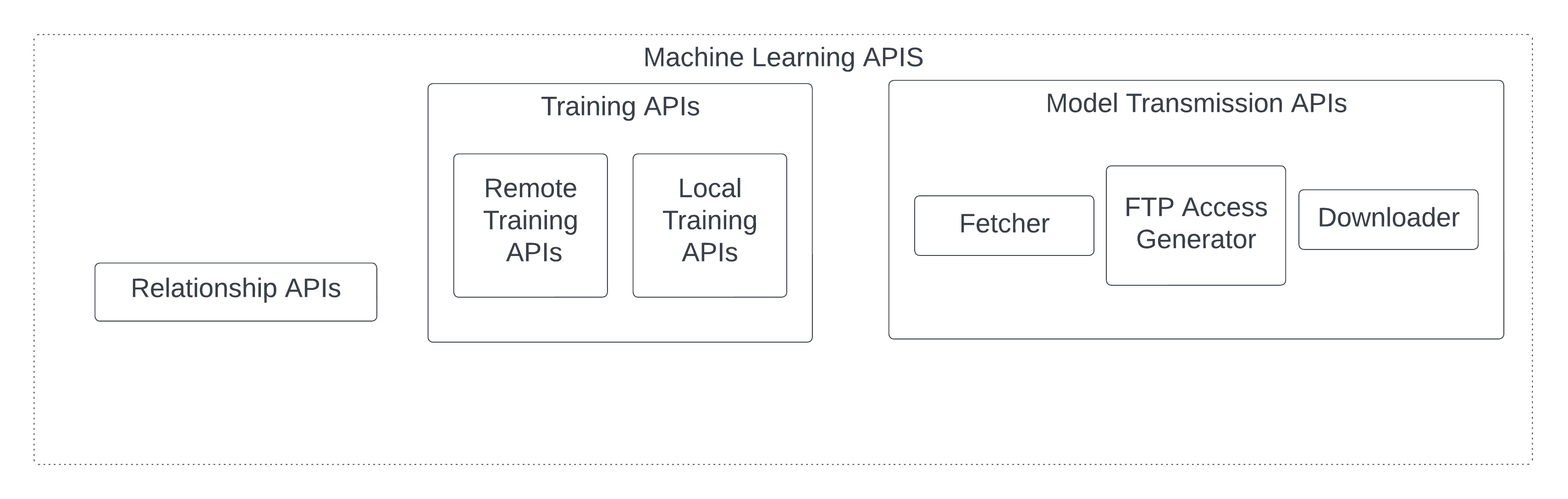}
    \caption{Machine Learning APIs}
    \label{fig:3-5}
\end{figure}

Figure \ref{fig:3-5} show the machine learning APIs sub-module. The federated learning APIs sub-module is a minimum set of functions that a machine learning model needs to define to be trained by different parties. All these functions are collected into a single class. Therefore, machine learning models can override those functions to allow themselves to be deployed for federated learning training. Machine learning APIs are further divided into relationships, training and communication APIs. 

Relationship APIs are functions to establish a relationship with other models. It includes functions to request remote sites to be the worker of itself or request being workers of a remote aggregation server. Functions related to relationships first send a message via the communication sub-module, and the relationship handler on other sites will handle the forwarded message. 

Training APIs are functions related to machine learning training, including remote and local training functions. Remote training APIs are functions used by aggregation servers to request a remote worker to perform specified rounds of training. Those functions need a pointer class that refers to a remote model as an argument. So the message sender within the federated learning communication sub-module can use the network address within the pointer to send the message to a remote message receiver, and the handler can use a unique ID within the pointer on the remote side to retrieve the model via getter of the data warehouse module. Local training APIs are functions used to conduct some calculations on a locally stored model. These include functions to conduct training based on data available, which is the same as regular machine learning training. Moreover, local training APIs include various functions to federate model weights from workers for the aggregation server based on different algorithms. 

Model transmission APIs are functions used to transfer model weights between different parties. Fetcher functions are responsible for sending messages requesting remote model weights. After receiving the fetch request, the FTP access generator on the remote side will generate a credential that the participant fetches the model can use that credential to download the model weights from the FTP server. The actual download is implemented in the downloader.

\section{Corporation Examples}
The machine learning APIs sub-module, federated learning communicator sub-module and data warehouse sub-module together enable federated learning mechanisms to be defined. This section will present how different components corporate together. Including an aggregation server adding a worker, model transmission between two different sites, an aggregation server asking a worker to conduct training, an aggregation server aggregating worker model weights and a complete federated learning training process.

\subsection{Adding a worker}
\begin{figure}[h]
    \centering
    \includegraphics[width=\textwidth,height=\textheight,keepaspectratio]{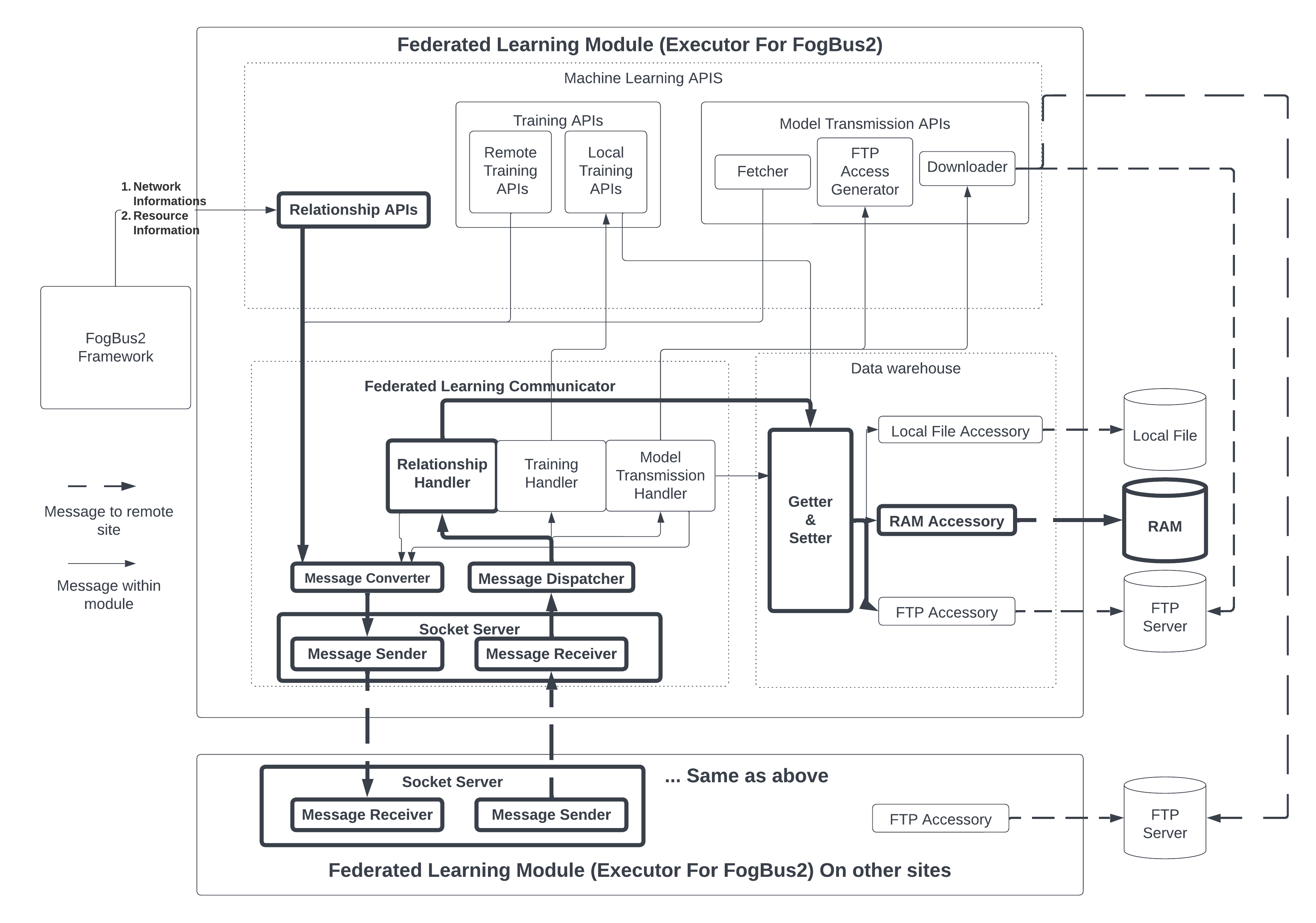}
    \caption{Components for adding worker}
    \label{fig:3-6-1}
\end{figure}

\begin{figure}[h]
    \centering
    \includegraphics[width=\textwidth,height=\textheight,keepaspectratio]{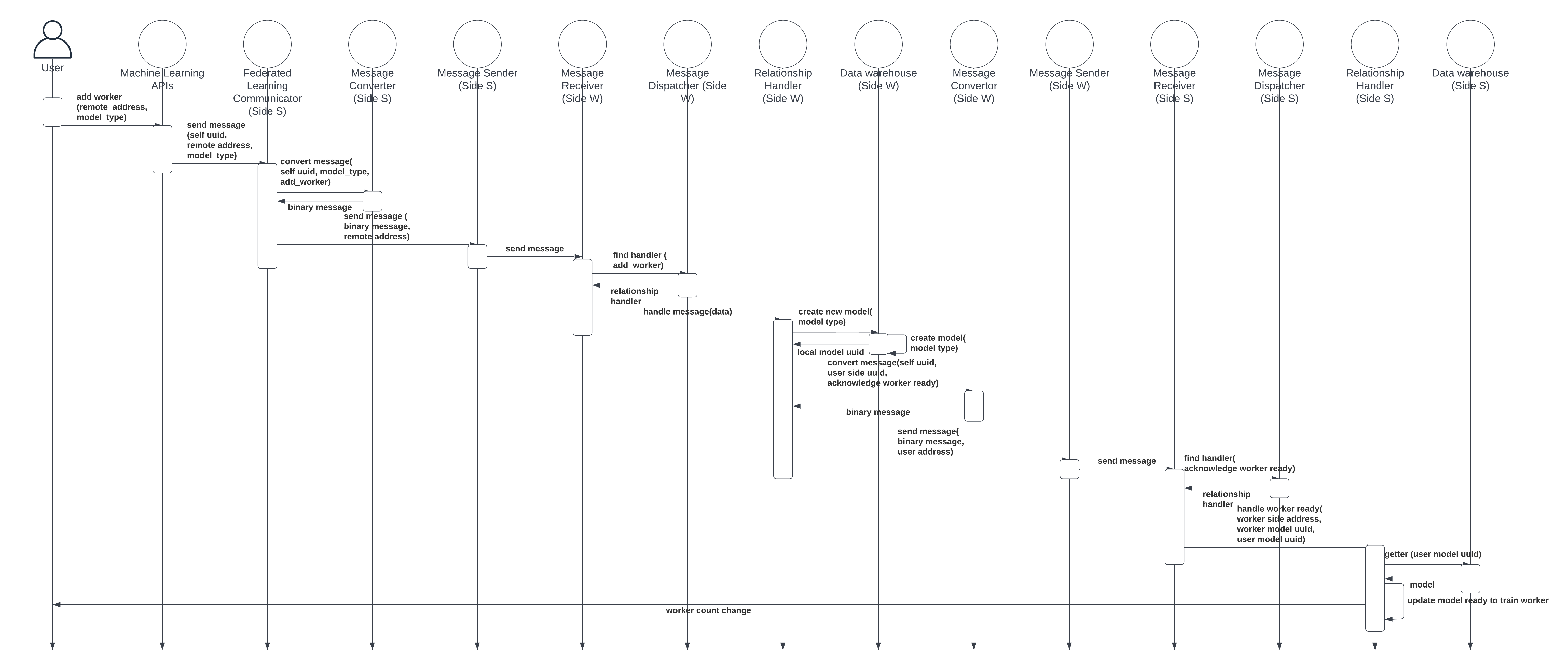}
    \caption{Adding worker sequence diagram}
    \label{fig:3-6-2}
\end{figure}

The components used in federated learning implementation by the aggregation server to add a remote site as a worker is shown in figure \ref{fig:3-6-1} and figure \ref{fig:3-6-2} demonstrate the message flows in between. For clarity of explanation, let all computation on the aggregation server-side, called side S and let all computation on the remote site, which is the site invited to be the aggregation worker, be side W. The function is called on side S initially. The steps involve:

1) Before calling the function, a machine learning model needs to be created on side S, which stands for the aggregation server model.

2) User first calls the function of the relationship function component in the machine learning APIs sub-module of adding a worker on side S. The function is implemented in the aggregation server model, requesting the side W to create a machine learning model with the same structure and be the worker model. Before the request, the worker machine learning model is not present on side W. Consequently, only the network address of the remote participant is required.

2) After calling the function of adding a worker, the function will call the message sender on side S to send an invitation to the remote participant corresponding to the network address. Function arguments also include the unique ID of the aggregation server model. Site W can then create a pointer class composed of the network address of the message sender on side S and a unique ID of the aggregation server model. The remote worker model can use this pointer to refer to its server model, the local aggregation server model.

3) The communicator will then call the message converter on side S to pack the message into a tuple and serialise it into binary data for socket transmission.

4) After receiving data in binary format, the message sender on side S will send the data over the socket to the message receiver on side W.

5) The message receiver on side W will then interpret the message as relating to the relationship by its dispatcher. Consequently, the remaining message is then directed to the relationship handler on side W.

6) The relationship handler on side W will create a model with an identical structure to the aggregation server model on side W. This model is served as the worker model. At the time of model creation, the model will also be added to the data warehouse module for later retrieval. This will provide the unique ID of the worker model.

7) The worker model on side W will save the unique ID of the aggregation server model and the side S network address as the server model's pointer. After this, the worker model is ready for further instructions, such as training.

8) The relationship handler on side W will inform the aggregation server that the worker model is ready. This is done by sending an informing message via message sender on side W. The message will contain the unique ID of the worker model and the aggregation server model, as well as the network address of the side S. Then the message will pass through the converter, which serialises the message and transmit the message through a socket connection.

9) After the message receiver on side S receives the acknowledgement from side W that the worker model is ready, it will let the relationship handler on side S handle the message.

10) The relationship handler on side S will then use the server ID side W provide to retrieve the aggregation server model from the data warehouse. After that, the pointer referring to the worker model, which consists of the unique ID of the worker model and the network address of side W, will be recorded into the aggregation server model class.

After the steps above finish, on side S, the aggregation server model has one extra worker model pointer stored. Moreover, on side W, the worker model has one server pointer stored, referring to the aggregation server model. Similar APIs include letting the remote site be the aggregation server or peer of itself.

\subsection{Sending a model}
\label{subsection:comm_model}
\begin{figure}[h]
    \centering
    \includegraphics[width=\textwidth,height=\textheight,keepaspectratio]{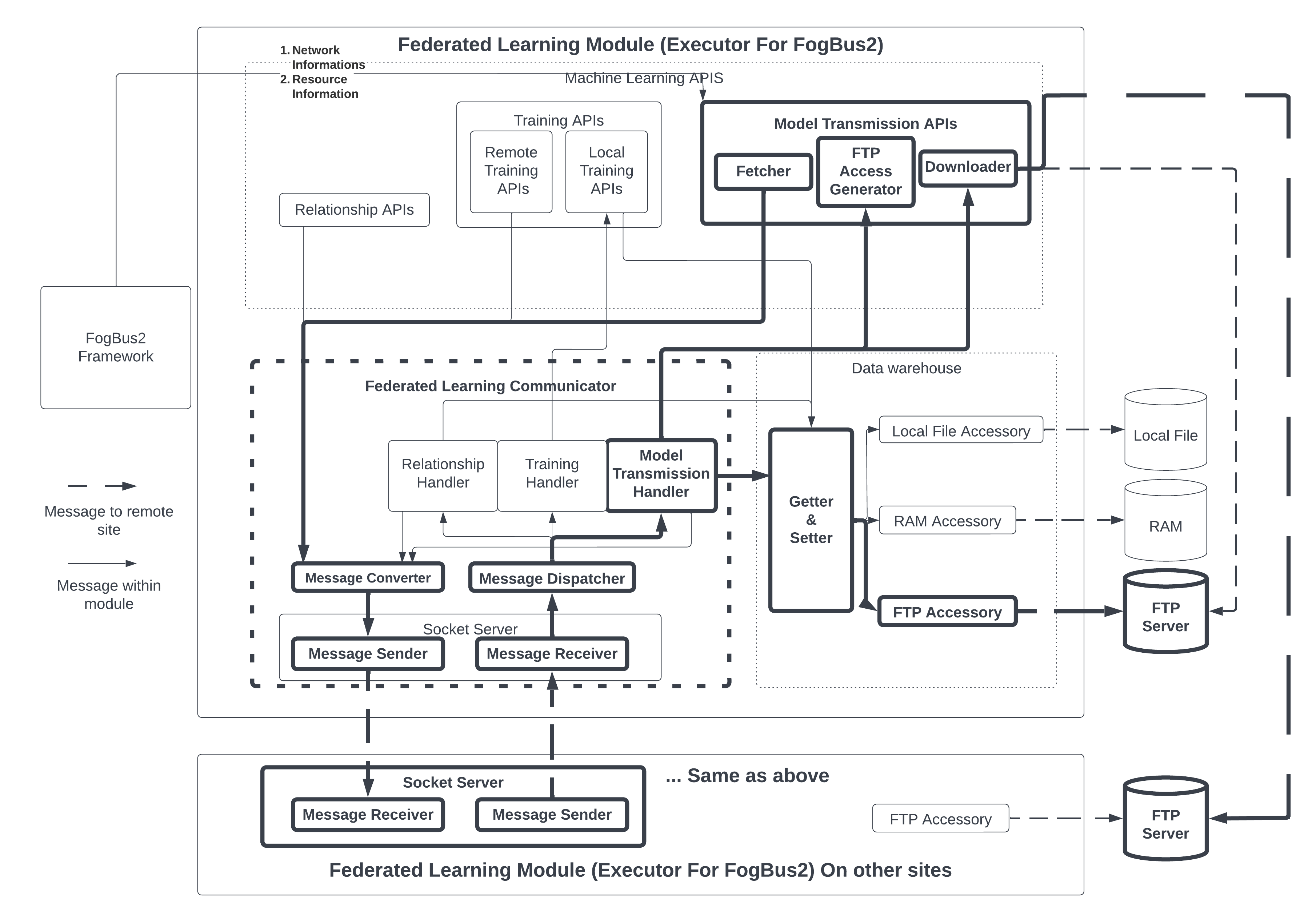}
    \caption{Components for sending model weights}
    \label{fig:3-7-1}
\end{figure}

\begin{figure}[h]
    \centering
    \includegraphics[width=\textwidth,height=\textheight,keepaspectratio]{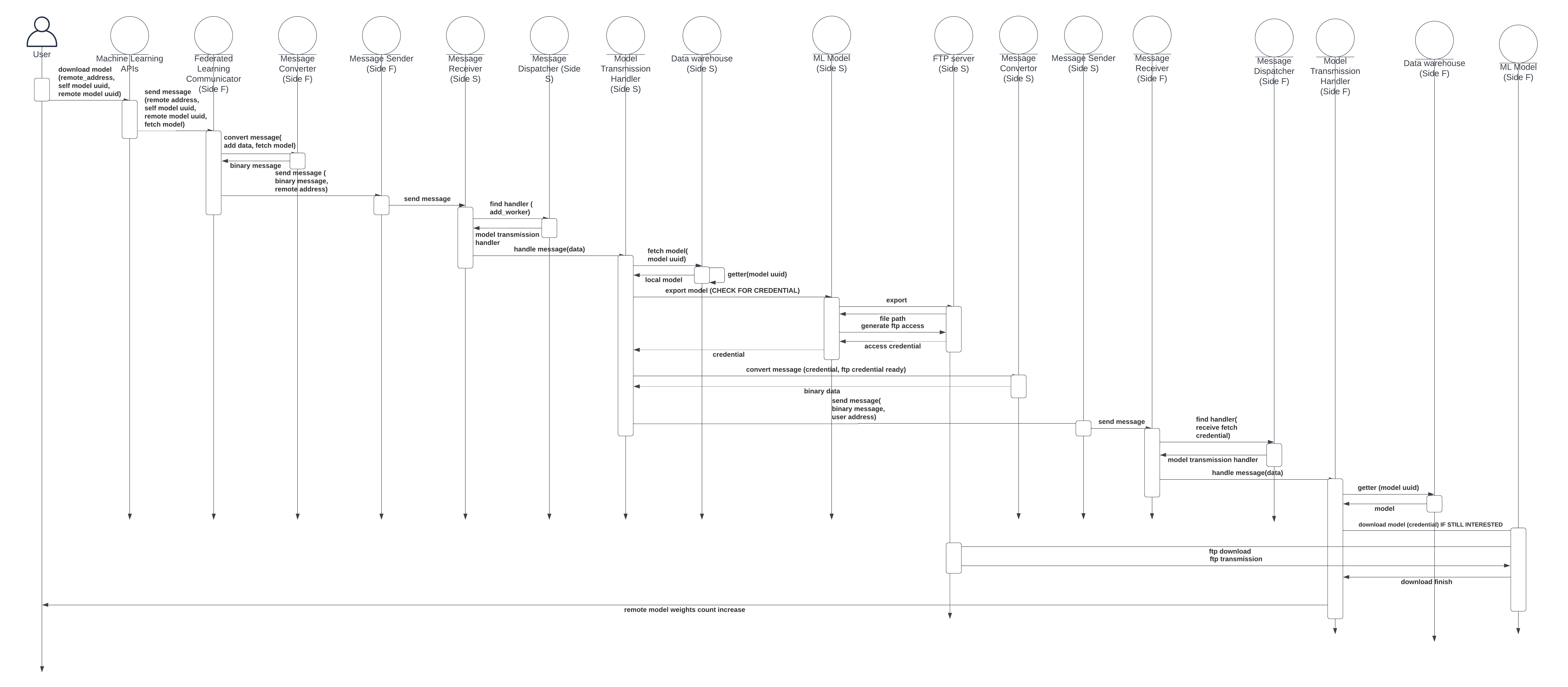}
    \caption{Communicate model sequence diagram}
    \label{fig:3-7-2}
\end{figure}

Figure \ref{fig:3-7-1} shows the component involved in communicating model weights between sites, while figure \ref{fig:3-7-2} shows the message flow between components. For this subsection, let the site which fetches the model weight be side F and the site which sends back model weights to be side S. The steps involve:

1) A model on side F calls fetching model function within the model transmission APIs. Arguments include a pointer referring to the remote model from which the local model wants to fetch weights. Moreover, identification information, such as the unique ID of the model on side F, is also provided.

2) Message is then serialised by a message converter and sent out by the message sender on side F. The message converter will also add an additional tag to indicate that the message is about fetching the model so the remote handler can react correspondingly.

3) When the message receiver on side S gets the message, the dispatcher on that side will forward the message to its own model transmission handlers. The handler will check the pointer side F sends to it and use the ID to retrieve the machine learning model from the data warehouse. 

4) If the model exists, side S will also check whether side S has the privilege to access weights of itself. Since model weights are not shared in public, it has to check for the identity of remote sites fetching it. This this where access control can be changed and extended. The default implementation in this research is by comparing the pointer referring to the model which sends out the fetching weights request against aggregation server model pointers and worker model pointers. In this way, if that remote model fetching the weight is a worker or aggregation server model, then the request to fetch weights will be approved and vice versa.

5) If the access check passes, the model transmission handler will export the model weights to a file in the FTP server. The current implementation uses a local folder as an FTP server drive. However, it can be extended to a remote FTP server by extending the data warehouse storage type as discussed in \ref{subsection:dw}.

6) The model transmission handler on side S will then collect the file name where model weights are stored and login credentials for downloading that file from the FTP server as a response. The login credential is a one-time username and password to the FTP server.

7) The credential will be sent back from side S to side F.

8) After the message receiver on side F get the message and forwards it to the model transmission handler. The handler will first get the local model, which sends out a fetch request. The model will check if it still wants the weights from that remote model. The check is necessary because the model fetching model weights from a remote site might not require that weights anymore when the weights credential comes back. For example, in synchronous federated learning training, the aggregation server will fetch model weights from all active workers when it decides to start the aggregation process. When the aggregation server receives enough responses, it will start aggregating. Suppose any credential for worker model weights arrives after the aggregation process, as the synchronous federated learning already moves on. In that case, the aggregation server will ignore those credentials to download model weights.

9) If the check suggests the local model still wants remote model weights, then the model transmission handler will use the downloader function to log in to the FTP server and download model weights to a local file.

After all steps above, a file containing the latest model weights of a remote model at the time of fetch is generated on side S.

\subsection{Asking a worker to train}
\begin{figure}[h]
    \centering
    \includegraphics[width=\textwidth,height=\textheight,keepaspectratio]{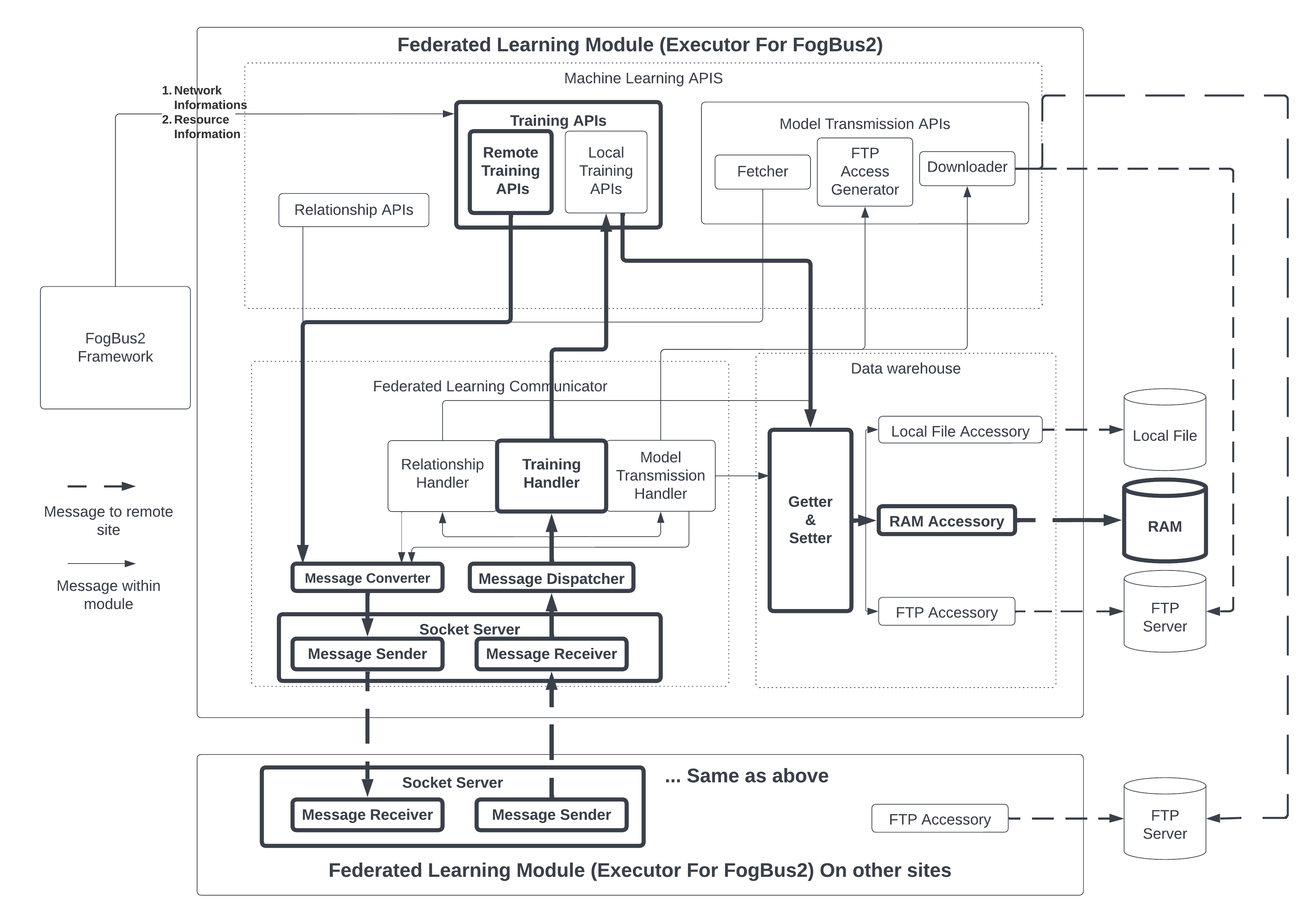}
    \caption{Components for letting worker train}
    \label{fig:3-8-1}
\end{figure}

\begin{figure}[h]
    \centering
    \includegraphics[width=\textwidth,height=\textheight,keepaspectratio]{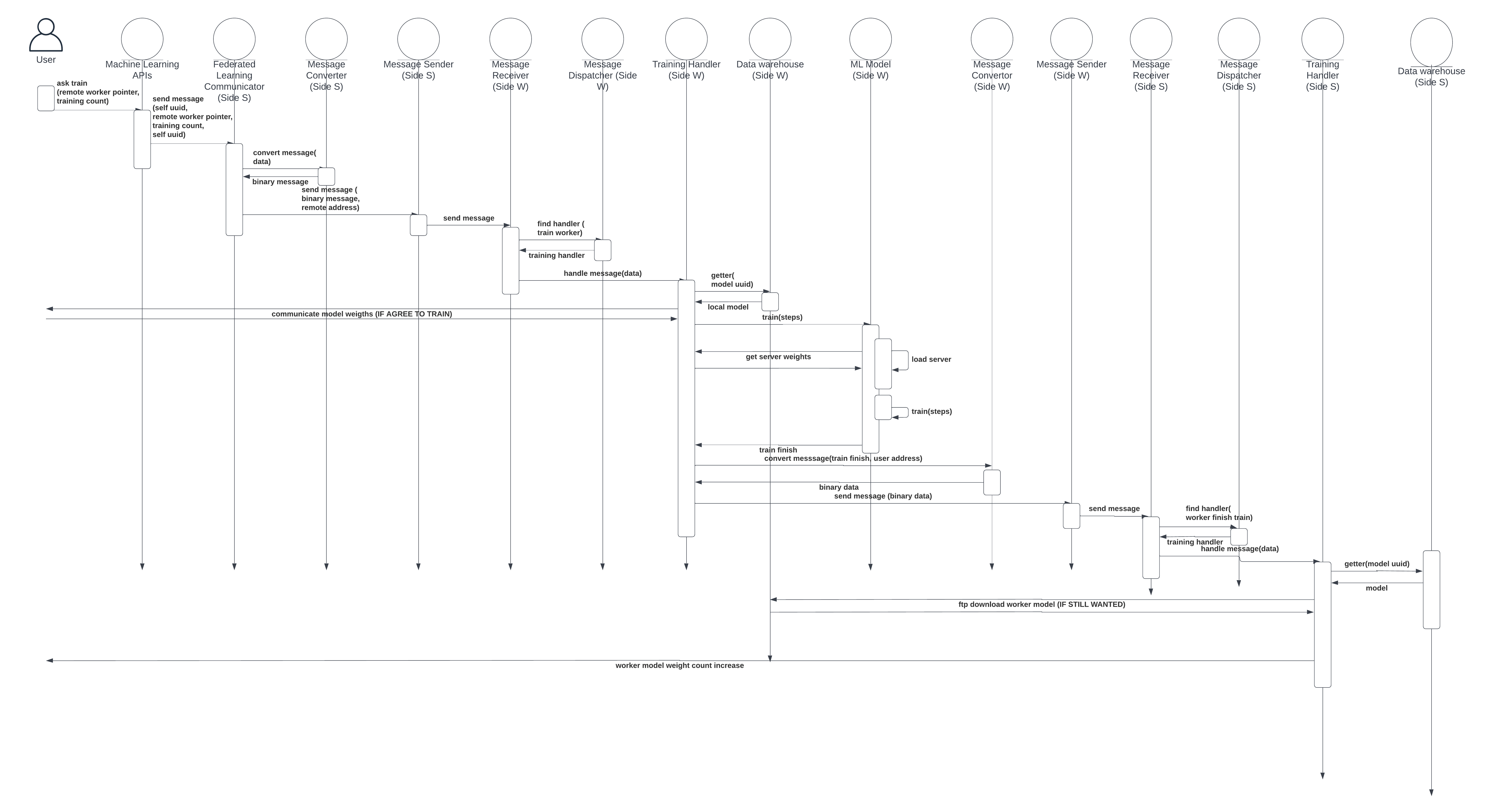}
    \caption{Remote worker train sequence diagram}
    \label{fig:3-8-2}
\end{figure}

Figure \ref{fig:3-8-1} shows the component involved when an aggregation server asks a remote worker to train, and the remote worker train correspondingly. Moreover, figure \ref{fig:3-8-2} shows the message flow between components. For this subsection, let the aggregation server side be side S and the remote worker side be side W. The steps involve:

1) Starting from side S, where the user calls the function from the Training APIs component in an aggregation server model. The argument, including a pointer, points to a remote worker model that asks to train and the number of training epochs.

2) The function call in APIs will then serialise the message by message converter on side S. The serialised message will then be forwarded to site W with a specified network address.

3) After receiving the message that asks the worker model to conduct training, the dispatcher on side W will return the training handler to handle the resting message.

4) The training handler will first retrieve the local worker model based on the pointer side S provided. After that, the training handler will check against the local model whether it agrees to conduct specified training or not. Reasons for rejecting training instructions from a remote site may include the remote site is not recognised by the worker model or there being insufficient computation resources at the moment. The default implementation in this research is that the worker model will check if the remote model on-site S belongs to one of its aggregation server models. This is done by checking if the pointer referring to the aggregation server model is stored in the aggregation server pointer collection of the worker model.

5) If the worker model on side W agrees to conduct specified training epochs, it will fetch the aggregation server model weights. The model weights communication process is described in subsection \ref{subsection:comm_model}.

6) After receiving the aggregation server model weights from side S, the original parameter weights of the worker model will be replaced by aggregation server model weights. Followed by training a specific amount of epochs as specified. The training data comes from local data files, and the implementation support reading from the database for real-time application or a local file for experiment purposes.

7) After completing training, side W will acknowledge that the training is done by sending the message back to side S. The acknowledgement message contains a pointer to the worker model itself and a pointer to the aggregation server.

8) When the acknowledgement is received as side S, the training handler will be responsible for handling the remaining message. It will first retrieve the aggregation server model from the data warehouse module and then check if the aggregation server model still wants results from side W. This is because the aggregation server can finish multiple rounds of aggregation while the worker model on side W conducts training. In this way, model weights from W are outdated and harmful to model accuracy if merged into aggregation server model weights. The default implementation in this research is that, if it is synchronous federated learning, the aggregation server will ignore training results from a worker if it conducts any rounds of aggregation between it ask that worker conduct training and receives an acknowledgement from that worker. If it is asynchronous federated learning, then the aggregation server will take the results no matter how many rounds of aggregation have already been conducted. The criteria for whether accepting results from a worker can be overridden by other logic, which provides an extension point.

9) If the model accepts the model weights from the worker model on side W, it will fetch the model weights, and the process for fetching is defined in subsection \ref{subsection:comm_model}.

After the steps above, the aggregation server side will have a new file containing the model weights of a remote worker model. The aggregation server model can use that weights to update local weights via aggregation.

\subsection{Aggregating worker model weights}
When the aggregation server model receives enough model weights or a reach time limit, it then aggregates the model weights received. The default implementation in this research for synchronous federated learning will wait until a specified amount of model weights are downloaded from workers. The default implementation will start aggregation for asynchronous federated learning once it receives any model weights from any worker. The aggregation process will load those model weights from local files to RAM via the data warehouse module, then federated results based on different aggregation algorithms. Noting that this step does not involve any communication with remote workers since all model weights have been downloaded beforehand. Moreover, during the aggregation process, if some workers respond with their updated model weights, the aggregation server will ignore it or keep those model weights for the next round of aggregation rather than the current aggregation round.

\subsection{A complete federated learning training example}
The code below provides an example of using APIs from the federated learning module to define federated learning behaviour. It defines a synchronous federated learning training process and records the time elapsed and efficiency as training progress. The site running the code is the federated learning server.
\begin{lstlisting}[language=Python, caption=Federated learning module usage example]
def minst_federated_learning_no_cs_even(client_addrs, amount):
    model = minst_classification()
    model.synchronous_federate_minimum_client = amount
    for i in range(3):
        model.add_client(client_addrs[0], (i,i+1))
    for i in range(3,6):
        model.add_client(client_addrs[1], (i,i+1))
    for i in range(6,amount):
        model.add_client(client_addrs[2], (i,i+1))
    while len(model.get_client()) < amount:
        time.sleep(WAITING_TIME_SLOT)
    time_stamp = [time.time()]
    time_diff = [0]
    accuracy = [model.model.accuracy]
    for i in range(100):
        for cli in model.get_client():
            model.step_client(cli, 10)
        while not model.can_federate():
            time.sleep(0.01)
        model.federate()
        time_stamp.append(time.time())
        time_diff.append(time_stamp[-1]-time_stamp[-2])
        accuracy.append(model.model.accuracy.item())
    return time_stamp, time_diff, accuracy
\end{lstlisting}

1) The function takes a bunch of network addresses, potentially the federated learning worker of the current site. The FogBus2 framework provides those network addresses. The other argument indicates the number of workers the aggregation server wants.

2) Line 2 of the code creates a MINST classification model. This model will serve as the aggregation server model.

3) Line 3 defines the number of model weights from workers the aggregation server model has to obtain to proceed to the next round of aggregation. Since this function is testing for synchronous federated learning, thus the amount of response required is set to the number of workers. This means the model will only proceed to aggregation and the next round of training until it receives all workers' responses.

4) Line 4 to 9 sends out relationship invitation which invites remote site to provide worker models of the current aggregation server model. Note that more than one worker model can be placed on the same computing resources. Consequently, for line 4 to 5, for example, the model request three worker models from the site with the network address as the first address in the network address list. Therefore, if the process goes flawless, three worker model will be constructed on a remote site with a network address specified by the first address in the list.

5) Line 10 to 11 wait for acknowledgement from all worker models. When all worker models are constructed successfully, and the acknowledgement reaches the aggregation server side, the worker count will be the same as the amount passed in as an argument.

6) After receiving the desired amount of workers, the aggregation server will start training. It will first call all workers to train in ten steps, as illustrated in lines 16 to 17.

7) Line 18 to 19, then constantly checking whether meeting the criteria to conduct the next round of aggregation. In this case, the criteria are whether the aggregation server obtains updated model weights from all workers.

8) If the aggregation criteria are satisfied, line 20 will conduct federation. Moreover, such a process will be repeated for a hundred rounds, as defined in line 15.

After the iteration, the aggregation server model will contain a trained machine learning model able to classify the MINST data set.

\section{Worker Selection Algorithm Design}\label{section:ws}
This research designed two heuristic algorithms to select workers participating in federated learning. The worker selection algorithm can be applied to synchronous and asynchronous federated learning. Both heuristic algorithm depends on the time required to complete training an entire batch of data for one epoch ${T_{one}}$, the time required to transmit the model ${T_{transmit}}$, however, they differ on the way to compute maximum training time allowed. 

\subsection{R-min R-max based worker selection}
Algorithm \ref{algo:rminmax} demonstrate the first heuristic algorithm.

\begin{algorithm}
\caption{R-min r-max based algorithm}\label{algo:rminmax}
\hspace{2pt} \textbf{Input}: 
\\ 1. $W$: A set of workers
\\ 2. $T_{one_w} \forall w \in W$: Training time required to go through all training data for one epoch
\\ 3. $T_{transmit_w} \forall w \in W$: Time required to communicate model weights
\\ 4. $rmin$: Minimum training epoch
\\ 5. $rmax$: Maximum training epoch
\\
\hspace{2pt} \textbf{Output}: $W_{selected} \in W$
\\1) $T_{min_w} \gets T_{one_w} * rmin + T_{transmit_w} \forall w \in W$
\\2) $T_{max_w} \gets T_{one_w} * rmax + T_{transmit_w} \forall w \in W$
\\3) $T_{minimum} \gets \min{T_{max_w}} \forall w \in W$
\\4) $W_{selected} \gets \forall w \in W: T_{min_w} >= T_{minimum}$
\end{algorithm}

Moreover, after each round of aggregation, $rmin$ and $rmax$ are updated based on average accuracy among all selected workers. Let $accuracy_n$ be the accuracy achieved at the aggregation server on round n, and $accuracy_{n-1}$ be the accuracy achieved last round. Then $rmin$ and $rmax$ are updated like:
\begin{equation}\label{equation:rmin}
rmin \gets rmin * (accuracy_n+1) / (accuracy_{n-1}+1)
\end{equation}

\begin{equation}
\label{equation:rmax}
rmax \gets rmax * (accuracy_{n-1}+1) / (accuracy_{n}+1)
\end{equation}

Firstly, this algorithm takes the training time required for each worker to go through their training data for one epoch as input. Also, time required for communicating the model weights between the aggregation server and a worker is considered. After that, $rmin$ and $rmax$ are two hyperparameters, which define the minimum and the maximum number of epochs a worker should train before sending back the worker model weights to the aggregation server. 

If a worker trains for an insufficient amount of epochs before responding to the aggregation server, then model weights from that worker will have limited differences compared to the original model weights obtained from the aggregation server. $rmin$ is then selected to let the worker contribute model parameter weights with a promising difference. Among different machine learning models and available data sizes, difference $rmin$ needs to be figured out to ensure a meaningful update from workers.

$rmax$, on the other hand, defines the maximum number of epochs a worker can train before responding. If a worker trains for too many epochs before aggregation, then the model weights will be biased to training data that the worker locally has. This will result in a large divergence between the aggregation server's update trend and that worker. In order to keep workers updating model weights in a similar direction as the average direction among all workers, regular communication with the aggregation server is necessary. This makes the algorithm introduce $rmax$ to limit workers from training too many epochs locally.

After selecting $rmax$ and $rmin$, the algorithm calculates the time required to train those amount of epochs plus transmission time for each worker. This time can also be interpreted as after the aggregation server sends out the instruction to conduct training, how long does the aggregation server have to wait until there is a response. Although this time can be varied since estimation for $T_{one_w}$ and $T_{transmit_w}$ has a difference with reality, it provides a heuristic suggesting which worker will respond faster.

The selection criteria intend to minimise the time that fast computing workers wait for slow computing workers. When there is a difference between the time required to finish specified training, fast computing workers can train for more epochs than slow computing workers, which allows them to respond to the aggregation server in a similar time. However, extra training rounds from fast computing workers cannot exceed $rmax$. For slow computing workers, the minimum requirement is complete $rmin$ rounds of training. This suggests the selection described in algorithm \ref{algo:rminmax} line $3) - 4)$. If a worker requires more time to train a minimum amount of epoch compared to the worker that can finish the maximum amount of training, then that worker is excluded. After excluding slow computing workers, it is guaranteed that within the time the fastest computing workers finish maximum epochs of training, all other selected workers can at least complete the minimum training requirement.

In order to achieve time efficiency of training, the worker selection algorithm lets fast computing workers participate in earlier training rounds. After the accuracy reaches a stable value, generally include slow computing workers to achieve higher accuracy. The initial worker selection process guarantee that only fast computing workers are selected. In order to generally include slow computing workers as training proceed, the update will decrease $rmin$ while increasing $rmax$. Increasing the upper limit and decreasing the lower limit has the following impact:

1. Since the maximum number of iterations a worker can train before aggregation increases, this also increases the time required to complete full training rounds for all workers. Consequently, increase the minimum value among those times.

2. In the same way, with decreasing $rmin$, the minimum requirement for workers decreases, decreasing the time required for each worker to complete minimum training requirements.

3. According to the selection criteria, a worker will be selected only when they can finish the minimum required training before the fastest worker completes the maximum allowed amount of training between aggregation on the server side. With decreasing time slow computing workers are required to finish minimum training, and with increasing time fast computing workers need to finish maximum training epochs. Slow computing workers can be included since they can finish minimum training before fast computing workers finish the maximum allowed training amounts.

Consequently, decreasing $rmin$ while increasing $rmax$ as the training progress can provide the effect of fast computing workers joining in an earlier round and slow computing workers joining later, which is time efficient. Training progress is expressed as an increase in the accuracy achieved by aggregated model weights against testing data. Since $rmin$ is updated by multiplying the accuracy of previous aggregation rounds and dividing by the accuracy in the current round,  the more significant increase there is between the accuracy achieved by the aggregation server model in two aggregation rounds, $rmin$ will drop faster and vice versa for $rmax$. 
Furthermore, the formula adjusts the numerator and denominator by adding one. The adjustment avoids the situation in which machine learning model accuracy surges in earlier training rounds. Without the adjustment, when there is a significant increase in accuracy, the factor deciding the decrease in $rmin$ is going to be massive. This will cause $rmin$ to decrease very fast and hit a low value in earlier training rounds. The same will happen on $rmax$ in terms of increase. If $rmin$ reaches a low value and $rmax$ reaches an enormous value, then a large proportion of workers is eligible based on the selection criteria. This causes slow computing workers to be included in training too early. Thus, the numerator and denominator are adjusted by adding one.

\subsection{Discussion of R-min R-max worker selection}
The algorithm addressed above has the potential to accelerate the federated learning training efficiency. However, the design has defects that can fail under specific scenarios.

The first scenario is due to improper initialisation of $rmin$ and $rmax$. If $rmin$ is initialised too low, it requires minimal time to satisfy the minimal training epochs. At the worker selection stage before the first round of training, since all workers will have the relatively low time required to satisfy minimal training epochs, many slow computing workers will be included. Including a large number of inefficient workers is harmful for training efficiency. On the other hand, initialising $rmin$ to a large value will result in a large training time required to satisfy minimum training epochs. This will exclude a large number of workers in the early stage. Although workers selected under large initial $rmin$ are fast responding, inadequate workers participating in early stage training can cause a large time before model accuracy starts to increase. Since $rmin$ only starts to drop once accuracy rises, slow accuracy growth in the early stage delays the time when more workers are included. This overall affects the training efficiency negatively. The same scenario happens when the initialisation of $rmax$ is chosen inappropriately. For every available machine learning model structure and training data, optimal initialisation of $rmin$ and $rmax$ can be derived by grid search. However, a close form solution cannot be derived before training. This makes the worker selection algorithm \ref{algo:rminmax} hard to be applied to large categories of models.

Secondly, the value of $rmin$ and $rmax$ drop and can increase too fast in the early stage. For machine learning training from scratch, model weights are initialised randomly. This cause the initial accuracy of the machine learning model to be relatively low compared to accuracy after a few epochs of training during the early stage of training. Since a significant accuracy increase leads to low $rmin$ and high $rmax$, the large difference between $rmin$ and $rmax$ arises when the accuracy surge happens in an earlier round of machine learning training. As a result, many slow workers will be selected in early rounds, which is time inefficient. If the machine learning model uses a pre-trained model, which can bypass the scenario that accuracy differences are too significant in earlier training rounds. However, this limits the types of machine learning models that can be extended so that only those with a pre-trained model can be considered. Furthermore, the issue of $rmin$ and $rmax$ diverging too quickly gets worse when the federated learning is conducted asynchronously. This is because asynchronous federated learning can aggregate results more frequently, which will cause a more frequent update of $rmin$ and $rmax$ and leads to a large diverge.

Consequently, algorithm \ref{algo:rminmax} has the issue of hard to initialise $rmin$ and $rmax$ such that they do not diverge too quickly. While during training, if the accuracy increases unstably, the value of $rmin$ and $rmax$ will be extremely low and large. This will cause a large proportion of slow workers to be selected, reducing time efficiency. Changing the federated learning from synchronous to asynchronous further worsened the situation.

\subsection{Training time based asynchronous federated learning}
A modified algorithm addressing issues of algorithm \ref{algo:rminmax} is shown in algorithm \ref{algo:tbased}.

\begin{algorithm}
\caption{Training time based worker selection}\label{algo:tbased}
\hspace{2pt} \textbf{Input}: 
\\ 1. $W$: A set of workers
\\ 2. $T_{one_w} \forall w \in W$: Training time required to go through all training data for one epoch
\\ 3. $T_{transmit_w} \forall w \in W$: Time required to communicate model weights
\\ 4. $r$: Worker training iteration
\\ 5. $T$: Time allowed for this round of training
\\
\hspace{2pt} \textbf{Output}: $W_{selected} \in W$
\\1) $T_{total_w} \gets T_{one_w} * r + T_{transmit_w} \forall w \in W$
\\2) $W_{selected} \gets \forall w \in W: T_{total} <= T$
\end{algorithm}

The algorithm selects workers based on the time required to complete a specified amount of training. Moreover, the algorithm will update $T$, which refers to the time allowed for each round of training. Let $W_{ns}$ be the group of workers not selected yet, $accuracy_n$ be the accuracy achieved at the current round of aggregation and $accuracy_{n-1}$ be the accuracy achieved last round. Let $A$ be the accuracy improvement threshold such that $T$ the training time allowed will only increase when the accuracy boost between two rounds of aggregation is less than that threshold. The update can be expressed in equation \ref{equation:tupdate}.
\begin{equation}\label{equation:tupdate}
T \gets \min{T_{total_w} \forall w \in W_{ns}} \text{  if  } accuracy_n - accuracy_{n-1} < A
\end{equation}
The idea of worker selection algorithm \ref{algo:tbased} is that unified epochs of training each worker have to perform before responding to the aggregation server. This allows calculating the total time required to conduct training and communicate model weights back as $T_{total}$. After that, a threshold time is manually selected, which excludes slow computing workers from training. As the training progresses, if the aggregation server model's accuracy stops increasing, more workers are included. This is achieved by increasing the time limit. The worker selection algorithm \ref{algo:tbased} solves issues for algorithm \ref{algo:rminmax}.

Firstly, algorithm \ref{algo:rminmax} has the issue that improper initialisation of hyperparameter $rmin$ and $rmax$ will affect the time efficiency of training. Moreover, close form solutions for optimal initial $rmin$ and $rmax$ cannot be computed. For algorithm \ref{algo:tbased}, the only hyperparameter that needs to be initialised is $T$. The initialisation is straightforward in that $T$ can be set to zero at the start. In this case, no worker can be selected, which will cause the accuracy not to increase. Consequently, the update mechanism in equation \ref{equation:tupdate} of $T$ is triggered. This allows more workers to be eligible for federated learning training. Initialise $T$ to zero or a small value have little impact on time efficiency. This is because if accuracy fails to increase for one epoch due to an insufficient worker selected, $T$ will increase and allow a more significant amount of workers to participate in federated learning training. Those initial training rounds that fail to gain an accuracy boost but just enable the algorithm to choose more workers to have little impact on time efficiency since earlier rounds only contain fast computing workers. Consequently, the worker selection algorithm \ref{algo:tbased} together with the update equation \ref{equation:tupdate} sacrifice little training time to allow the appropriate amount of workers to be selected, which has the potential to boost overall efficiency.

Secondly, $rmin$ and $rmax$ increase fast, causing slow computing workers to join federated learning training in earlier epochs in algorithm \ref{algo:rminmax}. The issue is improved in algorithm \ref{algo:tbased}. This is because algorithm \ref{algo:rminmax} makes more worker eligible to participate in training when accuracy increase, while algorithm \ref{algo:tbased} makes more worker eligible to participate in training only when accuracy stops increasing. Bringing slower workers in only when accuracy stops increasing means slow computing workers are selected only when a converged accuracy is achieved from faster computing workers. This prevents slow computing workers join in the early stage and only allows those slow workers to train when it is necessary to include them to achieve the desired accuracy, which improves time efficiency.

Thirdly, $rmin$ and $rmax$ also diverge fast if federated learning is conducted asynchronously since there is more frequent aggregation and hence more frequent update of $rmin$ and $rmax$. Each update increases the difference between them. Diverge between $rmin$ and $rmax$ leads to slow computing workers being included in early training epochs. Algorithm \ref{algo:tbased}, on the other hand, is compatible with asynchronous federated learning. This is because even when more frequent aggregation is conducted, as long as the model accuracy keeps increasing, update equation \ref{equation:tupdate} will not be triggered, preventing slow computing workers from joining.

The algorithm \ref{algo:tbased} is experimentally evaluated, and its effectiveness for synchronous and asynchronous federated learning will be demonstrated.

\subsection{Estimate time required for training}
Both worker selection algorithm depends on the time required to communicate model weights and conduct training for each worker. While after any worker is selected and the response model weights to the aggregation server, the actual time consumed for communication and training is updated. Initially, the value is estimated by a heuristic. The estimation for training is based on system parameters provided by FogBus2, while communication time is not. 

For the time required to transmit model weights, the estimation in this research transmits the randomly initialised model weights from the aggregation server to each worker to get the time required to transmit. This research does not use network statistics provided by FogBus2 because the implementation uses a separate channel for communication from FogBus2.

For training time, $T_{one}$ the time required to train one epoch is based on CPU availability, and the CPU frequency of all workers is estimated based on CPU frequency and availability termed as $CPU_{freq_w}$ and $CPU_{prop_w}$. The aggregation server will conduct training over one piece of data and record time consumed $T_{onedata}$, as well as CPU frequency allocated to conduct training $CPU_{freq_{server}}$. Moreover, $N_{w}$ the number of training data each worker contain is collected when worker acknowledge they are ready to train. Then $T_{one_w}$ is estimated based on equation \ref{equation:t_one_calculate}

\begin{equation}\label{equation:t_one_calculate}
T_{one} \gets T_{onedata} / CPU_{freq_{server}} * CPU_{freq_w} * CPU_{prop_w} * N_{w} \forall w \in W
\end{equation}

The equation first estimates the time required to train one piece of data on each worker based on the time to train one data on the aggregation server and the multiplier between CPU frequencies. After that, multiplying by the number of data each worker has estimated the time required for each worker to train. 

\lhead{\emph{Experiments}}  
\chapter{Experiments}\label{chapter:4}

\section{Experiment problem statement}
With the implementation described in section \ref{section:fl_implementation} and two worker selection algorithm designs described in section \ref{section:ws}, experiments are conducted to verify the effectiveness of the methods addressed. In particular, the following questions are investigated via the experiment:

1. Can the architecture successfully implement synchronous and asynchronous federated learning?

2. Is the architecture time efficient?

3. Can the worker selection algorithm be designed to improve the training time required to reach a specific accuracy?

4. Does asynchronous federated learning embedded in the architecture help to improve time efficiency compared to synchronous federated learning?

Successfully answering the following question can demonstrate if the architecture and worker selection algorithms are effective.

\section{Experiment Setup}

\subsection{Number of workers}
This research first conducts federated learning with only one worker model, which simulates sequential implementation. After that, the research conducted federated learning with 10 and 30 worker models separately. Those workers are evenly distributed into three different machines.

\subsection{Hardware specification}
For this research, all experiments are conducted on a MacBook Pro (2021) with an M1 pro chip. This research uses the Parallel Desktop to start four virtual machines. One virtual machine runs the aggregation server model, while the rest three runs worker models. Each virtual machine is allocated two gigabytes of RAM while CPU frequencies are evenly distributed into the four virtual machines. Three virtual machines evenly distribute all worker models for different numbers of worker models participating in federated learning. Such a worker virtual machine has three to four federated learning worker models when there are ten in total and ten worker models when there are thirty in total.

All communication is based on the network address each virtual machine has. Each virtual machine is allocated a separate network address for the Parallel Desktop software.

\subsection{Training data set}
This research use MINST data set \cite{deng2012mnist} and CIFAR-10 data set \cite{CIFAR}. Firstly, these two data sets have sufficient data such that both sets have 60000 training data. The rich amount of data can be split into pieces and distributed to different workers for experimental purposes. This ensures all workers have a sufficient amount of distinct training data. The rich amount of data further guarantees that each worker has a unique set of training data even when there are many workers. This supports experiment setup with various amounts of workers. Secondly, those two data sets are also commonly utilised in other research regarding federated learning. Thus, by using those two data set, comparison against other research can be easily conducted.

The amount of training data allocated to each worker model in each experimental setup when there are ten worker models are shown in table \ref{table:ten_worker}. At the same time, data distribution when there are thirty worker models is shown in table \ref{table:thirty_worker}.

\begin{table}[]
\centering
\caption{Batch of data each worker is allocated (10 worker)}\label{table:ten_worker}
\begin{tabular}{|l|l|l|l|l|l|l|l|}
\hline
 & Data set & W1 & W2/W3 & W4 & W5/W6 & W7 & W8/W9/W10 \\ \hline
1 & MINST & 10 & 0 & 0 & 0 & 0 & 0 \\ \hline
2 & MINST & 1 & 1 & 1 & 1 & 1 & 1 \\ \hline
3 & MINST & 1 & 0 & 3 & 0 & 0 & 2 \\ \hline
4 & CIFAR & 100 & 0 & 0 & 0 & 0 & 0 \\ \hline
5 & CIFAR & 10 & 10 & 10 & 10 & 10 & 10 \\ \hline
6 & CIFAR & 10 & 0 & 30 & 0 & 0 & 20 \\ \hline
\end{tabular}
\end{table}

\begin{table}[]
\centering
\caption{Batch of data each worker is allocated (30 worker)}\label{table:thirty_worker}
\begin{tabular}{|l|l|l|l|l|l|l|l|}
\hline
 & Data set & W1 & W2 - W10 & W11 & W12 - W20 & W21 & W22 - W30 \\ \hline
1 & MINST & 30 & 0 & 0 & 0 & 0 & 0 \\ \hline
2 & MINST & 1 & 1 & 1 & 1 & 1 & 1 \\ \hline
3 & MINST & 4 & 0 & 8 & 0 & 0 & 2 \\ \hline
4 & CIFAR & 300 & 0 & 0 & 0 & 0 & 0 \\ \hline
5 & CIFAR & 10 & 10 & 10 & 10 & 10 & 10 \\ \hline
6 & CIFAR & 40 & 0 & 80 & 0 & 0 & 20 \\ \hline
\end{tabular}
\end{table}

The experiment setup one and four in table \ref{table:ten_worker} and table \ref{table:thirty_worker} only allocate training batch of data to one worker model. This is used to simulate sequential training. In contrast, other setups distribute training data to different worker models. Set up two and set up five indicate the situation where each worker model holds an even amount of training data. On the other hand, set up three and set up six denote the case that training data is unevenly distributed. The total amount of data available for training among all workers is the same for setups one to three and setups four to six. With different amount of training data, the time required to complete training will various among workers.

All training is conducted asynchronously and synchronously for one hundred epochs. The long training epoch ensures sufficient time for the aggregation server model to achieve potential accuracy under available workers.

\subsection{Machine learning model}
\begin{lstlisting}[language=Python, caption=Machine Learning Model Used]
class MINST(nn.Module):
    def __init__(self):
        super(CNN, self).__init__()
        self.conv1 = nn.Sequential(
            nn.Conv2d(
                in_channels=1,
                out_channels=16,
                kernel_size=5,
                stride=1,
                padding=2,
            ),
            nn.ReLU(),
            nn.MaxPool2d(kernel_size=2),
        )
        self.conv2 = nn.Sequential(
            nn.Conv2d(16, 32, 5, 1, 2),
            nn.ReLU(),
            nn.MaxPool2d(2),
        )
        self.out = nn.Linear(32 * 7 * 7, 10)
        self.optimizer = optim.Adam(self.parameters(), lr=0.01)
        self.loss_func = nn.CrossEntropyLoss()
        self.accuracy = 0
        
class CIFAR(torch.nn.Module):
    def __init__(self):
        super(Net, self).__init__()
        self.conv1 = torch.nn.Conv2d(3, 16, 5)
        self.conv2 = torch.nn.Conv2d(16, 32, 5)
        self.pool = torch.nn.MaxPool2d(2, stride=2)
        self.fc1 = torch.nn.Linear(32 * 5 * 5, 120)
        self.fc2 = torch.nn.Linear(120, 84)
        self.fc3 = torch.nn.Linear(84, 10)
        self.loss_func = torch.nn.NLLLoss()
        self.optimizer = torch.optim.SGD(self.parameters(), lr=0.005)
        self.accuracy = 0
\end{lstlisting}
The machine learning model used for MINST and CIFAR-10 classification is shown in the code above. This is the model shared by all workers and the aggregation server participating in federated learning. The model structure is chosen in that data available from any worker individually is insufficient to achieve an accuracy of over fifty per cent, while data available from all workers together can produce more than fifty training accuracy.

\subsection{Data collected}
This research train the MINST and CIFAR-10 models under a federated learning mechanism. After each round of aggregation, the aggregation server tests the resulting model against testing data. This accuracy is collected. Moreover, the time required between the aggregation server asking selected workers to start to train and the aggregation server finishing merging results, which is the time needed for each aggregation rounds, is recorded. Thus the accuracy achieved over time under federated learning training is collected. This data is used to compare the time effectiveness of training between different set ups in table \ref{table:ten_worker} and table \ref{table:thirty_worker}.

\section{Experiment Results}
\subsection{Accuracy overtime}
\begin{figure}
\centering
\subfigure[MINST 10 workers]{\includegraphics[width=0.45\linewidth]{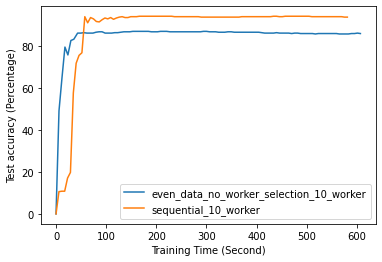}}\qquad
\subfigure[MINST 30 workers]{\includegraphics[width=0.45\linewidth]{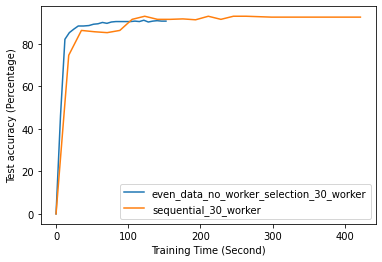}}\\
\subfigure[CIFAR 10 workers]{\includegraphics[width=0.45\textwidth]{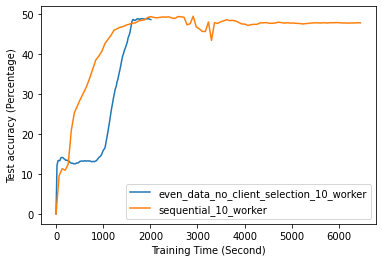}}\qquad
\subfigure[CIFAR 30 workers]{\includegraphics[width=0.45\textwidth]{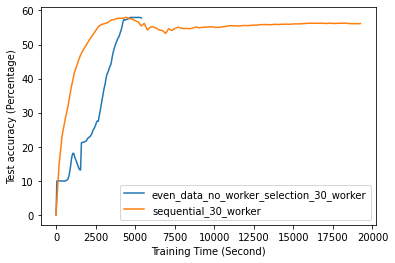}}
\caption{Sequential training VS FL training(even data distribution, no worker selection)}
\label{fig:sequential_vs_even}
\end{figure}

\begin{figure}
\centering
\subfigure[MINST 10 workers]{\includegraphics[width=0.45\linewidth]{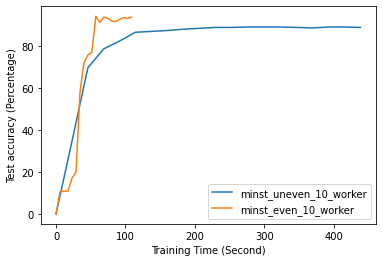}}\qquad
\subfigure[MINST 30 workers]{\includegraphics[width=0.45\linewidth]{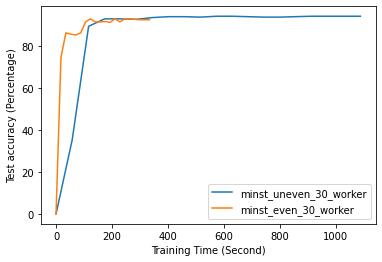}}\\
\subfigure[CIFAR 10 workers]{\includegraphics[width=0.45\textwidth]{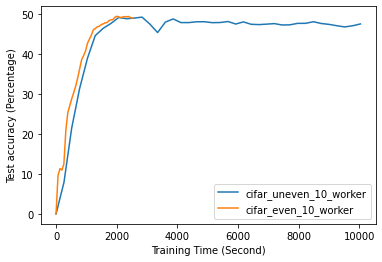}}\qquad
\subfigure[CIFAR 30 workers]{\includegraphics[width=0.45\textwidth]{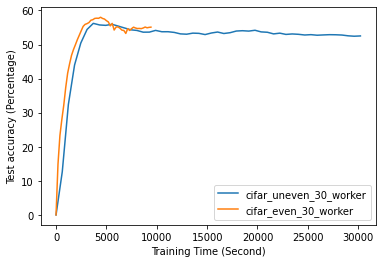}}
\caption{Even vs uneven data distribution}
\label{fig:even_vs_uneven}
\end{figure}

\begin{figure}
\centering
\subfigure[MINST 10 workers]{\includegraphics[width=0.45\linewidth]{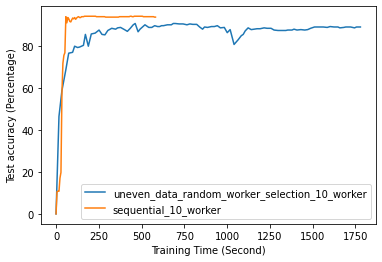}}\qquad
\subfigure[MINST 30 workers]{\includegraphics[width=0.45\linewidth]{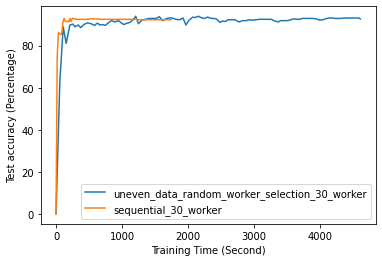}}\\
\subfigure[CIFAR 10 workers]{\includegraphics[width=0.45\textwidth]{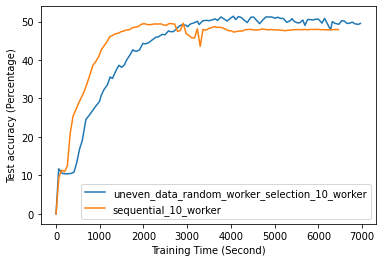}}\qquad
\subfigure[CIFAR 30 workers]{\includegraphics[width=0.45\textwidth]{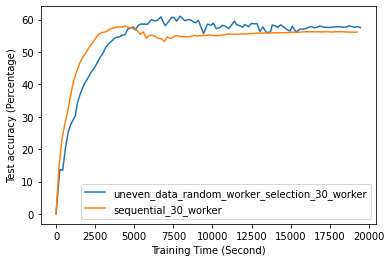}}
\caption{Random worker selection vs Sequential}
\label{fig:rws_vs_sequential}
\end{figure}

\begin{figure}
\centering
\subfigure[MINST 10 workers]{\includegraphics[width=0.45\linewidth]{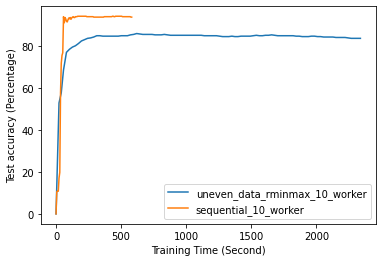}}\qquad
\subfigure[MINST 30 workers]{\includegraphics[width=0.45\linewidth]{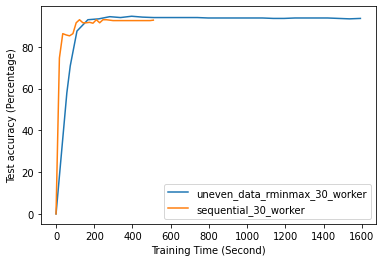}}\\
\subfigure[CIFAR 10 workers]{\includegraphics[width=0.45\textwidth]{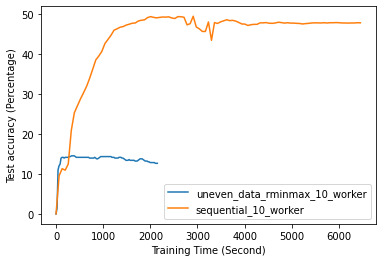}}\qquad
\subfigure[CIFAR 30 workers]{\includegraphics[width=0.45\textwidth]{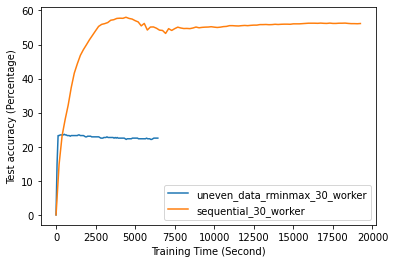}}
\caption{R-min r-max worker selection vs Sequential (rmin, rmax initialise to 5)}
\label{fig:rminmax_vs_sequential}
\end{figure}

\begin{figure}[h]
    \centering
    \includegraphics[width=\textwidth,height=\textheight,keepaspectratio]{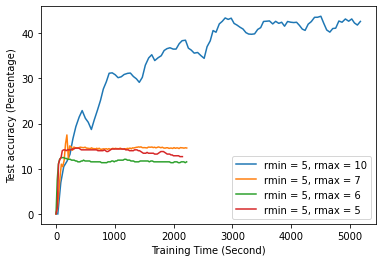}
    \caption{R-min r-max worker selection with different r-max initialisation}
    \label{fig:r-min-r-max-initialisation}
\end{figure}

\begin{figure}
\centering
\subfigure[MINST 10 workers]{\includegraphics[width=0.45\linewidth]{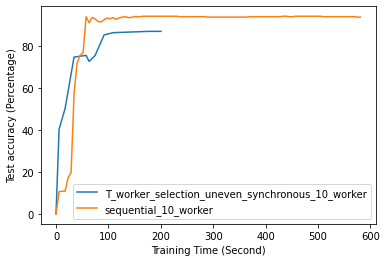}}\qquad
\subfigure[MINST 30 workers]{\includegraphics[width=0.45\linewidth]{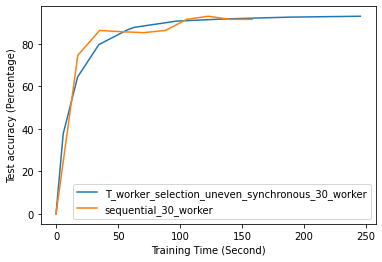}}\\
\subfigure[CIFAR 10 workers]{\includegraphics[width=0.45\textwidth]{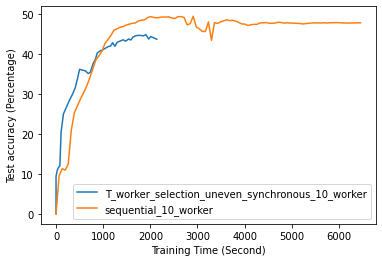}}\qquad
\subfigure[CIFAR 30 workers]{\includegraphics[width=0.45\textwidth]{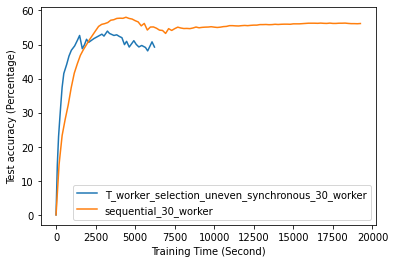}}
\caption{Algorithm \ref{algo:tbased} worker selection (synchronous) VS Sequential}
\label{fig:syn_t_ws_vs_s}
\end{figure}

\begin{figure}
\centering
\subfigure[MINST 10 workers]{\includegraphics[width=0.45\linewidth]{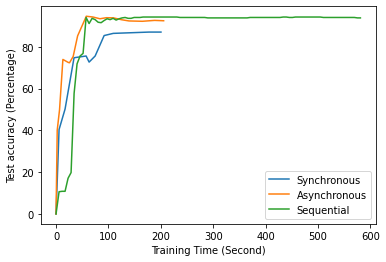}}\qquad
\subfigure[MINST 30 workers]{\includegraphics[width=0.45\linewidth]{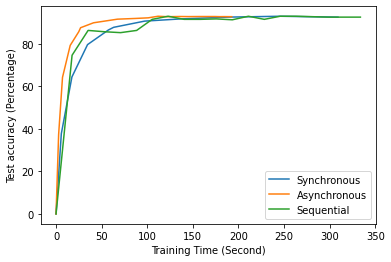}}\\
\subfigure[CIFAR 10 workers]{\includegraphics[width=0.45\textwidth]{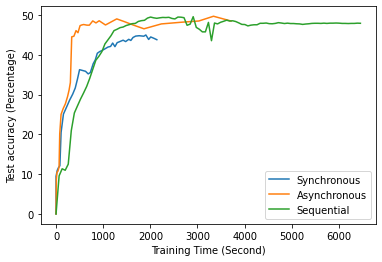}}\qquad
\subfigure[CIFAR 30 workers]{\includegraphics[width=0.45\textwidth]{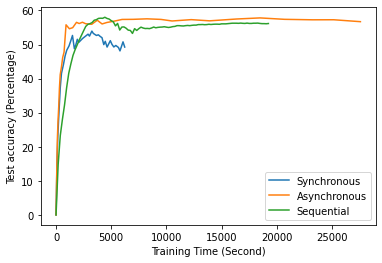}}
\caption{Algorithm \ref{algo:tbased} worker selection (synchronous) VS (asynchronous) VS Sequential}
\label{fig:asyn_vs_syn_vs_s}
\end{figure}

Figure \ref{fig:sequential_vs_even} to figure \ref{fig:asyn_vs_syn_vs_s} illustrate the results of federated learning training under setups in table \ref{table:ten_worker} and table \ref{table:thirty_worker}. In particular, the following findings are found based on experiment results:

1. Federated learning with even data distribution requires less time to reach stable accuracy, which means faster training at the initial stage. But sequential will eventually reach a better accuracy at last. This can be seen from figure \ref{fig:sequential_vs_even}, such that the blue line indicates federated learning training accuracy over time, and the orange line indicates improvements in the accuracy of sequential training over time. The blue line reaches a high value before the orange line. However, the orange exceeds the value of the blue line as time progress. This demonstrates the finding in the first point.

2. The time required for sets of workers that have even or uneven amounts of training data to each stable accuracy is similar. There is no particular increase or decrease in time efficiency. By comparing the rise in accuracy as time from figure \ref{fig:even_vs_uneven}, it can be seen that both lines have a similar growing trend.

3. Random worker selection will eventually reach the same accuracy level as sequential implementation. However, random worker selection requires a longer time to get the same accuracy as sequential implementation. This can be seen from figure \ref{fig:rws_vs_sequential} such that the blue line indicating random worker selection consistently achieve stable value after the orange line representing sequential implementation. Moreover, it can be seen that the accuracy growth of random worker selection is unstable compared to sequential execution.

4. The r-min and r-max worker selection algorithm failed to be more time efficient than sequential implementation. This can be seen from figure \ref{fig:rminmax_vs_sequential}. Moreover, incorrect initialisation of $rmin$ and $rmax$ can lead to failed training process such that accuracy never approaches the potential accuracy achievable based on all data from all workers. The failed training can be seen from figure \ref{fig:r-min-r-max-initialisation} that when $rmax$ is initialised to seven, six or five, the accuracy stays around fifteen per cent. In contrast, theoretical accuracy is around fifty per cent which is much higher.

5. The worker selection algorithm \ref{algo:tbased}, when combined with synchronous federated learning training, will beat random worker selection and sequential machine learning training during the early phase of training. This shows the effectiveness of the worker selection algorithm \ref{algo:tbased}. However, such lead only takes part in an early phase of training, while sequential training always reaches stable accuracy faster.

6. As shown in figure \ref{fig:asyn_vs_syn_vs_s}, the worker selection algorithm \ref{algo:tbased} combined with asynchronous federated learning training, has similar performance in the earlier phase. However, during a later stage of training, asynchronous federated learning has faster accuracy growth. This makes asynchronously federated learning and worker selection algorithm \ref{algo:tbased} more time efficient than synchronous federated learning or sequential machine learning training.

\subsection{Results evaluation}
Firstly, the first finding is that federated learning training is more time efficient than sequential training, given that the total amount of data held by all workers in federated learning training is the same as the amount of data used by sequential training. This demonstrates that federated learning architecture in this research is successful and time efficient. The federated learning training is faster because all workers conduct training simultaneously, which is more time efficient than sequential training that goes through all training data single-threaded. However, an interesting result is that when there are ten or thirty worker models, the training time is only reduced by a minor factor. This is because the accuracy increase is not linear related to the number of workers participating in federated learning. For example, when there are ten workers and each of them holds one piece of data, their local training will only increase the accuracy by a small per cent so does the merged model weights. However, for a worker holding ten pieces of data, the accuracy increase under a fixed round of training is much higher than the merged model weights since training benefits from a larger range of data. Thus, although federated learning requires less time for each round of training, the accuracy increases each round is smaller than sequential training that benefits from a more extensive range of data. This explains why the time efficiency boost does not grow linearly with number of workers added.

Secondly, the finding is that random worker selection will reach a similar level of accuracy as sequential training with a longer training time. This is because the random selection can select slow-computing workers in the earlier round of training, causing other fast-computing workers to wait for it to finish. The waiting time then influence the overall training efficiency. Moreover, unstable accuracy increases because the total amount of data available from selected workers is inconsistent under random worker selection.

Thirdly, the worker selection algorithm \ref{algo:rminmax} failed to outperform sequential federated learning. Moreover, the algorithm will fail under improper initialisation of $rmin$ and $rmax$. The algorithm is time inefficient because $rmin$ and $rmax$ diverge too fast in the early stage of machine learning training. Since model weights are randomly initialised, there will be significant accuracy growth in the earlier round of training. This accelerates the update of $rmin$ and $rmax$ based on update formula \ref{equation:rmin} and causes increasing diverge. When $rmin$ and $rmax$ separate, slow computing workers are also considered since the time required to complete minimal training is much less than the time necessary for fast computing workers to meet maximally allowed training epochs. Consequently, the worker selection algorithm \ref{algo:rminmax} quickly turns into selecting all workers after a few rounds of aggregation on the server. This causes slow computing workers to be considered when only asking fast computing workers to train can also increase accuracy fast, which is time inefficient. The scenario that training failed to approach theoretical accuracy demonstrated in figure \ref{fig:r-min-r-max-initialisation} further shows the constraint of the algorithm design. It can be seen that when initially $rmin$ and $rmax$ are closed to each other, the worker selection algorithm \ref{algo:rminmax} causes the federated learning to fail to train. This is because if $rmin$ and $rmax$ are close to each other, the time required for slow computing workers to finish training $rmin$ epochs is larger than the time necessary for fast computing workers to complete $rmax$ amount of training, causing a large number of workers not selected initially. However, as insufficient workers were included in the training, the accuracy failed to grow with limited data available from those chosen workers. If accuracy stays similar and does not grow, formula \ref{equation:rmax} and formula \ref{equation:rmin}, $rmin$ and $rmax$ will not update. This stops extra workers from joining the training, while selected workers fail to improve accuracy further and stop more workers from joining. Consequently, the training then stuck at a low accuracy range, which fails to train as illustrated in figure \ref{fig:r-min-r-max-initialisation}.

Lastly, synchronous and asynchronous federated learning, when combined with the worker selection algorithm \ref{algo:tbased}, shows better time efficiency in the earlier training phase. However, in the later stage of training, the training accuracy of synchronous federated learning is exceeded by sequential machine learning training, while asynchronous federated learning keeps efficient and becomes the most time efficient one. The worker selection algorithm \ref{algo:tbased} is more time efficient in the early phase compared to sequential machine learning training because only fast computing workers are selected to participate in training. When the federated learning training requires slow computing worker to join in later rounds of training, since synchronous federated learning requires fast computing workers to wait for slow computing workers, it becomes time inefficient as training progress to later rounds. This is also proved by the fact that the accuracy of sequential training exceeds it in later training rounds. On the other hand, asynchronous federated learning, even when including slow computing workers, does not require fast computing workers to wait. This causes asynchronous federated learning, including more workers while avoiding slow down training due to waiting for slow computing workers. That causes asynchronous federated learning outperforms sequential and synchronous training in terms of time efficiency. 

\lhead{\emph{Conclusions}}  
\chapter{Conclusions and Future Directions}\label{chapter:5}
\section{Contributions}
In this research, a federated learning framework is implemented which can work on top of FogBus2 or stand along. The framework enable new machine learning model extended easily on it and support different mechanism relating to worker selection access control and storage implementation be extended easily. Moreover, a worker selection algorithm is introduced, this worker selection algorithm improves the training time efficiency. Experiments through the federated learning implementation further verified that.
\subsection{Federated learning implementation}
\textbf{This research implement an highly extensible federated learning framework, where extension for new machine learning model, new worker selection strategy and new machine learning storage mechanism can be added by only extending single functions.} 
\\

The highly abstract framework introduced in this research enable easy extension of mechanism related to federated learning. New machine learning model can be extended as long as import and export model weights as well as merging model weights are defined. Moreover, worker selection strategy can be extended by overriding the worker selection function skeleton, where multiple system parameters relating to available workers are available for different algorithm design. Lastly, the flexible model storage design allowing the model weights being stored on different media, which supports deployment on various computing resources. The implementation also support asynchronous federated learning. The easy extension and lightweight property of the framework makes it a good tool for comparing different federated learning mechanism design.

\subsection{Worker selection algorithm}
\textbf{This research introduced a worker selection algorithm \ref{algo:tbased} which can improve training time efficiency. Moreover, with the asynchronous federated learning support, training time efficiency further improved.}
\\

In order to train a MINST classification model to a 80\% accuracy. The synchronous federated implementation with worker selection algorithm \ref{algo:tbased} is 33.9\% faster than sequential training where all data is collected into one place and trained. Moreover, when the worker selection algorithm \ref{algo:tbased} is combined with asynchronous federated learning, the training time further reduced 63.3\% which demonstrate the effectiveness of asynchronous federated learning as well as worker selection algorithm \ref{algo:tbased} in reducing training time. For CIFAR-10 classification, the worker selection algorithm \ref{algo:tbased} is 59.0\% faster than sequential training and training time reduced further by 36.4\% when combined with asynchronous federated learning.

\section{Future Directions}
\subsection{Extending more optimisation algorithms \ref{algo:tbased}}
Although this research experiment test the worker selection algorithm against sequential federated learning or random worker selection, it is worth comparing against other state of art federated learning optimisation mechanism. Since the federated learning framework in this research allow other mechanism be easily extended, mechanisms addressed other federated learning research can be implemented. Based on implementations on same framework, comparing training time efficiency can further demonstrate the effectiveness of worker selection algorithm \ref{algo:tbased} which is worth doing as next step. Moreover, the current framework provides the opportunity to integrate other machine learning techniques in a distributed manner, such as some state-of-the art distributed deep reinforcement learning techniques for dynamic scheduling of resources \cite{goudarzi2021distributedDDRL}.

\subsection{Extending security module to the framework}
Regarding the federated learning implementation, one thing ignored in this research is security measures related to federated learning. Although federated learning in nature protects data privacy as data never leave its origination, model weights shared to remote sites still can leak information about training data. Thus extra modification on model weights shared is necessary to prevent any sensitive information regarding training data is leaked out through the format of machine learning model weights. This is important since protect training data privacy is the key goal of federated learning which emphasise the important to extend a security module based on current federated learning framework.

\subsection{Further research on asynchronous federated learning grouping with worker selection}
This research proposed the worker selection algorithm \ref{algo:tbased} combined with asynchronous federated learning, which both demonstrate time effectiveness. However, the worker selection algorithm and asynchronous federated learning mechanism are separately developed and putted together without modification. It is worth exploring how worker selection can be tuned to achieve better time effectiveness given that it is going to be applied to asynchronous federated learning.

\subsection{Container Orchestration based on the proposed federated learning mechanisms}

When the number of requests for federated learning tasks increases, the framework requires high scalability to offer appropriate service. To enable the high scalability of the current implemented federated learning framework, it can be deployed on container orchestration platforms such as Kubernetes and K3S. Since FogBus2 is a containerized resource management framework, it can be deployed using container orchestration platforms such as Kubernetes or K3S \cite{wang2023container}. So, integration of this federating learning mechanism with container orchestration platforms is a meaningful next step to provide high scalability. 









\addtocontents{toc}{\vspace{2em}}  
\backmatter

\label{Bibliography}
\lhead{\emph{Bibliography}}  
\bibliographystyle{unsrtnat}  
\bibliography{Bibliography}  

\end{document}